\documentclass[12pt]{article}

\usepackage{amsfonts,latexsym}
\usepackage{epsfig}
\usepackage{amsfonts,latexsym,graphics}
\usepackage{float}
\usepackage[usenames]{color}
\input xy
\xyoption{all}
\def\bk{{\bf k}}
\def\br{{\bf r}}

\begin{document} 

\title{
Exact solution of the $p+ip$ pairing Hamiltonian and a hierarchy of integrable models.  
}

\author{Clare Dunning$^{(1)}$, Miguel Iba\~nez$^{(2)}$,  Jon Links$^{(3)}$,  \\
Germ\'an Sierra$^{(2)}$  and Shao-You Zhao$^{(3)}$  \\
~~\\
$^{(1)}$School of Mathematics, Statistics and Actuarial Science, \\
The University of Kent, CT2 7NZ, UK, \\
$^{(2)}$Instituto de F\'{\i}sica Te\'orica, UAM-CSIC, \\
Cantoblanco, 28049 Madrid, Spain \\
$^{(3)}$Centre for Mathematical Physics, School of Mathematics and Physics,\\
 The University of Queensland 4072,
 Australia,}

\maketitle
\begin{abstract}
Using the well-known trigonometric six-vertex solution of the Yang-Baxter equation we derive an integrable pairing Hamiltonian with anyonic degrees of freedom. The exact algebraic Bethe ansatz solution is obtained using standard techniques. From this model we obtain several limiting models, including the pairing Hamiltonian with $p+ip$-wave symmetry. An in-depth study of the $p+ip$ model is then undertaken, including a mean-field analysis, analytic and numerical solution of the Bethe ansatz equations, and an investigation of the topological properties of the ground-state wavefunction. Our main result is that the ground-state phase diagram of the $p+ip$ model consists of three phases. There is the known boundary line with gapless excitations that occurs for vanishing chemical potential, separating the topologically trivial strong pairing phase and the topologically non-trivial weak pairing phase.  We argue that a second boundary line exists separating the weak pairing phase from a  topologically trivial weak coupling BCS phase, which includes the Fermi sea in the limit of zero coupling. The ground state on this second boundary line is the Moore-Read state.     
\end{abstract}



\vfil\eject

\begin{tableofcontents}
\end{tableofcontents}

\vfil\eject

\section{Introduction}

Exactly solvable many-body models have played a very important role
in various branches of physics, especially in statistical mechanics and condensed matter. 
They have been fundamental in the understanding of phase transitions and non-perturbative
phenomena in low-dimensional and strongly correlated systems. Prominent examples
are provided by the 2D Ising model, the 1D anisotropic Heisenberg model, the interacting 1D Bose gas, and the 1D Hubbard model 
\cite{b82,kbi93}. 
The solvability  and integrability of these and related  models  emerges from a common algebraic
structure, due to the quantum Yang-Baxter equation satisfied by the so called $R$-matrix. 
The $R$-matrix is the basic object of the Quantum Inverse Scattering Method (QISM) 
\cite{kbi93}. It allows for the  construction of a one parameter family of commuting  transfer matrices whose
expansion,  in powers of the spectral parameter,  generates
 an infinite set of  conserved quantities including the Hamiltonian of the system. 
For the anisotropic Heisenberg model, also known as the XXZ model,   the $R$-matrix is the six-vertex trigonometric solution
of the Yang-Baxter equation and it depends on a spectral parameter $u$ and 
the quantum parameter $q$  related to the
anisotropy as $\Delta = (q^2+ q^{-2})/2$.  

A closely related family of exactly solvable models are the Richardson-Gaudin models,
where the most prominent examples are the Richardson model 
of BCS superconductivity \cite{crs97,lzmg03,r63,vp02,zlmg02}
and the Gaudin spin models \cite{g95}  (for a review see \cite{dps04}). 
The common algebraic structure of this family is the classical Yang-Baxter equation
 satisfied by the classical $r$-matrix. This latter equation is obtained from the
 quasi-classical limit of the quantum Yang-Baxter equation through $ R(u) \sim  1 + \hbar r(u)$. A feature of these
models, unlike the anisotropic Heisenberg model,  is the dependence of the Hamiltonian
on a large number of parameters. 
 In the Richardson model, they are the energies of the single particle levels
 and the BCS coupling constant, while in 
the Gaudin model they are the position of the spins.  
This abundance of parameters has its origin in the  transfer
matrix which is the product of $R$-matrices with shifted spectral parameters. This is called
an inhomogeneous transfer matrix.  
 This situation must be compared with the anisotropic Heisenberg  model
whose transfer matrix is homogeneous, which is the reason why the Hamiltonian of this model
only depends on $q$. 

Thus the majority of models constructed so far
are based  either on a homogeneous transfer matrix (e.g. anisotropic Heisenberg model),  or  in the quasi-classical limit
of an inhomogeneous transfer matrix (Richardson-Gaudin models). It is also of interest to
construct  physical models from  an inhomogenous transfer matrix before one takes 
the quasi-classical limit. The first result of this paper is to construct such a model
starting from the trigonometric six-vertex $R$-matrix discussed above. This model
describes the pairing interaction of  anyonic particles with  braiding statistics 
in momentum space parameterized by $q$,  and for that reason it can be termed an anyonic pairing model.
These anyons should be distinguished from the real space anyons  which satisfy
braiding statistics in 1D \cite{bgh07} or in 2D  \cite{lm77,w82}. 

\begin{figure}[t!]
\begin{center}
\includegraphics[height=3.5cm,angle= 0]{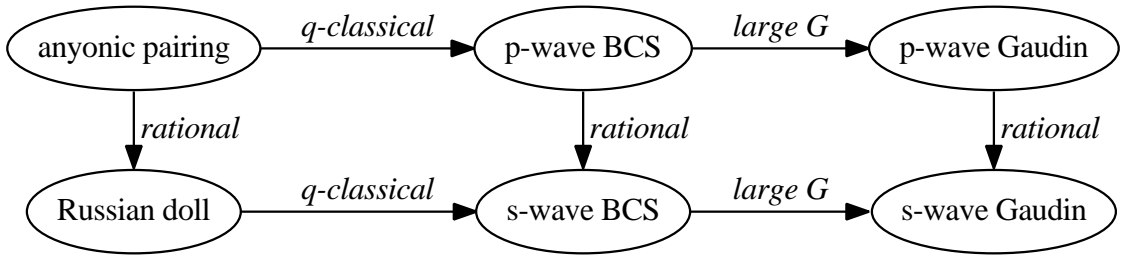}
\end{center}
\caption{\footnotesize{
Web of integrable pairing models.}}
\label{models}
\end{figure}

The one-body version of the anyonic pairing Hamiltonian we obtain coincides with the 
quantum mechanical Hamiltonian proposed by Glazek and Wilson  in 2002, 
which furnishes a simple example of a  renormalization group with limit cycles \cite{gw02,gw04}. 
The latter work  inspired the construction of  a  BCS model with $s$-wave symmetry
and time reversal symmetry-breaking where the renormalization group also has limit
cycles \cite{lrs04}. The mean-field gap equation of this BCS model has an infinite number 
of solutions related by discrete scale invariance, which explains  its name:
Russian doll BCS model. This model was shown in reference \cite{dl04}  to be exactly solvable
by the QISM using a rational $R$ matrix. The relation between the mean-field and the exact
solution of this model was established in reference \cite{als05}.  It then comes as no surprise
that the Russian doll BCS model can be obtained from the anyonic pairing model in the
limit where the trigonometric $R$-matrix becomes rational. However, if the  quasi-classical
limit of the anyonic pairing model is taken before the rational limit a BCS model with $p+ip$-wave symmetry is obtained, as has been reported in  \cite{ilsz09}. 
Taking a rational limit of the latter model or a quasi-classical limit of the Russian doll
model one arrives to the well known $s$-wave BCS model of Richardson \cite{r63}. Finally,
one can take the limit where the BCS coupling constant, of both the $s$- wave and $p+ip$-wave
models, goes to infinity obtaining two Gaudin models. 
The web of relations between all these models is shown in Fig. (\ref{models}).

The first part of this paper, Sect. 2,  is devoted to the construction of  
 the anyonic pairing model using the QISM and the derivation of the eigenvalues of the transfer matrix.
 In Sect. 3  the relationships of the models in Fig. \ref{models} are shown,  
 giving a unified picture of the pairing models that can be obtained from the trigonometric $R-$ matrix.
 In Appendix A we collect the more technical 
 computation of the wavefunction and scalar products of the anyonic model,
 as well as its relation with the aforementioned Glazek-Wilson model. 
 
 The bulk of the paper, Sect. 4,  is devoted to a detailed study of 
 the exactly solvable $p+ip$-wave model which was initiated in \cite{ilsz09}. 
 In recent years this model has received considerable
attention due to the existence of vortices with non-trivial braiding statistics \cite{rg00,i01,v88},
with the potential for implementing topological quantum computation \cite{nssfd08}.  
Furthermore,  there are some experimental indications
that $Sr_2 Ru O_4$  is  such a chiral p-wave superconductor \cite{kb09},  supporting topological
vortices \cite{dnt06,jbm09}. In the realm of cold atoms there are also possibilities to realize
$p+i p$-wave  paired phases  \cite{cs09,gr07,hetal09,lcg08} through experiments involving 
Feshbach resonaces \cite{fetal08,gsbj07,ietal08,n09,ztld08,zetal04}. 

We begin our study of the $p+ip$ model using  the 
  BCS mean-field theory,  to determine 
 the phase diagram in the plane $(1/g,x)$, where $1/g$ is 
the inverse of the  BCS coupling constant and 
$x$ is  the fraction of occupied energy levels. This phase diagram is divided into 
 three regions denoted as weak coupling, weak pairing and strong pairing in the terminology
introduced by Read and Green \cite{rg00}. The strong and weak pairing regions are separated by 
a line in the $(1/g,x)$ plane where the chemical potential vanishes,  giving rise
to a second-order phase transition and a topological
transition described by a winding number  in momentum space  introduced by Volovik \cite{v88}. 
The latter two regions are related by a strong-weak duality which connects two ground
states having the same total energy, and value of the gap, while the chemical potential
differs in the sign.  On the other hand, 
the weak pairing  and weak coupling regions are separated by another line in the $(1/g,x)$ 
plane which can be called the Moore-Read line because the projected BCS state
coincides precisely with the Moore-Read pfaffian state related to the Fractional Quantum Hall
state at filling fration 5/2. All the points of this line have zero total energy
and are mapped, under the duality transformation,  into the vaccum state. However
the gap does not vanish on this line and the nature of the transition is rather subtle since,
as we show below, it is  associated to a discontinuity of the ground state energy, or in another words
a zeroth order phase transition. Volovik has discussed the existence of a Higgs like phase transition
for the $p+ip$ model,  meaning a qualitative change in the dispersion relation of the quasi-particles \cite{v07}.
We find that this type of transition occurs on a line located in the weak coupling region, which
is close, but not identical,  to the Moore-Read line.  A peculiarity of the Read-Green and the Moore-Read
lines is that they are insensitive to  the particular choice of the energy levels, and in that sense
they are topological. This feature is not shared by the Volovik line whose position in the phase diagram 
depends on the energy level distribution. 

The analysis of Sect. 4 continues with the computation
of the fidelity susceptibility which confirms, as expected, the singular nature of the Read-Green line,
while the Moore-Read line does not show any special  feature. Next we analyze the continuum limit
of the Bethe ansatz equations of the $p+ip$ model using an electrostatic analogy which allows
us to uncover the complex structure of the arcs formed by the roots of the Bethe ansatz equations. We also recover
the mean-field gap and chemical potential equations as was done in previous works for the BCS model
with $s$-wave symmetry \cite{g95,rsd02,r77}. The structure of the arcs, the values of the ground state energy, gaps, chemical
potential and occupation numbers  are compared with the numerical results
obtained by solving the Bethe ansatz equations for systems with finite size, finding a resonable agreement. 
A detailed  analysis of the exact solution allow us to understand the weak-strong duality 
in terms of a dressing operation that maps exact eigenstates between those two 
regions. The physical picture that emerges is that the ground state of the weak pairing region
is formed by strongly localized pairs and by delocalized pairs with zero energy.  
We then discuss the zeroth order phase transition and the winding number of the mean-field
and exact solutions. For the mean-field solution the winding number vanishes
in the strong pairing region while it does not in the weak pairing and weak coupling regions.
However in the exact solution the winding number only vanishes in the weak pairing region
where it counts the number of Bethe ansatz equation  roots that  are equal to zero. In this sense
the winding number plays the role of a topological order parameter for the weak pairing region. 
Another qualitative distinction between the weak pairing and weak coupling regions 
is derived by studying the structure of the vortex wavefunction by means of the 
Bogolioubov-de Gennes equations. We find that 
 the vortex wavefunction  has an oscillatory behaviour in the weak coupling region
which does not appear in the weak pairing region.

Finally,  in  Appendices A and B we discuss several technical issues: 
 relation between  the Glazek-Wilson model 
and the anyonic model, 
computation of wavefunction scalar  products and correlation functions of the anyonic pairing and the $p+ip$ models,
a  $q$-deformed version of the Moore-Read state, solution of the gap and chemical potential
equations.

\section{Construction of an integrable anyonic pairing model}

\subsection{The Yang--Baxter equation}
A systematic method to construct integrable quantum systems is through the Quantum Inverse Scattering Method (QISM) \cite{kbi93}. The key element in this approach is a solution of the Yang--Baxter equation \cite{b82,j90}, commonly known as an $R$-matrix. This is a linear operator  acting   on a two-fold tensor product space which is dependent on $x\in\mathbb{C}$ known as the spectral parameter. Denoting the $R$-matrix as $R(x)\in{\rm End}(V\otimes V)$, the Yang--Baxter equation reads 
 \begin{eqnarray}
 R_{12}(x/y)R_{13}(x)R_{23}(y)=R_{23}(y)R_{13}(x)R_{12}(x/y)
 \label{ybe}
 \end{eqnarray}
which acts on the three-fold space $V\otimes V\otimes V$. The subscripts above refer to the spaces on which the operators act. E.g.
$$R_{12}(x)=R(x)\otimes I. $$
The specific example we will use is the well-known $XXZ$ solution associated with the quantum algebra $U_q[sl(2)]$\footnote{It is convenient for our purposes to express the deformation parameter as $q^2$ rather than the more familiar $q$.} \cite{jm95}
\begin{eqnarray*} 
R(x)
&=& \left(\begin{array}{ccccc} 
(q^2x-q^{-2}x^{-1}) &0&|&0&0 \\
0&(x^{}-x^{-1} )&|&{(q^2-q^{-2})}&0 \\ 
-&-&~&-&- \\ 
0&{(q^2-q^{-2})}&|&(x^{}-x^{-1} )&0 \\
0&0&|&0&(q^2x-q^{-2}x^{-1})  \end{array} \right).  
\end{eqnarray*}
We define the local $L$-operator as $L(x)=R(xq^{-1})$ which is a solution of 
\begin{eqnarray}
R_{12}(x/y)L_{13}(x)L_{23}(y)=L_{23}(y)L_{13}(x)R_{12}(x/y)
\label{lax}
\end{eqnarray} 
as a consequence of (\ref{ybe}).  
It is customary to consider $L(x)\in{\rm End}(V\otimes V)$ as acting on the tensor product of an ``auxiliary'' space and a local ``quantum'' space. We use the subscript $a$ to denote the auxiliary space and the labels $j=1,...,L$ for the local quantum spaces of a many-body system. 

To apply the QISM to construct a pairing Hamiltonian we next introduce the hard-core boson 
(Cooper pair) operators in terms of fermion operators. Let $\{c^\dagger_{j\sigma},c_{j\sigma}:j=1,...,L,\sigma=\pm\}$, 
denote a set of creation and annihilation operators with the usual canonical anticommutation relations and 
where the labels $\pm$ refer to time-reversed pairs. The hard-core boson operators are defined 
\begin{eqnarray*}
b_j = c_{j-}c_{j+}, \qquad b_j^\dagger = c^\dagger_{j+}c^\dagger_{j-}.
\end{eqnarray*}
We also set $N_j=b^\dagger_jb_j$ as the Cooper pair number operator such that the total number of Cooper pairs in a system is given by the eigenvalue of 
$$N=\sum_{j=1}^L N_j. $$
Throughout we will use $M$ to denote the eigenvalues of $N$. 

For each $j$ the local Hilbert space of states is four-dimensional. Excluding  the subspace of states which contain unpaired fermions reduces the local Hilbert spaces to dimension two, spanned by the vacuum $|0\rangle$ and the paired state $b^\dagger_j|0\rangle$.   
Now expressing the linear operators of the quantum space of the $L$-operator in terms of hard-core boson operators we have   
\begin{eqnarray*}
L_{aj}(x)&=& x\left(\begin{array}{cc} 
q^{2N_j-I} & 0 \\
0 & { q^{I-2N_j}} 
\end{array} \right) 
+ (q^{2}- q^{-2} )
\left(\begin{array}{cc} 
0 & b_j \\
b^{\dagger}_j &  0 
\end{array} \right) 
-x^{-1} \left(\begin{array}{cc} 
q^{I-2N_j} & 0 \\
0 &  q^{2N_j-I} 
\end{array} \right) 
\end{eqnarray*}    
noting that
$$q^{\gamma N_j}= (q^{\gamma}-1)N_j + I. $$
Using the $L$-operator in this form, we will next apply the QISM to construct an integrable pairing Hamiltonian with anyonic degrees of freedom.
 
\subsection{Transfer matrix, Hamiltonian, and exact solution}

The most common use of the QISM is to construct a transfer matrix on a closed chain (to adopt the language of spin models). Here we use this approach with a ``twist'' in the boundary conditions, which will be chosen in a sector dependent manner as described below. The quantum space for such a model is the $L$-fold tensor product space $W=V^{\otimes L}$. We introduce the monodromy matrix $T(x)\in{\rm End}(V\otimes W)$ defined by 
$$T(x) = U_a L_{a1}(xz_1^{-1})L_{a2}(xz_2^{-1})....L_{aL}(xz_L^{-1}) $$
where 
$$
U= \left( \begin{array} {cc} 
\exp(-i\alpha') & 0 \\
0 & \exp(i\alpha') 
\end{array}  \right). 
$$
The monodromy matrix is commonly expressed in a matrix form as
\begin{eqnarray*}
T(x)= \left( \begin{array} {cc} 
T^1_1(x) & T^1_2(x) \\
T^2_1(x) & T^2_2(x) 
\end{array}  \right). 
\end{eqnarray*} 
where the entries are operators acting on $W$.  
As a result of (\ref{lax}) and 
$$\left[U\otimes U,\,R(x)\right]=0$$
it follows that 
\begin{eqnarray}
R_{ab}(x/y)T_{a}(x)T_{b}(y)=T_{b}(y)T_{a}(x)R_{ab}(x/y)
\label{mono}
\end{eqnarray}
which is an operator equation on $V\otimes V \otimes W$, with the two auxiliary spaces  labelled by $a$ and $b$. 

Defining the transfer matrix to be 
\begin{eqnarray*}
t(x)={\rm tr} \left(T(x)\right)=T^1_1(x)+T^2_2(x). 
\label{tm}
\end{eqnarray*}
it follows from (\ref{mono}) that the transfer matrices form a commutative family; i.e.
\begin{eqnarray}
\left[t(x),\,t(y)\right]=0\qquad\forall\,x,\,y \,\in\,\mathbb{C}. 
\label{tmcommute}
\end{eqnarray}  
The transfer matrices can be expanded in a Laurent series
$$t(x)=\sum_{j=-L}^L t^{(j)}x^j$$ 
and because of (\ref{tmcommute}) the co-efficients commute
\begin{eqnarray*}
\left[t^{(j)},\,t^{(k)}\right]=0, \qquad \,-L\leq j,\,k \leq L.
\end{eqnarray*} 
One may check directly that the transfer matrix commutes with $N$, so the Hilbert space of states can be decomposed into sectors labelled by the eigenvalues $M$ of $N$.  The leading order terms in the expansion are 
\begin{eqnarray*}
t^{(L)}&=& \left(\prod_{j=1}^L z_j^{-1} \right)
\left(\exp(-i\alpha')q^{2N-L} + \exp(i\alpha') q^{L-2N}\right) \\
t^{(L-1)}&=&0. 
\end{eqnarray*}
We will construct a Hamiltonian in terms of the operator $t^{(L-2)}$, where the choice of the variable $\alpha$ will depend on the sector labelled by $M$. Setting
\begin{eqnarray*}
q&=&\exp(i\beta) \\
\alpha&=& \alpha'+\beta(L-2M+2)
\end{eqnarray*}
we define
\begin{eqnarray*}
H&=& \frac{1}{4\sin(2\beta)\sin(\alpha)}\left[t^{(L-2)}\prod_{j=1}^Lz_j+2\cos(\alpha)\sum_{j=1}^L z_j^2\right].
\end{eqnarray*}
The Hamiltonian may be expressed as 
\begin{eqnarray}
H&=&   \sum_{j=1}^L z_j^2 N_j -\frac{\sin(2\beta)}{\sin(\alpha-2\beta) }\sum_{k>r}^L z_k z_r \left(\exp(-i\alpha)
 d_r^\dagger d_k + {\rm h.c.} \right)    
\label{ham}
\end{eqnarray}
with 
\begin{eqnarray}
d_j&=& b_{j} \prod_{k=j+1}^{L} q^{4N_k-2I},  \nonumber  \\
d^\dagger_j &=& b_{j}^\dagger \prod_{k=j+1}^{L} q^{2I-4N_k},  
\label{string}
\end{eqnarray}
$\alpha,\,\beta$ are assumed to be real and h.c. denotes hermitian conjugate.
Due to the non-local action of the operators $d_j^\dagger,\,d_j$
they satisfy
\begin{eqnarray*}
d_{j} d_{k} &=& q^{4} d_{k} d_{j} \qquad\qquad\,\,\, j>k \nonumber \\
d^\dagger_{j} d_{k} &=& q^{-4} d_{k} d^\dagger_{j} \qquad\qquad j>k 
\end{eqnarray*} 
amongst other relations, and therefore can be viewed as anyonic creation and annihilation operators. 
By the nature of the above construction, the anyonic Hamiltonian is integrable as it commutes with all terms in the expansion of the transfer matrix. That is 
\begin{eqnarray*}
\left[H,\,t^{(j)}\right]=0,\qquad j=-L,...,L
\end{eqnarray*}
establishing that the $t^{(j)}$ provide a set of conserved operators\footnote{In principle for integrability to hold one would need to check that the number of independent conserved operators is not less that the number of degrees of freedom.}.

Recall that the derivation of the Hamiltonian was made using a restricted subspace of the Hilbert space obtained by excluding unpaired states. By noting that the second term in (\ref{ham}), which describes the scattering of Cooper pairs, vanishes on unpaired states\footnote{In standard BCS theory this phenomenon is well-known and goes by the name of the {\it blocking effect}, e.g. see \cite{vr01}).}   it is straightforward to extend the action of the Hamiltonian to the full $4^L$-dimensional Hilbert space. 
We define anyonic creation and annihilation operators $\{a_{j\pm}^\dagger,\,a_{j\pm}: j=1,...,L\}$ which are defined in terms of the canonical fermion operators through 
\begin{eqnarray*}
a_{j\sigma}&=& c_{j\sigma} \prod_{k=j+1}^{L} q^{2n_{k\sigma}-I},  \nonumber  \\
a^\dagger_{j\sigma} &=& c_{j\sigma}^\dagger \prod_{k=j+1}^{L} q^{I-2n_{k\sigma}}  
\end{eqnarray*}
where $n_{j\sigma}=c^\dagger_{j\sigma} c_{j\sigma}$. These operators satisfy the relations
\begin{eqnarray}
\{a_{j\sigma},\,a_{j\rho}\}=\{a_{j+},\,a_{k-}\}&=&0, \\
\{a_{j\sigma}, \, a^\dagger_{j\rho}\}&=&\delta_{\sigma\rho}I, \\
a_{j\sigma} a_{k\sigma} &=& - q a_{k\sigma} a_{j\sigma} \qquad\qquad j>k \\
a^\dagger_{j\sigma} a_{k\sigma} &=& - q^{-1} a_{k\sigma} a^\dagger_{j\sigma} \qquad\quad j>k 
\label{braid-anyons}
\end{eqnarray} 
and those relations obtained by taking hermitian conjugates. The usual fermionic commutation relations are recovered in the limit $q\rightarrow 1$.  
Then the Hamiltonian 
\begin{eqnarray*}
H&=&   \frac{1}{2}\sum_{j=1}^L z_j^2 \left(n_{j+}+n_{j-}\right) -\frac{\sin(2\beta)}{\sin(\alpha-2\beta) }\sum_{k>r}^L z_k z_r \left(\exp(-i\alpha)
 a_{r+}^\dagger a_{r-}^\dagger a_{k-}a_{k+} + {\rm h.c.} \right)    
\end{eqnarray*}
is an extension to the $4^L$-dimensional Hilbert space which has the same action as (\ref{ham}) when the unpaired states are excluded.

Having constructed the Hamiltonian through the QISM it is a standard calculation to obtain the exact solution through the algebraic Bethe ansatz. We simply present the key results. Explicit details relevant to the derivation of the formulae below may be found in \cite{grs96,kmt99,kbi93,vp02}. 

The first step is to identify that the vacuum state $\left|0\right>$ admits the properties 
\begin{eqnarray*}
T^1_1(x)\left|0\right>&=&a(x)\left|0\right>, \\
T^1_2(x)\left|0\right>&=&0, \\
T^2_1(x)\left|0\right>&\neq&0, \\
T^2_2(x)\left|0\right>&=&d(x)\left|0\right>
\end{eqnarray*} 
where 
\begin{eqnarray*}
a(x)&=& \exp(-i\alpha')\prod_{k=1}^L (q^{-1}xz_k^{-1}-qx^{-1}z_k),   \\
d(x)&=& \exp(i\alpha')\prod_{k=1}^L(qxz_k^{-1}-q^{-1}x^{-1}z_k). 
\end{eqnarray*}
We then look for eigenstates of $t(x)$ in the form
\begin{eqnarray*}
\left|\Phi(Y)\right>=\prod_{j=1}^M T^2_1(y_j)\left|0\right>
\end{eqnarray*}
where $Y=\{y_j\}$. Note that 
\begin{eqnarray*}
\left[T^2_1(y_j),\,T^2_1(y_k)\right]=0
\end{eqnarray*}
so the operators may be ordered arbitrarily. Using the algebraic relations amongst the $T^i_j(x)$ operators, this leads to the eigenvalues $\Lambda(x)$ of the transfer matrix being given by  \cite{grs96,kmt99,kbi93,vp02} 
\begin{eqnarray*}
\Lambda(x) &=& \exp(-i\alpha')\prod_{k=1}^L (q^{-1}xz_k^{-1}-qx^{-1}z_k) 
\prod_{j=1}^M \frac{ q^2 x^2 - q^{-2}y_j^2  }{ x^2 -  y_j^2 } \\
&&  ~~~+\exp(i\alpha')\prod_{k=1}^L(qxz_k^{-1}-q^{-1}x^{-1}z_k)\prod_{j=1}^M \frac{ q^{-2} x^2 - q^2y_j^2   }{  x^2 -y_j^2 } \\
&=& \exp(-i\alpha)\prod_{k=1}^L (xz_k^{-1}-q^2x^{-1}z_k) 
\prod_{j=1}^M \frac{ x^2 - q^{-4}y_j^2  }{ x^2 -  y_j^2 } \\
&&  ~~~+\exp(i\alpha)\prod_{k=1}^L(xz_k^{-1}-q^{-2}x^{-1}z_k)\prod_{j=1}^M \frac{ x^2 - q^4y_j^2   }{  x^2 -y_j^2 }
\end{eqnarray*}
such that the parameters $\{y_j\}$ satisfy the Bethe ansatz equations
\begin{eqnarray}
\exp(-2i\alpha)\prod_{k=1}^L\frac{1-q^2 y_m^{-2}z^2_k}{1-q^{-2}y_m^{-2}z^2_k}
=\prod_{j\neq m}^{M} \frac{ 1- q^4 y_m^{-2}y_j^2 }{1- q^{-4} y_m^{-2}y_j^2 }
\qquad\qquad m=1,...,M. 
\label{bae}
\end{eqnarray} 
To obtain the energy expression $E$ for the Hamiltonian (\ref{ham}), it is a matter of expanding the transfer matrix eigenvalues in the Laurent series
$$\Lambda(x)=\sum_{j=-L}^L \Lambda^{(j)} x^j$$ 
giving 
\begin{eqnarray}
E&=& \frac{1}{4\sin(2\beta)\sin(\alpha)}\left[\Lambda^{(L-2)}\prod_{j=1}^Lz_j+2\cos(\alpha)\sum_{j=1}^L z_j^2\right] \nonumber \\
&=& \frac{\sin(\alpha)}{\sin(\alpha-2\beta)}\sum_{j=1}^M y^2_j.
\label{nrg}
\end{eqnarray}
We leave to Appendix \ref{scalarproduct} the computation of the scalar product and correlators
 for this model.   

From the form of the Hamiltonian (\ref{ham}) and the definition of the operators $d_j$, eq. (\ref{string})
one can see that the model parameters can be restricted to the domain $0 \leq \alpha, \beta \leq \pi$.
If  $\alpha =0$ the Hamiltonian (\ref{ham}) reduces to 
\begin{eqnarray*}
H&=&   \left(\sum_{r=1}^L z_r d_r^\dagger \right)\left(\sum_{k=1}^L
 z_k d_k \right).    
\end{eqnarray*}
Formally  eq. (\ref{nrg}) yields that  the energies are zero. However in this limit the roots $y_j$ can be divergent, such that non-zero energies do occur.  

It is possible to generalise the above construction in a number of ways. Instead of using the closed chain transfer matrix construction, one can adopt the open chain version developed by Sklyanin \cite{s88}. In this manner a Hamiltonian is obtained which not only involves scattering interactions between the anyonic pair operators given by (\ref{string}), but also their complex conjugates. A second route is to use different solutions of the Yang--Baxter equation. An obvious choice would be to employ Baxter's eight-vertex solution \cite{b71,b82}. But this solution suffers from the fact that it is not $U(1)$ invariant which prohibits the construction of a Hamiltonian which conserves particle number. Beyond this there are however  many known solutions of the Yang-Baxter equation which do possess $U(1)$ symmetries, particularly those associated with representations of the quantum algebras $U_q[g]$ where $g$ is a classical simple Lie algebra (e.g. see \cite{j86}). In principle these can be applied for the construction of models, which in particular cases will provide anyonic generalisations of integrable pairing models based on bosonic or fermionic degrees of freedom.

\section{Limiting cases}

\subsection{Russian doll and $s$-wave BCS models}

An example of a many-body system which admits a cyclic Renormalization Group (RG) equation analogous to the Glazek-Wilson Hamiltonian (see (\ref{gw}) of Appendix \ref{cyclic}) was studied in \cite{lrs04}. The many-body Hamiltonian was taken to be the $s$-wave BCS model with complex-valued coupling parameter, later to become known as the Russian Doll (RD) BCS model. The cyclic RG was discovered through a mean-field analysis. Subsequently in \cite{dl04} it was proved that the RD BCS model is integrable, and in \cite{als05} the connection between the Bethe ansatz solution and the RG equation was exposed. Next we will show that the RD BCS model can be obtained as a limiting case of the anyonic pairing model (\ref{ham}). Moreover the exact solution for the usual $s$-wave BCS model, which was first found in \cite{r63}, is obtained by performing a second limiting procedure.  

We introduce a new variable $\eta$ and define the parameters $\varepsilon_k,\,k=1,...,L$ and $v_j,\,j=1,...,M$ through 
\begin{eqnarray*}
z_k&=& \exp(2\beta \varepsilon_k/\eta)  \\
y_j&=&\exp( 2\beta v_j/\eta )
\end{eqnarray*}    
and redefine the Hamiltonian (\ref{ham}) by a simple rescaling and additional conserved term 
\begin{eqnarray*}
H \mapsto \frac{\eta}{\sin(2\beta)}(H - N)
\end{eqnarray*}
so the energy expression (\ref{nrg}) becomes 
$$E=\frac{\eta \sin(\alpha)}{\sin(\alpha-2\beta)\sin(2\beta)}\sum_{j=1}^M y^2_j -\frac{\eta}{\sin(2\beta)}M. $$ 
Now we take what is known as the {\it rational limit} $\beta \rightarrow 0$ to obtain the following expressions for the Hamiltonian, Bethe ansatz equations, and energy respectively
\begin{eqnarray}
H&=&   2\sum_{j=1}^L \varepsilon_j N_j -G\sum_{k>r}^L  \left(\exp(-i\alpha)
 b_r^\dagger b_k + {\rm h.c.} \right) \label{rd1}      \\ 
\exp(-2i\alpha)\prod_{k=1}^L \frac{v_j-\varepsilon_k-i\eta/2}{v_j-\varepsilon_k+i\eta/2} &=& \prod_{m\neq j}^M\frac{v_j-v_m-i\eta}{v_j-v_m+i\eta},\qquad j=1,...,N \label{rd2}\\ 
E&=&2\sum_{j=1}^M v_j +GM\cos(\alpha). \label{rd3}
\end{eqnarray} 
where $G=\eta/\sin(\alpha)$. This is the exact solution for the Russian Doll BCS model in terms of the parameterisation used in \cite{dl04}. Applying the same limiting procedure to the expressions (\ref{slavnov},\ref{averagen}) derived in Appendix \ref{scalarproduct} yields the wavefunction scalar product and one-point functions given in \cite{dl04}. 

Performing a second limiting procedure to the Russian Doll Hamiltonian yields the $s$-wave Hamiltonian. To make this result explicit we implement the following change of variable $\alpha \rightarrow \eta \alpha$ and add the conserved quantity $-GN$ to the Hamiltonian (\ref{rd1}). Taking the {\it quasi-classical limit} $\eta \rightarrow 0$ we obtain from (\ref{rd1},\ref{rd2},\ref{rd3})  
\begin{eqnarray}
H&=&   2\sum_{j=1}^L \varepsilon_j N_j -G\sum_{k,r=1}^L  
 b_r^\dagger b_k  \label{r1}      \\ 
\frac{2}{G}+\sum_{k=1}^L \frac{1}{v_j-\varepsilon_k} &=& \sum_{j\neq m}^M\frac{2}{v_m-v_j},\qquad j=1,...,M  \label{r2}\\ 
E&=&2\sum_{j=1}^M v_j  \label{r3}
\end{eqnarray} 
 where $G=1/\alpha$. Up to a change in notational conventions this is precisely the exact solution first given in \cite{r63}.

\subsection{The $p+ip$-wave BCS model} \label{pwave}

The Hamiltonian of the BCS pairing model with $p+ip$-wave symmetry was introduced
in reference \cite{ilsz09} 
\begin{eqnarray}
&H& = \sum_\bk \frac{\bk^2}{2 m} \;  c^\dagger_\bk c_\bk - \frac{G}{4 m} \sum_{\bk \neq \pm \bk'}  (k_x - i k_y) (k'_x + i k'_y) \; 
 c^\dagger_\bk  c^\dagger_ {-\bk }  c_{- \bk'} c_{ \bk'}
\label{pipham}
\end{eqnarray}
where $c_\bk, c^\dagger_\bk$ are destruction and creation operators
of  two-dimensional polarised fermions with momentum $\bk=(k_x,k_y)$, 
$m$ is their mass and $G$ is a dimensionless coupling constant
which is positive for an attractive interaction.  We remark that we impose no constraint on the choice for the ultraviolet cut-off, which we denote as $\omega$. Likewise the distribution of the momenta $\bk$ is arbitrary other than the assumption that all momentum states arise in time-reversed pairs $\bk$ and $-\bk$. In particular this means that a one-dimensional system is obtained by simply setting all $k_y=0$. 

We shall  discuss now how this Hamiltonian can be obtained from 
another limiting case of the Hamiltonian (\ref{ham}), which  establishes the integrablity of the $p+ip$-wave BCS model. The strategy here is to take the quasi-classical limit first, rather than the rational limit. While this general approach has previously appeared in the literature to construct integrable Hamiltonians in the context of Gaudin algebras \cite{admor02,ado01,dps04,des01}, the connection with the $p+ip$ model has thus far not been made apparent for reasons we will discuss below.       

Introducing the parameterisation
\begin{eqnarray*}
\alpha &=& -i\gamma  t \\
\beta  &=&  -i\gamma p \\
\end{eqnarray*}
and taking the limit $\gamma\rightarrow 0$ leads to the Hamiltonian
\begin{eqnarray*}
H&=&   \sum_{j=1}^L z_j^2 N_j -\frac{2p}{t-2p }\sum_{k> r}^L z_k z_r 
 \left(b_r^\dagger b_k +{\rm h.c.}\right)      
\end{eqnarray*}
with the Bethe ansatz equations and energies given respectively by
\begin{eqnarray}
-2t-4p\sum_{k=1}^L \frac{z_k^2}{y_m^{2}-z_k^2} &=& -8p\sum_{j\neq m}^M\frac{y_j^2}{y_j^2-y_m^2}, \label{triggaudin} \\
E&=& \frac{t}{t-2p}\sum_{j=1}^M y_j^2 \nonumber
\end{eqnarray} 
Next we set 
\begin{eqnarray}
G=\frac{2p}{t-2p} =  \frac{2}{\alpha/\beta -2}
\label{gcoupling}
\end{eqnarray}
 in terms of which we have  
\begin{eqnarray}
H&=&   \sum_{j=1}^L z^2_j N_j -G\sum_{k>r}^L {z_k z_r}\left( 
 b_r^\dagger b_k + {\rm h.c.} \right),  \label{pham} \\ 
\frac{G^{-1}-L+2M-1}{y_m^2}+\sum_{k=1}^L \frac{1}{y_m^2-z^2_k} &=& \sum_{j\neq m}^M\frac{2}{y_m^2-y_j^2},
\label{pbae}  \\
E&=& (1+G)\sum_{j=1}^M y_j^2. \label{pnrg}
\end{eqnarray}
To show that (\ref{pham}) is equivalent to (\ref{pipham}), we first 
define the  Cooper pair operators 
\begin{eqnarray*}
\tilde{b}^\dagger_\bk = c^\dagger_\bk  c^\dagger_ {-\bk }, \qquad 
\tilde{b}_\bk = c_{-\bk}  c_ {\bk }, \qquad
\tilde{N}_\bk =\tilde{b}^\dagger_\bk \tilde{b}_{\bk},
\label{a2} 
\end{eqnarray*}
which have odd symmetry (consistent with $p$-wave pairing): 
\begin{eqnarray*}
\tilde{b}_{-\bk} = - \tilde{b}_{\bk}, \qquad 
\tilde{N}_\bk = \tilde{N}_{-\bk}.
\label{a3}
\end{eqnarray*}
We let ${\bf K}_+$ denote the 
set of momenta where $k_x > 0$ and no restriction placed on $k_y$, 
so that we avoid issues with double counting. Excluding the  unpaired states and setting $m=1$, the Hamiltonian
(\ref{pipham}) takes the form  
\begin{eqnarray}
{H} = \sum_{\bk\in{\bf K}_+} z_\bk^2  \;  \tilde{N}_\bk  
- G  \sum_{\bk\neq\bk'\in{\bf K}_+}  z_\bk z_{\bk'} \; 
\exp({ -i \phi_{\bk} }) \exp({ i \phi_{\bk'}}) 
  \tilde{b}^\dagger_\bk \tilde{b}_{\bk'}
\label{transham}
\end{eqnarray}
where $z_\bk=|\bk|$ and $\exp(i \phi_\bk)=(k_x+ik_y)/\bk^2$.
Performing the unitary transformation
\begin{eqnarray}
\tilde{b}^\dagger_\bk  = \exp({ i \phi_{\bk} })
b^\dagger_\bk , \qquad  \tilde{b}_\bk  = \exp({ -i \phi_{\bk} })
b_\bk 
\label{unitary} 
\end{eqnarray}
and using the integers rather than $\bk$ to enumerate the momentum states in the half plane brings (\ref{transham}) into (\ref{pipham}).

Note that the main results derived for the exact solution of the $p+ip$ model, viz.  (\ref{pbae},\ref{pnrg}),
are functions of the squares of the Bethe roots. Throughout the subsequent analyses (with the exception of Appendices \ref{1cf} and  \ref{2cf}) we will simplify the notation by making the substitution
\begin{eqnarray}
y_j^2 \mapsto y_j. 
\label{maps}
\end{eqnarray}

In terms of the original parameterisation, we may now write that the exact eigenstates of the Hamiltonian (\ref{pipham}) with $M$ fermion
pairs are given by 
\begin{eqnarray}
|\psi \rangle = \prod_{j=1}^M C(y_j) |0\rangle, 
\,\,\, \quad C(y) = \sum_{\bk \in {\bf K}_+} \frac{k_x - i k_y}{
\bk^2 - y} c^\dagger_{\bk} \; c^\dagger_{- \bk}
\label{pipstates}
\end{eqnarray}
where the rapidities $y_j,\,j=1,...,M$ satisfy the Bethe ansatz
equations 
\begin{eqnarray}
\frac{q}{y_j} + \frac{1}{2} \sum_{\bk \in {\bf K}_+} 
\frac{1}{y_j - {\bf k}^2} - \sum_{l \neq j}^M \frac{1}{y_j - y_l} = 0,
\qquad\;  m = 1, \dots, M 
\label{pipbae}
\end{eqnarray}
with $ 2 q= 1/G - L + 2 M - 1$. 
The total energy of the state (\ref{pipstates}) is given by
\begin{eqnarray}
E = ( 1 + G) \sum_{j=1}^M y_j . 
\label{pipnrg}
\end{eqnarray}

This limiting case is closely related to other models which have been constructed using the hyperbolic Gaudin algebras \cite{admor02,ado01,dps04,des01}. In fact the conserved operators for the Hamiltonian (\ref{pham}) which are obtained by taking the quasi-classical limit of the operators $\{\tilde{t}(qz_j):j=1,...,L\}$ are equivalent to those which are discussed in \cite{admor02,ado01,dps04,des01} with an important difference due to the coupling parameter of the overarching anyonic model being defined in a sector-dependent manner. To make this explicit, expressing the leading term expansion  as
\begin{eqnarray*}
\tilde{t}(qz_k)\sim I-2p\gamma  \tau_k
\end{eqnarray*}
we find 
\begin{eqnarray}
\tau_k = (2G^{-1}-L+2M)N_k -\sum_{l\neq k}^L\left(  \frac{2z_kz_l}{z_k^2-z_l^2} \left(b_k^\dagger b_l +b_k b_l^\dagger\right) 
+ \frac{z_k^2+z_l^2}{z_k^2-z_l^2}(2 N_k N_l-N_k-N_l)\right). 
\label{tauk} 
\end{eqnarray}
Besides a constant term, the difference with the conserved operators in \cite{admor02,ado01,dps04,des01} is the $M$- and $L$-dependence on the co-efficient of $N_k$.  
Through the conserved operators we can write for each sector with fixed $M$
\begin{eqnarray*}
H&=&  \frac{G}{2}\sum_{k=1}^L z_k^2 (\tau_k-M) 
\end{eqnarray*}
The eigenvalues $\lambda_k$ of the $\tau_k$ are 
\begin{eqnarray*}
\lambda_k
&=& \sum_{j=1}^M \frac{z_k^2+y_j^2}{z_k^2-y_j^2}
\end{eqnarray*}
which follows from taking the leading terms in the expansion of (\ref{consevalues}) in Appendix \ref{scalarproduct}.  
One may check, with the help of (\ref{pbae}), that
\begin{eqnarray*}
E&=& (1+G)\sum_{j=1}^M y_j^2 \\
&=& \frac{G}{2}\sum_{k=1}^L z_k^2 (\lambda_k-M)
 \end{eqnarray*}
 as required. 

It is important to clarify here the reason why the conserved operators are said to be associated with hyperbolic Gaudin algebras. 
Introducing the scaling parameter $\nu$ we set 
\begin{eqnarray}
\varepsilon_k =\frac{1}{\nu}\ln(z_k), \qquad \qquad v_j=\frac{1}{\nu}\ln(y_j). \label{change}  
\end{eqnarray}    
Through this change of variables the Bethe ansatz equations (\ref{pbae}) can be expressed as 
\begin{eqnarray}
\frac{2}{G}-L+2M+\sum_{k=1}^L \coth(\nu(v_m-\varepsilon_k)) &=& 2\sum_{j\neq m}^M\coth(\nu(v_m-v_j))
\label{hyperbae}
\end{eqnarray} 
and in a similar way the expressions for $\tau_k$ and $\lambda_k$ can be written in terms of hyperbolic functions.    
In this form,  we can now take the rational  limit to recover the $s$-wave BCS model from the $p+ip$ model. 
We redefine the Hamiltonian (\ref{pham}) as  
\begin{eqnarray*}
H &\mapsto& \frac{1}{\nu}(H - (1+G)N).
\end{eqnarray*}  
Then setting $G \mapsto \nu G$ and taking the limit $\nu\rightarrow 0$, (\ref{pham},\ref{pbae},\ref{pnrg}) reproduce (\ref{r1},\ref{r2},\ref{r3}).

\subsection{The Gaudin models}

The Gaudin model can be considered as the  $G \rightarrow \infty$ limit of the corresponding 
$s$- wave and $p$-wave models introduced above. In the $s$-wave model  
the $U(1)$ symmetry is promoted to 
a $SU(2)$  symmetry and the Gaudin model can be 
  regarded  as a rotational invariant  spin chain
with local spin 1/2 operators ${\bf S}_i$ whose dynamics is described 
 by a set of commuting operators 
\begin{eqnarray*}
R_i = \sum_{j \neq i}^L  \frac{  {\bf S}_i \cdot {\bf S}_j}{ \varepsilon_i - \varepsilon_j} , \qquad
[ R_i, R_j] =0 
\end{eqnarray*}
These operators are simultaneously diagonalized by the Richardson ansatz in terms
of a set of rapidities $\varepsilon_j$  satisfying  the $G \rightarrow \infty$ limit of the Bethe ansatz equations 
(\ref{bae}), namely 
\begin{eqnarray*}
\sum_{k=1}^L \frac{1}{v_j-\varepsilon_k} &=& \sum_{j\neq m}^M\frac{2}{v_m-v_j},\qquad j=1,...,N 
\end{eqnarray*} 
The $p$-wave Gaudin model corresponds to the trigonometric version considered
earlier with the commuting operators given by  the $G \rightarrow \infty$ limit 
of the operators $\tau_k$ defined in eq. (\ref{tauk}). The corresponding Bethe ansatz equations  are given by
the $G \rightarrow \infty$ limit of (\ref{pipbae}),  or equivalently (\ref{hyperbae}).

\section{Analysis of the $p+ip$ model} \label{pwavegs}

\subsection{Ground-state phase diagram from BCS mean-field theory}

To undertake the mean-field analysis of the $p+ip$ model it is more convenient to work with the Hamiltonian
\begin{eqnarray}
&\mathcal{H}& = \sum_\bk \frac{\bk^2}{2 m} \;  c^\dagger_\bk c_\bk - \frac{{G}}{4 m} \sum_{\bk, \bk'}  (k_x - i k_y) (k'_x + i k'_y) \; 
 c^\dagger_\bk  c^\dagger_ {-\bk }  c_{- \bk'} c_{ \bk'}
\label{mfpipham}
\end{eqnarray}
which differs from (\ref{pipham}) only in that the double sum is now not restricted to ${\bk \neq  \pm \bk'}$. The differences between using (\ref{pipham}) and (\ref{mfpipham}) will be discussed at the end of Subsection \ref{phasediagram}.

The derivation of the gap and chemical potential equations which give the mean-field solution of (\ref{mfpipham}) follows now standard techniques which are a straightforward extension of the original methods of the BCS paper \cite{bcs57}. Here we will only outline the main steps. First we assume that the number of fermions is even and that in the ground state of the system all fermions are paired. Our convention will again be to use $N$ to denote the Cooper pair number operator and $M$ for the eigenvalues of $N$. The  BCS order parameter associated to
(\ref{mfpipham}) is 
\begin{eqnarray*}
\hat{\Delta} = \frac{G}{m} \sum_\bk (k_x + i k_y) 
\langle  c_{- \bk} c_{ \bk} \rangle
\end{eqnarray*}
in terms of which the Hamiltonian (\ref{mfpipham}) can be approximated as 
\begin{eqnarray*}
\mathcal{H} \approx \sum_\bk \xi_{\bk}
\;  c^\dagger_\bk c_\bk 
- \frac{1}{4} \sum_{\bk } \left(  \hat{\Delta} \; (k_x - i k_y) 
 c^\dagger_\bk  c^\dagger_ {-\bk }   + h.c. \right) +\frac{|\hat{\Delta}|^2}{4G}+\mu M
\end{eqnarray*}
where $\xi_{\bk}= {\bk^2}/{2 m} - {\mu}/{2} $ and  $\mu/2$ is the 
chemical potential. This Hamiltonian can be
diagonalized by a Bogoliubov transformation. By minimising the energy, and fixing the expectation value of the number of Cooper pairs to be $M$, it is found that the gap $\Delta=|\hat{\Delta}|$
and chemical potential are the solutions of the equations 
\begin{eqnarray}
\sum_{\bk \in {\bf K}_+ } \frac{\bk^2}{\sqrt{ (\bk^2 - \mu)^2 
+ \bk^2 \Delta^2}} & = & \frac{1}{G}
\label{gap} \\
\mu \; \sum_{\bk \in {\bf K}_+  } \frac{1}{\sqrt{ (\bk^2 - \mu)^2 
+ \bk^2 \Delta^2}} & = 
& 2 M - L + \frac{1}{G} \label{chempot} 
\end{eqnarray}
where we have set $m=1$ and $2L$ is the total number of momentum states below the cut-off $\omega$.  The mean-field expression for the ground-state energy is 
\begin{eqnarray}
E_0=\frac{1}{2}\sum_{\bk \in {\bf K}_+} \bk^2\left(1-\frac{2\bk^2+\Delta^2-2\mu}{2\sqrt{(\bk^2 - \mu)^2 + \bk^2 \Delta^2}}\right).    
\label{mfnrg}
\end{eqnarray} 
The normalised ground state is given by 
 \begin{eqnarray}
|\Psi\rangle=\prod_{\bk \in {\bf K}_+}(u_\bk I +v_\bk c^\dagger_{\bk}c^\dagger_{-\bk})|0 \rangle
\label{Psi}
\end{eqnarray} 
 where 
 \begin{eqnarray*}
 |u_\bk|^2&=&\frac{1}{2}\left(1+\frac{\bk^2-\mu}{\sqrt{(\bk^2-\mu)^2+\bk^2\Delta^2}}  \right)  \\
 |v_\bk|^2&=&\frac{1}{2}\left(1-\frac{\bk^2-\mu}{\sqrt{(\bk^2-\mu)^2+\bk^2\Delta^2}}\right) 
 \end{eqnarray*}
and the phases must be chosen such that 
$$\frac{(k_x+ik_y)\hat{\Delta}^*v_\bk}{u_\bk}=2E(\bk) - \bk^2 + \mu$$ 
is real. Above, $E(\bk)$ is the quasiparticle energy spectrum 
\begin{eqnarray}
E(\bk) = \frac{1}{2}\sqrt{ (\bk^2 - \mu)^2 + \bk^2 \Delta^2 }. 
\label{spectrum}
\end{eqnarray}

To obtain an approximation for the ground state with a fixed number of Cooper pairs $M$, one can take a projection of (\ref{Psi}). Writing (\ref{Psi}) as 
$$|\Psi\rangle = \left(\prod_{\bk\in{\bf K}_+} u_\bk^{-1}\right) \exp\left(\sum_{\bk\in{\bf K}_+}\frac{v_\bk}{u_\bk} c^\dagger_{\bk}c^\dagger_{-\bk}\right)|0\rangle, $$ 
projection onto a fixed number of $M$ pairs gives  
\begin{eqnarray}
|\psi \rangle = \left[ \sum_{\bk \in {\bf K}_+}  {\mathfrak g}(\bk)  
c^\dagger_ {\bk } c^\dagger_ {-\bk } 
\right]^M 
|0 \rangle 
\label{pgs}
\end{eqnarray}
where
\begin{eqnarray}
{\mathfrak g}(\bk) = \frac{v_\bk}{u_\bk}= \frac{2E(\bk) - \bk^2 + \mu}{(k_x + i k_y)\hat{\Delta}^*} .
\label{gees}
\end{eqnarray}

\subsubsection{The Moore-Read and Read-Green lines} \label{phasediagram}

By analysing the above mean-field results we can piece together the various phases of the model. To some extent this has been undertaken in \cite{rg00}, and we will first review their findings. Then we will extend the analysis to expose a duality that exists in the ground-state phase diagram. 

Following \cite{rg00},   
note that the spectrum is gapless at $\mu=0$ as $|\bk|\rightarrow 0$. 
Furthermore, the behaviour of ${\mathfrak g}(\bk)$ as 
$|\bk| \rightarrow 0$ depends on the sign of $\mu$,
\begin{eqnarray*}
{\mathfrak g}(\bk) \sim \left\{
\begin{array}{cc}
k_x - i k_y, & \mu < 0, \\
1/(k_x + i k_y), & \mu > 0. \\
\end{array}
\right.
\label{} 
\end{eqnarray*}
It was argued in \cite{rg00} that $\mu=0$ corresponds to a topological phase transition, a subject we will return to later in Subsection \ref{topo}. From (\ref{chempot}) it is seen that $\mu=0$ necessarily implies 
\begin{eqnarray*}
 M=\frac{1}{2}\left(L-\frac{1}{G}\right) 
\end{eqnarray*}
Introducing the filling fraction $x=M/L$ and defining $g=GL$ we equivalently have 
\begin{eqnarray}
 x=\frac{1}{2}\left(1-\frac{1}{g}\right). 
\label{rgline}
\end{eqnarray}
We will refer to (\ref{rgline}) as the Read-Green (RG) line of the phase diagram.

In real space the state (\ref{pgs}) takes the form of a pfaffian 
\begin{eqnarray*}
\psi(\br_1, \dots, \br_{2 M}) =
{\cal A}[ \hat{{\mathfrak g}}(\br_1- \br_2) \dots  \hat{{\mathfrak g}}(\br_{2M-1}- \br_{2M})]
\label{10}
\end{eqnarray*}
where ${\cal A}$ denotes the antisymmetrization of the 
positions and $\hat{{\mathfrak g}}(\br)$ is the Fourier transform of ${\mathfrak g}(\bk)$.
We will refer to the case $\mu=0$ as the Read-Green (RG) state. 
For $\mu > 0$ the large distance behaviour is $\hat{{\mathfrak g}}(\br) \sim 1/(x + i y)$, 
which asymptotically reproduces the Moore-Read (MR) state found in studies of the Fractional Quantum Hall Effect 
\cite{mr91} (see Appendix \ref{mfnum} for the  derivation of this result and its comparison
with the exact wavefunctions in real space).  In momentum space co-ordinates the (unnormalised) MR state of $M$ Cooper pairs is 
\begin{eqnarray}
|MR\rangle =\left[ \sum_{\bk \in {\bf K}_+}  \frac{1}{k_x+ik_y}  
c^\dagger_ {\bk } c^\dagger_ {-\bk } 
\right]^M 
|0 \rangle . 
\label{mrstate}
\end{eqnarray}  

Upon closer inspection of the mean-field equations we find that the MR state (\ref{mrstate}) coincides (up to normalisation) with the projected ground state (\ref{pgs}) whenever 
\begin{eqnarray}
\Delta^2=4\mu 
\label{mrconstraint}
\end{eqnarray}
The result is easily verified by simply substituting (\ref{mrconstraint}) into (\ref{spectrum},\ref{gees}). Furthermore, substituting (\ref{mrconstraint}) into (\ref{gap},\ref{chempot}) and adding these equations yields   
\begin{eqnarray*}
M=L-\frac{1}{G},
\end{eqnarray*}
or in terms of the filling fraction
\begin{eqnarray}
x=1-\frac{1}{g}.
\label{mrline}
\end{eqnarray}
We will refer to (\ref{mrline}) as the MR line of the phase diagram. Finally, we substitute (\ref{mrconstraint}) into (\ref{mfnrg}) to obtain
\begin{eqnarray}
E_0= 0. 
\label{mrnrg}
\end{eqnarray}  
This means that the condensation energy of the state
equals the energy of the Fermi sea! 

From the above considerations we gain an initial insight into topological properties of the model. Both the RG line given by (\ref{rgline}), for which the excitation spectrum becomes gapless as $|\bk|\rightarrow 0$, and the MR line given by (\ref{mrline}), for which the ground-state energy is zero,  
hold {\it independent} of the choice of distribution for the momenta $\bk$. These are properties of the model that are protected (in a topological sense to be discussed in Subsection \ref{topo}) from perturbations of the system which are reflected by changes in the momentum distribution.

Both the RG and MR lines can be viewed as boundary lines in the ground-state phase diagram relative to a duality property of the mean-field solution, which is our next item to discuss. 
First we introduce parameters $a$ and $b$ related to the chemical
potential and the gap by
\begin{eqnarray}
a, b = \mu - \frac{{\Delta}^2}{2} \pm
{\Delta} \sqrt{ \frac{ {\Delta}^2}{4} - \mu } 
\label{ab}
\end{eqnarray}
which simply allows us to write the quadratic term under the square root in (\ref{gap},\ref{chempot}) in terms of its roots:
$$\sqrt{(\bk^2-\mu)^2+\bk^2\Delta}=\sqrt{(\bk^2-a)(\bk^2-b)}. $$ 
The parameters $a$ and $b$ 
can either both be real, or form a complex pair in which case we denote them as
$\epsilon \pm i \delta$. In either case they satisfy the relation
\begin{eqnarray*}
\mu^2  = a b .
\end{eqnarray*}
We see from (\ref{mfnrg}) that in terms of $a$ and $b$ the ground-state energy reads
\begin{eqnarray}
E_0=\frac{1}{2}\sum_{\bk \in {\bf K}_+} \bk^2\left(1-\frac{2\bk^2-a-b}{2\sqrt{(\bk^2 -a)(\bk^2-b)}}\right)    
\label{abnrg}
\end{eqnarray} 
which is independent of the sign of $\mu$.
On the other hand, the sign of the
$\mu$ coincides, by 
eq. (\ref{chempot}), with that of 
\begin{eqnarray}
q_0 =  L\left(x - \frac{1}{2} + \frac{1}{2 g}\right). 
\label{qnought}
\end{eqnarray}
This implies the existence of a duality between phases which we will denote as weak pairing and strong pairing, adopting the terminology of \cite{rg00}. Specifically, the two ground states with filling fractions $x_W$ (with $\mu>0$) and $x_S$ (with $\mu<0$) satisfying 
\begin{eqnarray}
x_W + x_{S}  = 1 - \frac{1}{g}
\label{mfduality}
\end{eqnarray}
share the same values of  $a,b,g, \mu^2$. Besides differing in the sign of $\mu$, they also differ in the values of $\Delta$. One can check that the required map which preserves (\ref{ab}) is  
\begin{eqnarray*}
\Delta^2&\mapsto& \Delta^2 -4\mu  \\
\mu &\mapsto& -\mu
\end{eqnarray*}
 Finally, from (\ref{abnrg}) we see these ground states have the same mean-field energy, and from (\ref{spectrum}) the same excitation spectrum above them. 

Taking into account all these 
considerations, the ground state of the model
consists of three phases detailed in Table 1. In the weak
coupling BCS phase where $ x > 1 - 1/g$, 
$a$ and $b$ form a complex
pair parameterized by $\epsilon$ and $\delta$. 
When $x = 1 - 1/g$, the MR line, both $a$ and $b$ are equal
and real. This line is dual to the vacuum. From (\ref{spectrum}) the mean-field solution predicts this line is gapped, so it is not clear at this stage in what sense the MR line might be considered a phase transition. We shall come back to this issue later on when we discuss
the solution of the model in terms of the Bethe ansatz
equations. 
The two regions denoted weak pairing and strong pairing, for which the parameters $a$ and $b$ are real and negative, are separated by the critical RG
line where the chemical potential $\mu=0$ and the 
excitation gap vanishes. The RG line is self-dual.

\begin{center}
\begin{small}
\begin{tabular}{|c|c|c|c|}
\hline
Phase & $a, b$ &  $\mu , \Delta$  & $x$ \\
\hline 
\hline
Weak coupling BCS& $\epsilon \pm i \delta$  &  $\mu > {\Delta^2}/{4}$  
& $ x > 1 -{g^{-1}}$  \\
\hline
Moore-Read line & $a = b = - \mu$  & $\mu =   {\Delta^{2}}/{4}  $ 
& $ x_{MR} = 1 - {g^{-1}}$  \\
\hline
Weak pairing & $ a < b < 0$ & $ 0 < \mu <  {\Delta^2}/{4}$ 
& $  (1 - {g^{-1}})/2 < x < 1 - {g^{-1}} $  \\
\hline
Read-Green line & $ a < b = 0$ & $ \mu = 0$  
& $  x_{RG} =  (1 - {g^{-1}})/2$  \\
\hline
Strong pairing  & $ a < b < 0$ & $ \mu < 0$  
& $  x <  (1 - {g^{-1}})/2$  \\
\hline
\end{tabular}
\end{small}

\vspace{0.1 cm}

Table 1: \footnotesize{Phases of the $p+ip$ model in terms of the mean-field variables $\mu,\,\Delta,\,a,\,b$.} 
\end{center}

To solve the mean-field gap (\ref{gap}) and chemical potential (\ref{chempot}) equations for the case of large $L$ it is useful to undertake a continuum limit whereby the sums are replaced by integrals. Making the change of variable $\bar{\varepsilon}= \bk^2/\omega$ we introduce a density function $\bar{\rho}(\bar{\varepsilon})$ with the choice of normalisation 
\begin{eqnarray*}
\int_0^1 d\bar{\varepsilon} \,\bar{\rho}(\bar{\varepsilon}) = 1. 
\end{eqnarray*}
The densities are chosen as 
\begin{eqnarray}
\bar{\rho}(\bar{\varepsilon})= \frac{1}{2\sqrt{\bar{\varepsilon}}}
\label{dens1}
\end{eqnarray}
which corresponds to free fermions in one dimension (1D), and 
\begin{eqnarray}
\bar{\rho}(\bar{\varepsilon})= 1
\label{dens2}
\end{eqnarray}
for free fermions in two dimensions (2D).
The integral approximations of (\ref{gap},\ref{chempot}) are 
\begin{eqnarray}
\frac{1}{g} &=&\int^1_0 d\bar{\varepsilon} \frac{\bar{\varepsilon} \,\bar{\rho}(\bar{\varepsilon}) }{\sqrt{ (\bar{\varepsilon}-\bar{a}) (\bar{\varepsilon}-\bar{b}) }}, 
\label{gapcont} \\
2 \left|x-\frac{1}{2}+\frac{1}{2g}\right| &=&\sqrt{\bar{a}\bar{b}}\int^1_0 d\bar{\varepsilon} \frac{\bar{\rho}(\bar{\varepsilon}) }{\sqrt{ (\bar{\varepsilon}-\bar{a}) (\bar{\varepsilon}-\bar{b}) }},  
\label{chempotcont}
\end{eqnarray}
where $\bar{a}=a/\omega,\,\bar{b}=b/\omega$. From (\ref{mfnrg}) we obtain the ground-state energy per pair as 
\begin{eqnarray}
e_0 &=&\frac{\omega}{2x} \int^1_0 d\bar{\varepsilon}\, \bar{\varepsilon}\, \bar{\rho}(\bar{\varepsilon}) \left( 1-\frac{2\bar{\varepsilon}-\bar{a}-\bar{b} }{2\sqrt{ (\bar{\varepsilon}-\bar{a}) (\bar{\varepsilon}-\bar{b}) }}\right). 
\label{nrgcont}
\end{eqnarray} 
The integral equation approximations become exact in the thermodynamic limit $L\rightarrow \infty$, which requires that $G\rightarrow 0$, $M\rightarrow \infty$ in order for $g=GL$, $x=M/L$ to remain finite. Solutions of (\ref{gapcont},\ref{chempotcont}) are given in Appendix \ref{mfnum}. Numerical results are presented in Figs. \ref{mffig} and \ref{mfgapfig} below.

To conclude here, we turn to the comparison between the Hamiltonians (\ref{pipham}) and (\ref{mfpipham}). 
Excluding the subspace of unpaired states we have 
\begin{eqnarray*}
H(G)=(1+G)\mathcal{H}(G/(1+G))
\end{eqnarray*}
which  shows that $H$ is equivalent to $\mathcal{H}$ up to a scaling factor $1+G$ and the redefintion of the coupling $G \mapsto G/(1+G)$. For systems with an even number of fermions, the mean-field ground states of $H$ and $\mathcal{H}$ are equal in the thermodynamic limit since we take $G\rightarrow 0$.

\subsubsection{The Volovik line}

\begin{figure}[t!]
\begin{center}
\scalebox{0.8}{\input{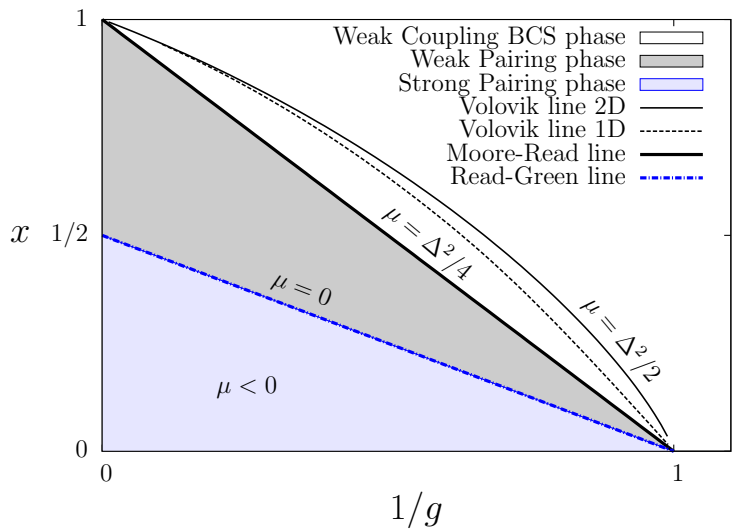}}
\end{center}
\caption{\footnotesize{
Ground-state phase diagram of the $p + i p$ 
model in terms of the inverse coupling 
$1/g$ and filling fraction $x=M/L$. Phase boundaries 
are given by the Read-Green line ($\mu=0$) and the 
Moore-Read line ($\mu=\Delta^2/4$). The phase boundary conditions are independent of the choice of the momentum distribution, and independent of the ultraviolet cut-off.
Also shown are the Volovik lines corresponding to 1D and 2D momentum distributions.
}}
\label{volovik-phase}
\end{figure}

The energy gap between the ground and first excited states is the minimum of the quasiparticle excitation energies as given by (\ref{spectrum}). In order to make comparison with later results for finite-sized systems where the total fermion number is fixed, we will consider an elementary excitation to be associated with two quasiparticle excitations. From (\ref{spectrum}) we find that the gap $E_{{\rm gap}}$ is  
\begin{eqnarray*}
E_{{\rm gap}}(\bk)= \left\{ 
\begin{array}{cc} 
\displaystyle |\mu|, & \displaystyle \mu\leq {\Delta^2}/{2}, \\
\displaystyle \frac{1}{2}\sqrt{\Delta^2(4\mu-\Delta^2)}, &  \displaystyle \mu\geq {\Delta^2}/{2}.
\end{array}
\right.
\end{eqnarray*} 
For  $\mu > \Delta^2/2$ the quasiparticle energy has a minimum at a non zero momentum ${\bf k}$ 
different from zero, while for $\mu  \leq  \Delta^2/2$ the minimum is achieved at ${\bf k}=0$. This situation
is reminiscent of the Higgs transition that takes places for a scalar field with a quartic potential \cite{v07}.
For this reason we  call the case $\mu = \Delta^2/2$ the Volovik line in the $p+ip$ phase diagram. 
 In this instance eliminating $\Delta$ from the  
gap and chemical potential equations (\ref{gap},\ref{chempot}) gives
\begin{eqnarray*}
\sum_{\bk \in {\bf K}_+ } \frac{\bk^2}{\sqrt{ \bk^4 
+  \mu^2}} & = & \frac{1}{G},
\\
\mu \; \sum_{\bk \in {\bf K}_+  } \frac{1}{\sqrt{ \bk^4+\mu^2}}& =  
& 2 M - L + \frac{1}{G}. 
\end{eqnarray*}
Further eliminating the $1/G$ term we obtain the equation of the Volovik line 
\begin{eqnarray}
x= \frac{1}{2}-\frac{1}{2L} \sum_{\bk \in {\bf K}_+ } \frac{\bk^2-\mu}{\sqrt{ \bk^4 
+  \mu^2}}  
\label{volovikline}
\end{eqnarray}
which lies in the weak coupling BCS phase. In contrast to the RG (\ref{rgline}) and MR (\ref{mrline}) line equations, the Volovik line is dependent on the distribution of the momenta $\bk$ and consequently is not topologically protected.
In the continuum limit the equation (\ref{volovikline}) reads
\begin{eqnarray}
x= \frac{1}{2}-\frac{1}{2} \int_0^1 d \bar{\varepsilon} \;  \bar{\rho}(\bar{\varepsilon})   
\frac{ \bar{\varepsilon} - \bar{\mu} }{ \sqrt{  \bar{\varepsilon}^2 + \bar{\mu}^2} }
\label{voloviklineb}
\end{eqnarray}
which relates the filling fraction $x$ with the value of the normalized  chemical
potential $\bar{\mu}$. In 2D the gap equation (\ref{gapcont}) becomes
\begin{eqnarray*}
\frac{1}{g}= - \bar{\mu} + \sqrt{ 1 + \bar{\mu}^2 } 
\end{eqnarray*}
which can be easily inverted to yield $\bar{\mu}$ as a function of $g$. 
Plugging that result into the integral of (\ref{voloviklineb}) 
gives the Volovik line for the 2D model
\begin{eqnarray}
x= \frac{1}{2}   \left( 1 - g^{-1}  + \frac{1}{2} ( g - g^{-1}) \log
\frac{ g + g^{-1} + 2}{g - g^{-1}}  \right). 
\label{volovikline2D}
\end{eqnarray}
In 1D the gap equation (\ref{gapcont}) yields
\begin{eqnarray*}
\frac{1}{g} = \frac{1}{3 \bar{\mu}} F_{2,1} \left( \frac{1}{2}, \frac{3}{4}, \frac{7}{4}, - \frac{1}{\bar{\mu}^2} \right) 
\end{eqnarray*}
which together with the integration of (\ref{voloviklineb}) gives
\begin{eqnarray*}
x = \frac{1}{2} \left( 1 - g^{-1} + F_{2,1} \left( \frac{1}{4}, \frac{1}{2}, \frac{5}{4}, - \frac{1}{\bar{\mu}^2}\right) \right) 
\end{eqnarray*}
which is the equation of the Volovik line in 1D.  In Fig. \ref{volovik-phase}
 we depict the Volovik lines, of the 1D and 2D models,
on the the phase diagram of the $p+ip$ model.  One can see that they lie close to the MR line
although do not coincide with it. As explained earlier the Volovik line depends on the choice
of the energy level density $\rho(\varepsilon)$, while the MR  and RG lines are independent of it.

\begin{figure}[h!]
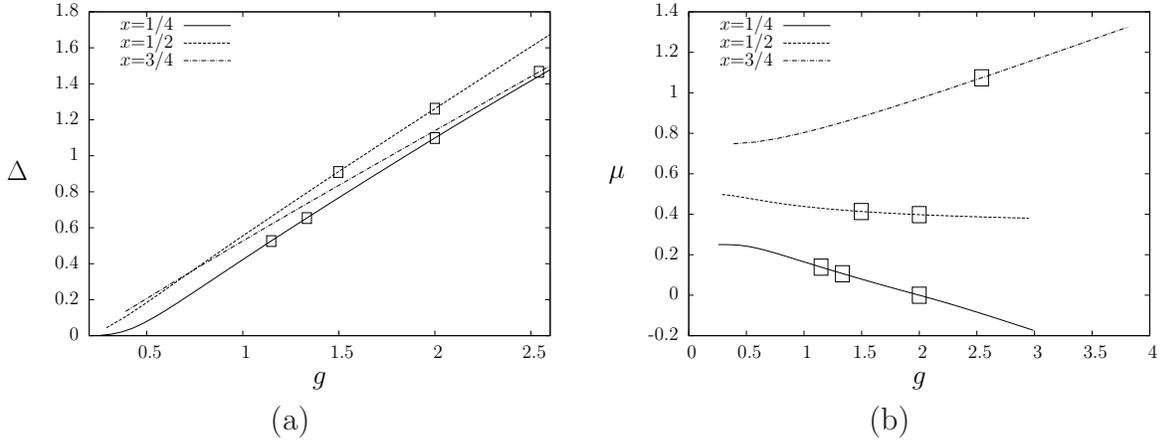

\begin{center}
$$
\begin{array}{cc} 
\scalebox{0.6}{\input{Deltas.tex}} &
\scalebox{0.6}{\input{mus.tex}} \\
({\rm a}) & ({\rm b})
\end{array}
$$
\end{center}
\caption{\footnotesize{Mean-field order parameter, $\Delta$, and chemical potential, $\mu$, solution of the gap and chemical potential equations in 2D for filling fractions $x=1/4,1/2$ and 3/4, vs. $g$. The rectangles over the curves indicate, from left to right, the Volovik, Moore-Read and Read-Green points ($x=1/4$); the Volovik and Moore-Read points ($x=1/2$), and the Volovik point ($x=3/4$).}}
\label{mffig}
\end{figure}

\begin{figure}[h!]
\begin{center}
\scalebox{0.7}{\input{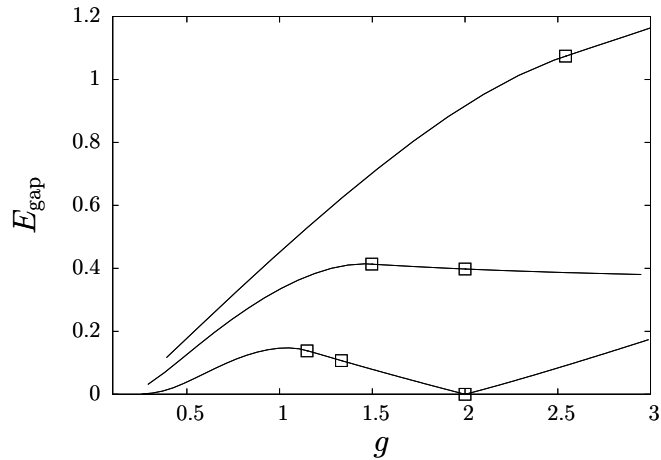}}
\end{center}
\caption{\footnotesize{ The mean-field energy gap $E_{\rm gap}$ versus coupling constant $g$ for a 2D momentum density distribution as given by (\ref{dens2}). From top to bottom the curves shown correspond to the filling fractions $x=3/4,\, 1/2,\, 1/4$. The squares over the curves indicate, from left to right, the Volovik, Moore-Read, and Read-Green points respectively.
}}
\label{mfgapfig}
\end{figure}

\subsubsection{Fidelity susceptibility}

The notion of fidelity has been adopted in a variety of ways to characterise quantum phase transitions \cite{zp06,zb08}. Here we focus on one such formulation, which is the fidelity susceptibility \cite{g08}. 
 
The fidelity susceptibility $\chi_{ F}$ is defined as 
\begin{eqnarray*}
\chi_{ F}&=& \left< \frac{d\Psi}{dG} \,\,\right|\left. \frac{d\Psi}{dG}\right> -\left< \left.\frac{d\Psi}{dG}\,\,\right| \Psi\right> \left< \Psi \left|\,\,\frac{d\Psi}{dG}\right. \right>.
\end{eqnarray*}
For the mean-field wavefunction this reduces to 
\begin{eqnarray*}
\chi_F&=& \sum_{\bk\in{\mathbf K}_+} \left< \left.\frac{d\psi_\bk}{dG}\right|\frac{d\psi_\bk}{dG}\right> = \sum_{\bk\in{\mathbf K}_+} \left(\frac{du_\bk}{dG}\right)^2+\left(\frac{dv_\bk}{dG}\right)^2.
\end{eqnarray*}
The derivatives of $u_\bk$ and $v_\bk$ with respect to $G$ can be computed in terms of 
$$\frac{d\mu}{dG}=c,\qquad\qquad \frac{d\Delta}{dG}=d. $$
which never vanish for $G >0$,  as can be seen from Fig. \ref{mffig}. 
Now
\begin{eqnarray*}
\left(\frac{d u_\bk}{dG}\right)^2
+\left(\frac{d v_\bk}{dG}\right)^2
&=&\frac{\epsilon_\bk}{4}\frac{(c \Delta +d (\epsilon_\bk-\mu))^2}{((\epsilon_\bk-\mu)^2+\epsilon_\bk\Delta^2)^{2}}
\end{eqnarray*}

\begin{figure}[t!]
\begin{center}
\scalebox{0.7}{\input{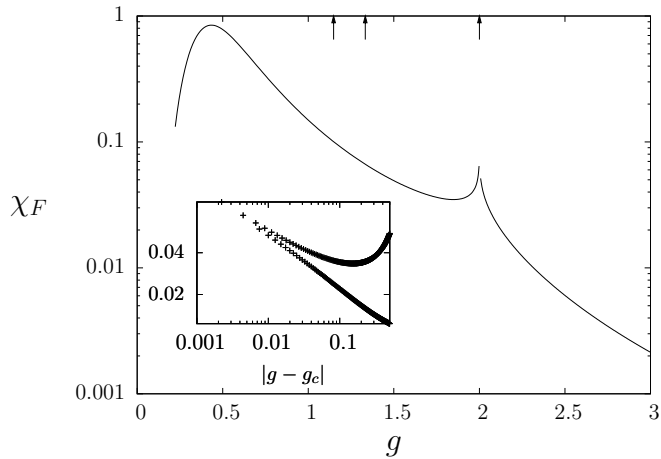}} 
\end{center}
\caption{\footnotesize{
Plot of the mean-field  fidelity susceptibility (\ref{chimf}) as a function of the coupling $g$ for $x=1/4$. The 2D momentum distribution (\ref{dens2}) is used and the fidelity susceptibility is  
 in units of $L^3$. The singularity at $g_{c}=2$  is apparent. Arrows indicate from left to right the Volovik, Moore-Read, and Read-Green points respectiveley. The inset emphasises the logarithmic character of this singularity, with the distance to the critical point $g_{c}$ shown on a logarithmic scale. The two data sets correspond to both cases $g>g_c$ and $g<g_c$.
}}
\label{suscep}
\end{figure}

In the mean-field theory gapless excitations occur when $\epsilon_{\mathbf 0}=\mu=0$. Now
\begin{eqnarray*}
\lim_{\epsilon_{\mathbf 0}\rightarrow 0} \lim_{\mu\rightarrow 0} \left[\left(\frac{d u_{\mathbf 0}}{dG}\right)^2
+\left(\frac{d v_{\mathbf 0}}{dG}\right)^2\right] &=& \infty.
\end{eqnarray*}
This indicates the fidelity susceptibility is divergent on the Read-Green line.
The expression of $\chi_F$ in the continuum limit  is given by

\begin{eqnarray}
\chi_F  = \frac{L^3}{4}  \int_0^1 d \bar{\varepsilon} \; \bar{\rho}( \bar{\varepsilon}) 
\, \varepsilon  \left[   \frac{  \bar{c} \bar{ \Delta} + \bar{d} ( \bar{\varepsilon} - \bar{\mu})}{
 ( \bar{\varepsilon} - \bar{\mu})^2 + \bar{ \varepsilon} \bar{\Delta}^2 } \right]^2
 \label{chimf}
 \end{eqnarray}
 where 
$$\frac{d \bar{\mu}}{dg}=\bar{c},\qquad\qquad \frac{d \bar{\Delta}}{dg}=\bar{d}. $$
 In Fig. \ref{suscep}  we plot $\chi_F$ as a function of $g$ for $x=1/4$. The results show a logarithmic divergence
 at the  Read-Green point $g_c = 2$, reflecting that the model is critical here.

\subsection{Analysis of the Bethe ansatz equations in the continuum limit} \label{tdl}

\subsubsection{The electrostatic analogy}

For the $s$-wave BCS model, the relationship between the mean-field gap and chemical potential equations and the exact Bethe ansatz solution was first explored by Gaudin \cite{g95}. This was achieved by use of an electrostatic analogy, whereby the Bethe ansatz equations represent the equilibrium conditions for a two-dimensional system of fixed and mobile charges. Further studies along this line for the $s$-wave case can be found in \cite{rsd02,r77}. Our aim below is to extend this approach for the $p+ip$ model. In the general framework of exactly solvable models based on hyperbolic Gaudin algebras this type of study has previously been conducted in \cite{admor02}. There the mapping from the hyperbolic form of the Bethe ansatz equation such as (\ref{hyperbae}) to a set of algebraic equations was implemented by a change of variable. However this change of variable is not the same as (\ref{change}), so the algebraic form considered in \cite{admor02} is different from (\ref{pbae}).     

Taking into account the change of notational conventions as described by (\ref{maps}), 
let us write the Bethe ansatz equation (\ref{pbae}) as
\begin{eqnarray}
 - \frac{1}{2} 
\sum_{k=1}^L \frac{1}{y_m - z_k^2} - \frac{q_0}{y_m}
+ \sum_{j \neq m }^M \frac{1}{y_m - y_j} = 0, \qquad
m =1, \dots, M 
\label{a34}
\end{eqnarray}
where $q_0$ is given by (\ref{qnought}).
This equation admits an electrostatic analogy
associating  $-1/2$ charges at the fixed positions $z_k^2$, a $-q_0$ charge at the origin, and mobile $+1$ charges with positions $y_j$. With these
assignments (\ref{a34}) amounts to the vanishing of the
total electric field acting on the charge $y_m$. 
The continuum limit of (\ref{a34}) is given by 
(we shall follow the 
conventions of references \cite{g95,rsd02}): 
\begin{eqnarray}
\int_\Omega d \varepsilon \frac{ \rho(\varepsilon)}{
\varepsilon - y} - \frac{q_0}{y} - P \int_\Gamma |dy'| 
\frac{r(y')}{ y'- y} = 0,\;\; \forall y \in \Gamma
\label{a35}
\end{eqnarray}
where
$
\Omega = (0, \omega)
$
denotes the interval on the real line where
lie the energy levels which can be associated to a
negative charge density $- \rho(\varepsilon)$ such that 
\begin{eqnarray*}
\int_0^\omega d\varepsilon\,\rho(\varepsilon)=\frac{L}{2}.
\end{eqnarray*}
 The function
$r(y)$ denotes the charge density associated to the 
roots $y_m$ which lie on an arc $\Gamma$ of the complex plane. 
It satisfies
\begin{eqnarray*}
\int_\Gamma |d y| \; r(y)  = M 
\end{eqnarray*}
while the energy (\ref{pnrg}) in the continuum limit is given by 
\begin{eqnarray}
\int_\Gamma |d y| \; y \; r(y)  = E.  
\label{a38}
\end{eqnarray}

\noindent 
For the ground-state roots the topology  of the arc $\Gamma$ depends crucially
on the domain of the phase diagram one is exploring. 
In each phase $\Gamma$ consists of the union of several types
of arcs whose properties are given in Table 2.  
Table 3 displays the arcs associated 
to the different phases of the model.
We shall first comment on the results contained in these tables 
and will give the derivation later. 

\begin{figure}[t!]
\begin{center}
\scalebox{0.65}{\input{phases.tex}}
\includegraphics[height= 5.9 cm,angle=0]{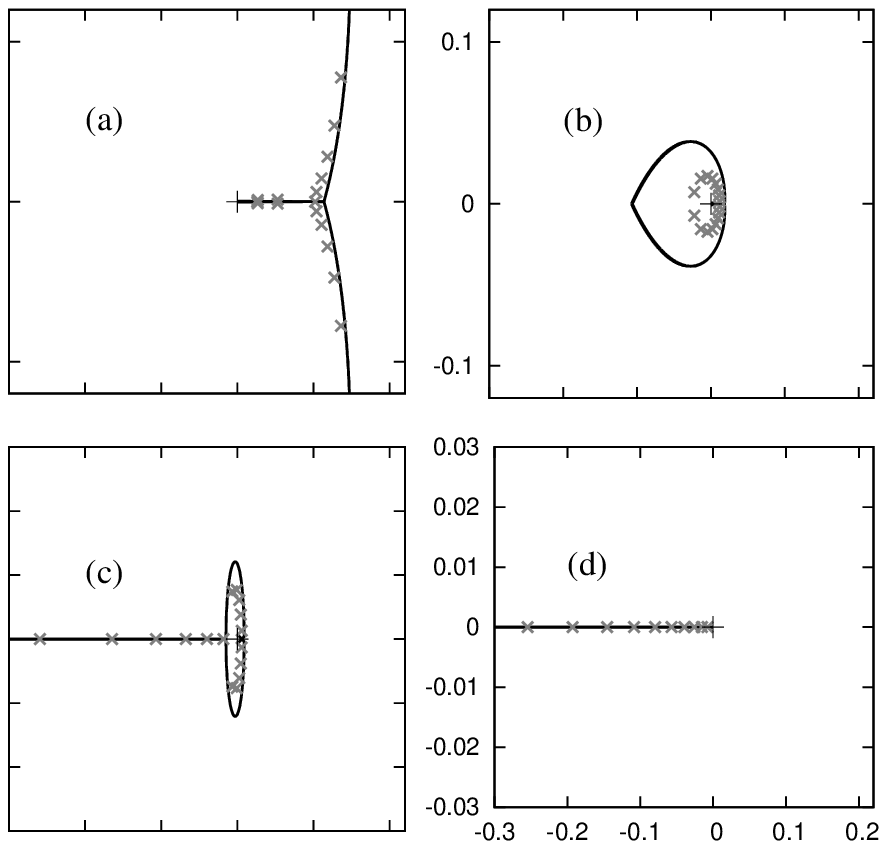}
\end{center}
\caption{\footnotesize{Representation of the electrostatic arcs in several regions of the phase diagram. The roots of the Bethe ansatz equations for $L=64$ and $M=16$ are also shown. For the four smaller figures, the same horizontal scale is used which is shown explicitly for case (d). Cases (a) and (b) have the same vertical scale, as do cases (c) and (d). We note that while the roots lie close to the  electrostatic arc in cases (a), (c) and (d), the agreement is not as good in case (b) which lies near the Moore-Read line.   
}}
\label{arcsplot}
\end{figure}

An arc of type $\Gamma_A$ is a segment of the real
line $(0, \varepsilon_A)$ which is contained in the
real interval $\Omega$. At $G=0$, all the roots
$y_m$ associated to the ground state lie on 
$\Gamma_A$, where   $\varepsilon_A = \mu$ only depends on the filling fraction $x$. 
When the coupling is switched on, $g > 0$, the roots closer to the Fermi
level form a complex arc $\Gamma_B$  parameterized
by its end points $a,b=\epsilon \pm i \delta$, and which cuts
the real axis at $\varepsilon_A$ whose value has also changed slightly. 
In the weak coupling BCS phase, $\Gamma$ is the union of the arcs
$\Gamma_A$ and $\Gamma_B$. This type of arc is the familiar
one encoutered in the solution of the $s$-wave model \cite{g95,rsd02}. As $g$
increases, the arc  $\Gamma_B$ enlarges and bends towards
the negative real axis. At the Moore-Read line
the complex arc closes and is denoted by $\Gamma_C$. 
This arc cuts
the real axis at 
 $\varepsilon_A >0$ and $a = b < 0$. Some 
real roots may still remain, so that  $\Gamma$ at the Moore-Read line
is the union $\Gamma_A \cup  \Gamma_C$. In the weak pairing phase 
the closed  arc $\Gamma_C$ remains but there appear
some new roots on the negative real axis forming the open
arc $\Gamma_D  = (a, b)$.  Hence 
$\Gamma =\Gamma_A \cup  \Gamma_C \cup \Gamma_D$. As $g$ increases
further, the closed arc $\Gamma_C$ shrinks and $\Gamma_D$ becomes
larger until $g$ reaches 
the Read-Green line, where $\Gamma_C$ disappears. 
Finally in the strong pairing phase there are only negative real roots.
In Fig. \ref{arcsplot} we display the types of arcs depending on the region
of the phase diagram together with the exact numerical roots.

\begin{center}
\begin{tabular}{|c|c|c|c|c|}
\hline
Type &  Reality & Topology &  Parameters &  Meaning  \\
\hline
\hline
$\Gamma_A$ &  real &  open &  $(0, \varepsilon_A) \subseteq \Omega $ &  Fermi pairs  \\
\hline
$\Gamma_B$ & complex & open &  $\epsilon \pm i  \delta $ & BCS pairs \\
\hline
$\Gamma_C$ & complex & closed &   $\varepsilon_A, a=b$ & BCS pairs \\
\hline
$\Gamma_D$ &  real & open &  $(a, b) \not\subseteq \Omega  $ & BEC pairs \\
\hline
\end{tabular}

\vspace{0.5 cm}

Table 2: \footnotesize{Classification of the arcs containing the roots 
of the Bethe ansatz equation (\ref{a35}).} 
\end{center}

\begin{center}
\begin{tabular}{|c|c|c|c|}
\hline
Phase & $\Gamma$ \\
\hline 
\hline
Weak coupling BCS&   $\Gamma_A \cup \Gamma_B $ \\
\hline
Moore-Read line &  $\Gamma_A \cup \Gamma_C$ \\
\hline
Weak pairing &  $\Gamma_A \cup \Gamma_C \cup \Gamma_D$ \\
\hline
Read-Green line &  $\Gamma_D$ \\
\hline 
Strong pairing  &  $\Gamma_D$ \\
\hline
\end{tabular}

\vspace{0.5 cm}

Table 3: \footnotesize{Structure of the electrostatic arcs in the different phases of the 
$p+ip$ model.} 
\end{center}
Next we consider the solution of (\ref{a35})
in the different regions of the phase diagram.

\subsubsection{Weak coupling BCS phase}

In this region $\Gamma = \Gamma_A \cup \Gamma_B$, 
where $\Gamma_A = (0,\varepsilon_A) , \;\; \varepsilon_A < \omega$
and $\Gamma_B$ is a complex arc with end points 
$\epsilon \pm i \delta$. 
In the continuum limit, 
the roots belonging to $\Gamma_A$  lie
between the $z_k^2$. 
Recalling that the 
charge associated to the $y_m$ is $+1$, while that
associated to the $z^2_k$ is $- 1/2$, one finds that the effect of
the arc $\Gamma_A$ can  be taken into account by writing (\ref{a35})
as
\begin{eqnarray}
- \int_0^{\varepsilon_A}  d \varepsilon \frac{ \rho(\varepsilon)}{
\varepsilon - y} +  \int_{\varepsilon_A}^\omega  d \varepsilon \frac{ \rho(\varepsilon)}{
\varepsilon - y} 
- \frac{q_0}{y} - P \int_{\Gamma_B} |dy'| 
\frac{r(y')}{ y'- y} = 0,\;\; \forall y \in \Gamma_B.
\label{a39}
\end{eqnarray}

\noindent
To solve this equation we introduce a
function $h(y)$ which is analytic outside $\Gamma_B$, $\Omega$ and 0,  
and  has a branch cut on the arc $\Gamma_B$, such that 

\begin{eqnarray}
r(y) \; |d y| = \frac{1}{2 \pi i} (h_+(y) - h_-(y) ) \; dy, \qquad
y \in \Gamma_B 
\label{a40}
\end{eqnarray}

\noindent
where $h_+(y)$ and $h_-(y)$ are the limiting
values of $h(y)$ to the right and left of $\Gamma_B$. 
The appropiate ansatz for $h(y)$ is

\begin{eqnarray} 
h(y) = R(y) \left[ 
\int_0^\omega  d \varepsilon \frac{ \varphi(\varepsilon)}{ \varepsilon - y} -
\frac{Q}{y} 
 \right]
\label{a41}
\end{eqnarray}

\noindent 
where 

\begin{eqnarray}
R(y) = \sqrt{ (y -a) (y - b)} 
= \sqrt{ (y - \epsilon)^2 + \delta^2}
\label{a42}
\end{eqnarray}

\noindent
such that $R(\varepsilon)$ is positive
on the interval $(\omega_A, \omega)$ and 
negative on $(0, \omega_A)$. In particular $R(0) < 0$. 
Using (\ref{a40}) one can write the last term of (\ref{a39}) as

\begin{eqnarray}
P \int_{\Gamma_B} |dy'| 
\frac{r(y')}{ y'- y} =  \int_{L_B} \frac{dy'}{2 \pi i} 
\;  \frac{h(y')}{ y'- y}
\label{a43}
\end{eqnarray}

\noindent
where $L_B$ is a counterclockwise contour surrounding $\Gamma_B$. 
Plugging (\ref{a41}) into (\ref{a43})
and deforming the contour $L_B$ around $\Omega, 0$ and $\infty$ one finds

\begin{eqnarray*}
 \int_{L_B} \frac{dy'}{2 \pi i} 
\;  \frac{h(y')}{ y'- y} = 
- \int_0^{\omega_A} d \varepsilon \; \frac{ \varphi(\varepsilon) | R( \varepsilon)|}{ \varepsilon - y} 
+  \int_{\omega_A}^{\omega} d \varepsilon \; \frac{ \varphi(\varepsilon) R( \varepsilon)}{ \varepsilon - y}
- \frac{Q \; R(0)}{y} - \int_0^{\omega} d \varepsilon \varphi(\varepsilon) - Q.  
\end{eqnarray*}

\noindent
The solution of the electrostatic equation (\ref{a39})
is obtained with the choices

\begin{eqnarray}
\varphi(\varepsilon) = \frac{\rho(\varepsilon)}{ |R(\varepsilon)|},
\qquad Q = \frac{q_0}{R(0)}, \qquad 
\int_0^\omega  d \varepsilon \; 
 \frac{\rho(\varepsilon)}{ |R(\varepsilon)|}  = - Q = \frac{q_0}{|R(0)|}.
\label{a45}
\end{eqnarray}

\noindent
Observe that the last equation in (\ref{a45}) coincides, up to a rescaling of the variables,
with the chemical potential equation (\ref{chempotcont}) (use
 $|R(0)| = \sqrt{a b} = | \mu|$).  In the $s$-wave model
the analogue of (\ref{a45}) provides the gap equation.
However for the $p+ip$ model the roles of the gap
and chemical potential equations are reversed. 
To find the gap equation from the electrostatic model
one has to   
compute the number of roots on the arc $\Gamma$:

\begin{eqnarray}
M = \int_{\Gamma_A} |dy| r(y) + \int_{\Gamma_B} |dy| r(y)  
= 2 \int_0^{\varepsilon_A} d \varepsilon \; \rho(\varepsilon) + 
 \int_{L_B} \frac{dy}{2 \pi i} 
\; h(y)
\label{a46}
\end{eqnarray}

\noindent 
Plugging (\ref{a41}) into (\ref{a46}) and deforming 
the contour $L_B$ as done above,  one finds

\begin{eqnarray*}
\frac{1}{2 G} = \int_0^\omega  d \varepsilon \;  \frac{ \varepsilon \;  \rho( \varepsilon)}{
\sqrt{ (\varepsilon- \epsilon)^2 + \delta^2}}
\end{eqnarray*}

\noindent
where we have also used (\ref{a45}). This is the gap equation 
(\ref{gapcont}).  The energy of the solution can be computed 
from (\ref{a38})

\begin{eqnarray}
E = \int_{\Gamma_A} |dy| \;  y  \;  r(y) + \int_{\Gamma_B} |dy|
\;  y \;  r(y)  
= 2 \int_0^{\varepsilon_A} d \varepsilon \; \varepsilon \;  \rho(\varepsilon) + 
 \int_{L_B} \frac{dy}{2 \pi i} \;  y 
\; h(y).
\label{a48}
\end{eqnarray}

\noindent 
As in the case of equation (\ref{a45}), we  
substitute (\ref{a41}) into (\ref{a48}) and after the contour
deformation of $L_B$ one finds

\begin{eqnarray}
E = \int_0^\omega d \varepsilon \; \rho(\varepsilon) \; \varepsilon \left(
1 - \frac{ \varepsilon - \epsilon}{
\sqrt{ (\varepsilon- \epsilon)^2 + \delta^2}}
\right)  
\label{a49}
\end{eqnarray}
in agreement with (\ref{nrgcont}). 
\noindent
This formula is similar to the one found for the $s$-wave model 
 except for the fact that is missing 
a constant term $- \delta^2/4$ corresponding to the
condensation energy \cite{g95,rsd02}. Finally, the equation for the arc
$\Gamma_B$ is given by the equipotential equation

\begin{eqnarray*}
{\rm Re} \left[\int_{\epsilon - i \delta}^y dy' \; 
h(y')\right] = 0, \qquad  y \in \Gamma_B 
\end{eqnarray*}

\noindent 
while the density of roots is

\begin{eqnarray*}
r(y) = \frac{1}{\pi} |h(y)|, \qquad  y \in \Gamma_B. 
\end{eqnarray*}

\subsubsection{Moore--Read line}

This case is reached when the 
arc $\Gamma_B$ studied in the previous
subsection closes, i.e. $\delta = 0$, so that
$a = b = \epsilon$. The function $R(y)$ defined in eq. (\ref{a42})
becomes $R(y) = y -a$, so it no longer has a branch cut. 
To find the closed arc $\Gamma_C$ one introduces
an analytic function $s(y)$ satisfying

\begin{eqnarray*}
r(y) \; |d y| = s(y) \; dy, \qquad
y \in \Gamma_C. 
\end{eqnarray*}

\noindent 
The Bethe anstaz equation is similar to (\ref{a39}), i.e.

\begin{eqnarray}
- \int_0^{\varepsilon_A}  d \varepsilon \frac{ \rho(\varepsilon)}{
\varepsilon - y} +  \int_{\varepsilon_A}^\omega  d \varepsilon \frac{ \rho(\varepsilon)}{
\varepsilon - y} 
- \frac{q_0}{y} - P \int_{\Gamma_C} |dy'| 
\frac{r(y')}{ y'- y} = 0,\;\; \forall y \in \Gamma_C
\label{a53}
\end{eqnarray}

\noindent
where $\varepsilon_A$ is the point where $\Gamma_C$ cuts
the positive real axis. The analogue of (\ref{a43}) is

\begin{eqnarray}
P \int_{\Gamma_C} |dy'| 
\frac{r(y')}{ y'- y} =  \int_{\Gamma_C} dy' 
\;  \frac{s(y')}{ y'- y},
\label{a54}
\end{eqnarray}

\noindent
where $\Gamma_C$ is the counterclockwise contour
containing the complex roots. To solve (\ref{a53})
we choose the following ansatz

\begin{eqnarray}
s(y) = \frac{1}{i \pi} \left[ 
\int_\Omega d \varepsilon \frac{ \rho(\varepsilon)}{ \varepsilon - y} + 
\frac{q_0}{y} 
 \right]
\label{a55}
\end{eqnarray}

\noindent
which is motivated by the two level 
$s$-wave model where the roots  form
a closed loop when the coupling constant is less
than half the distance between the two levels 
\cite{g95,rsd02}. Introducing (\ref{a55}) into
(\ref{a54}) one can check that (\ref{a53}) holds (we
use that $\int_{\Gamma_C} dy'/(y' - y) = i \pi$ for
$y \in \Gamma_C$).  The total number of roots $M$ 
is given by 

\begin{eqnarray*}
M = 2 \int_0^{\varepsilon_A} d \varepsilon \; \rho(\varepsilon) + 
 \int_{\Gamma_C} d y \; s(y) = 2 q_0 \,\,\Longrightarrow\,\,
x = 1 - \frac{1}{g} 
\end{eqnarray*}
which shows that the criterion for the Moore--Read line is reproduced.   
The chemical potential equation is equivalent 
to the vanishing of the  potential
$s(y)$ at the point $a$ where $\Gamma_C$ cuts the real axis, i.e.

\begin{eqnarray*}
q_0 = - a \int_0^\omega d \varepsilon \frac{\rho(\varepsilon)}{ \varepsilon - a} 
\Longleftrightarrow s(a) = 0.
\end{eqnarray*}

\noindent
The latter condition follows from continuity of the weak
coupling case since the density of roots on the arc $\Gamma_B$
also vanishes at the end points. 

The equation of the arc $\Gamma_C$ is given by

\begin{eqnarray}
{\rm Im} \left[ \int_{y_0}^y d y' \; s(y') \right] =0, \qquad y \in \Gamma_C 
\label{a58}
\end{eqnarray}

\noindent 
where $y_0$ is a point of $\Gamma_C$ which can be taken as $a$. 
Doing the integral one finds

\begin{eqnarray}
{\rm Re} \left[ 
\int_\Omega d \varepsilon \rho(\varepsilon) 
\log \left( \frac{  \varepsilon - y}{ \varepsilon - y_0} \right) - 
q_0  \log \frac{y}{y_0} \right]  = 0. 
\label{a59}
\end{eqnarray}

In 1D this equation becomes, 
 in normalized variables (i.e. $y \rightarrow \omega y$)

\begin{eqnarray*}
{\rm Re} \left[ 
\sqrt{y} 
\log \left( \frac{ \sqrt{y} + 1}{ \sqrt{y} -1 } \right) 
- \sqrt{y_0} 
\log \left( \frac{ \sqrt{y_0} + 1}{ \sqrt{y_0} -1 } \right)
+ \log \left( \frac{ y - 1}{y_0 -1} \right) -  
(1 - \frac{1}{g})  \log \frac{y}{y_0} \right]  = 0, 
\end{eqnarray*}

\noindent 
while in 2D it  is 

\begin{eqnarray*}
{\rm Re} \left[ 
- y  \log \left( 1 - \frac{1}{y} \right) 
+  y_0  \log \left( 1 - \frac{1}{y_0} \right) 
+ \log \left( \frac{ y -1}{y_0 -1} \right) - 
(1 - \frac{1}{g})  \log \frac{y}{y_0} \right]  = 0. 
\end{eqnarray*}

These eqs. have two real solutions, corresponding to the
intersections points of $\Gamma_c$ with the real axis: 
$y_0 = a/\omega$ and $y = \varepsilon_A/\omega$. The total energy
can be computed in a way similar to eq. (\ref{a48})
and the result is

\begin{eqnarray*}
E = 2 \int_0^{\varepsilon_A} d \varepsilon \; \varepsilon \;  \rho(\varepsilon) + 
 \int_{L_C} dy \; y \;  s(y)   = 0, 
\end{eqnarray*}

\noindent 
which agrees with the mean-field result (\ref{mrnrg}). This result
can also be obtained as a limit of  eq.(\ref{a49}) by setting $\delta = 0$
and $\epsilon = a < 0$.

\subsubsection{Weak pairing phase}

This is the most complicated phase concerning the structure
of the arc $\Gamma$ which consists now of three pieces, 
$ \Gamma = \Gamma_A \cup  \Gamma_C \cup \Gamma_D $, 
where $ \Gamma_A = (0, \varepsilon_A)$, 
$\Gamma_D = (a,b)$ and $\Gamma_C$ is a closed
arc that touches the points $\varepsilon_B = b$ and $\varepsilon_A$ on the real line.
Combining the results of the two previous cases
we introduce a function $h(y)$ associated to the open arc 
$\Gamma_D$ 

\begin{eqnarray}
h(y) = R(y) \left[ 
\int_0^\omega d \varepsilon \frac{ \varphi(\varepsilon)}{ \varepsilon - y} -
\frac{Q}{y} 
 \right]
\label{a63}
\end{eqnarray}

\noindent 
and a function $s(y)$ associated to the closed
arc $\Gamma_C$, i.e. 

\begin{eqnarray}
s(y) = \frac{1}{i \pi} R(y) \left[ 
\int_0^\omega d \varepsilon \frac{ \varphi(\varepsilon)}{ \varepsilon - y} -
\frac{Q'}{y} 
 \right],
\label{a64}
\end{eqnarray}

\noindent
where

\begin{eqnarray*}
R(y) = \sqrt{ ( y -a) ( y - b)}, \qquad a < b < 0 
\end{eqnarray*}

\noindent
which is always positive for $\varepsilon \in \Omega$. 
The electrostatic equation (\ref{a35}) reads,

\begin{eqnarray*}
- \int_0^{\varepsilon_A}  d \varepsilon \frac{ \rho(\varepsilon)}{
\varepsilon - y} +  \int_{\varepsilon_A}^\omega  d \varepsilon \frac{ \rho(\varepsilon)}{
\varepsilon - y} 
- \frac{q_0}{y} =  \int_{L_D} \frac{dy'}{ 2 \pi i}  
\frac{h(y')}{ y'- y}  +  \int_{L_C} dy' 
\frac{s(y')}{ y'- y}, \qquad  y \in \Gamma_C \; \;   {\rm or} \; \; 
\Gamma_D 
\end{eqnarray*}

\noindent
and it is solved by the ansatze (\ref{a63}, \ref{a64}) provided

\begin{eqnarray}
\varphi(\varepsilon) = \frac{\rho(\varepsilon)}{ R(\varepsilon)},
\qquad 
Q = - \frac{q_0}{R(0)},   \qquad Q = Q'
\label{a67}
\end{eqnarray}

\noindent
and

\begin{eqnarray}
\int_0^\omega  d \varepsilon \frac{ \rho(\varepsilon)}{R(\varepsilon)}  = - Q = \frac{q_0}{R(0)}.
\label{a68}
\end{eqnarray}

The last equation coincides with the chemical potential equation
(\ref{chempotcont}) since $\mu =  R(0) = a b >0$. On the other hand
the gap eq.(\ref{gapcont}) is obtained from the counting of the 
number of roots $M$

\begin{eqnarray}
M = 2 \int_0^{\varepsilon_A} d \varepsilon \; \rho(\varepsilon) + 
 \int_{\Gamma_C} d y \; s(y) + \int_{L_D} \frac{dy}{2 \pi i}  \;
h(y)  \Longrightarrow \int_0^\omega d \varepsilon \; \frac{ \varepsilon \rho(\varepsilon)}{
\sqrt{(\varepsilon -a) ( \varepsilon - b)}} = \frac{1}{2 G} .
\label{a69}
\end{eqnarray}

\noindent
The total energy is given by

\begin{eqnarray}
E &=& 2 \int_0^{\varepsilon_A} d \varepsilon \; \varepsilon \;  \rho(\varepsilon) + 
 \int_{\Gamma_C} d y \; y \;  s(y) + 
 \int_{L_D} \frac{dy}{2 \pi i} \;  y 
\; h(y) \nonumber \\
 &=& \int_0^\omega d \varepsilon \; \rho(\varepsilon) \; \varepsilon \left(
1 - \frac{ \varepsilon - (a + b)/2}{
\sqrt{ (\varepsilon- a)(\varepsilon -b)}}
\right)  
\label{a70}
\end{eqnarray}

\noindent
which coincides with (\ref{nrgcont}). 
Finally the equation for the closed arc $\Gamma_C$ is the same
as (\ref{a58}) with $s(y)$ replaced by (\ref{a64}).
The results obtained so far must reproduce those found ealier
for the Moore-Read line. Indeed, if $ a \rightarrow b$, one
can check that the function $s(y)$ given in 
(\ref{a64}) coincides with (\ref{a55}).

\subsubsection{Read--Green line}

This case can be obtained from the previous one taking the limit
$b \rightarrow 0$. One can easily prove that the number of roots
on $\Gamma_A$ and $\Gamma_C$ is zero so that all the roots
lie in the open arc $\Gamma_D$. The gap and chemical potential
equations and energy are given by (\ref{a68},\ref{a69},\ref{a70})
setting $b = 0$.  The criticality of this line can be seen
from the fact that the function $R(\varepsilon)/2$, which is the mean-field
energy of the excitations, 
vanishes as $\sqrt{\varepsilon} \propto |k|$ when $|k| \rightarrow 0$.
Note that this also means that the density of roots is divergent at the origin, as can be deduced from (\ref{a63}).

\subsubsection{Strong pairing phase} \label{spp}

In this case all the roots lie on the open arc 
 $\Gamma_D = (a,b)$  belonging to the negative real axis. 
As in the weak pairing phase we introduce the function $h(y)$
given by (\ref{a63}), so that the electrostatic equation
becomes

\begin{eqnarray*}
\int_0^{\omega}  d \varepsilon \frac{ \rho(\varepsilon)}{
\varepsilon - y} 
- \frac{q_0}{y} =  \int_{L_D} \frac{dy'}{ 2 \pi i}  
\frac{h(y')}{ y'- y}, \qquad y \in \Gamma_D.  
\end{eqnarray*}

\noindent 
The solution of this equation is provided by 

\begin{eqnarray*} 
\varphi(\varepsilon) = \frac{\rho(\varepsilon)}{ R(\varepsilon)},
\qquad Q = \frac{q_0}{R(0)} 
\end{eqnarray*}

\noindent
and 

\begin{eqnarray*}
\int_\Omega \varphi(\varepsilon) = - Q = \frac{|q|}{R(0)}.
\end{eqnarray*}

\noindent
Notice the change of sign of the equation for $Q$
as compared to (\ref{a67}). Computing the number of roots on the arc gives
the gap equation:

\begin{eqnarray*}
\frac{1}{2 G} = \int_0^\omega d \varepsilon \;  \frac{ \varepsilon  \rho(\varepsilon)}{
\sqrt{(\varepsilon -a) ( \varepsilon -b)}}.  
\end{eqnarray*}

\noindent
The total energy is given by

\begin{eqnarray*}
E =  
 \int_{L_D} \frac{dy}{2 \pi i} \;  y 
\; h(y) = \int_0^\omega d \varepsilon \; \rho(\varepsilon) \; \varepsilon \left(
1 - \frac{ \varepsilon - (a + b)/2}{
\sqrt{ (\varepsilon- a)(\varepsilon -b)}}
\right) . 
\end{eqnarray*}

\subsection{Topological properties of the ground state}  \label{topo}

\subsubsection{Dressing and duality}

Above we established, through use of the electrostatic analogy, that the continuum limit of the Bethe ansatz equations for the ground state is in agreement with the mean-field solution. However with reference to Fig. \ref{arcsplot}, we see a noticeable discrepancy between the theoretical arc and the position of the roots near the Moore-Read line. Here we re-examine the Bethe ansatz equations in the case of finite-sized systems to clarify some subtle issues about the nature of the ground state. Using an algebraic approach we will introduce the concept of dressing of states, which gives a different perspective on the duality that was seen in the mean-field analysis.

For a generic splitting of the set of roots $Y$ into nonintersecting sets $Y'$ and $Z$ such that $Y=Y' \cup Z$, we can express the Bethe ansatz equations (\ref{pbae}) as    
\begin{eqnarray}
&&{G^{-1}-L+2M-1} +\sum_{k=1}^L \frac{y_m}{y_m-z_k^2}\nonumber \\
&&\qquad \qquad  = \sum_{y_j\in Y', y_j\neq y_m} \frac{2y_m}{y_m-y_j}+ \sum_{y_j\in Z} \frac{2y_m}{y_m-y_j},   \quad y_m \in Y'  \label{A}
\\
&&{G^{-1}-L+2M-1} +\sum_{k=1}^L \frac{y_m}{y_m-z_k^2} \nonumber \\
&&\qquad \qquad = \sum_{y_j\in Z, y_j\neq y_m} \frac{2y_m}{y_m-y_j}+\sum_{y_j\in Y'} \frac{2y_m}{y_m-y_j},    
\quad y_m \in Z. \label{B}
\end{eqnarray}    
Setting $|Y'|=M'$ and $|Z|=P$ we take the sum over elements in $Z$ in (\ref{B}) to give
\begin{eqnarray}
&&P({G^{-1}-L+2M-1}) +\sum_{y_m\in Z}\sum_{k=1}^L \frac{y_m}{y_m-z_k^2} \nonumber\\
&& \qquad \qquad = \sum_{y_j\in Z,y_m\in Z,y_j\neq y_m} \frac{2y_m}{y_m-y_j}+\sum_{y_j\in Y' ,y_m\in Z} \frac{2y_m}{y_m-y_j}
\nonumber \\
&& \qquad \qquad = P(P-1)+\sum_{y_j\in Y' ,y_m\in Z} \frac{2y_m}{y_m-y_j}    
\label{C}
\end{eqnarray}
Suppose that at some limiting value of $G$ we have 
\begin{eqnarray*}
y_m &\neq& 0 \qquad{\rm for\,all}\,\, y_m \in Y' , \\
y_m &=& 0 \qquad{\rm for\,all}\,\, y_m \in Z .
\end{eqnarray*}
Taking note that $M'+P=M$ 
equation (\ref{C}) informs us that 
\begin{eqnarray}P=G^{-1}-L+ 2M\quad \Rightarrow \quad P=L-2M'-G^{-1},
\label{em}
\end{eqnarray} while from (\ref{A}) we obtain
\begin{eqnarray}
{G^{-1}-L+2M-1} +\sum_{k=1}^L \frac{y_m}{y_m-z_k^2} &=& \sum_{y_j\in Y', y_j\neq y_m} \frac{2y_m}{y_m-y_j}+ 2P ,   \quad y_m \in Y' \nonumber \\ 
\Rightarrow {G^{-1}-L+2M'-1} +\sum_{k=1}^L \frac{y_m}{y_m-z_k^2} &=& \sum_{y_j\in A, y_j\neq y_m} \frac{2y_m}{y_m-y_j},   \quad y_m \in Y'. \label{new}
 \end{eqnarray}
Equation (\ref{new}) is simply the set of Bethe ansatz equations for a system of $M'$ Cooper pairs. The above calculations suggest that given such a solution set $Y'$, we can augment it with $P$ additional roots which all have zero value to obtain the solution set $Y$, provided that $P$ is given by (\ref{em}). It is important to check that this solution set is indeed valid for there are known examples where the roots of the Bethe ansatz equations must be distinct, most noticibly the repulsive Bose gas \cite{ik82}.  

A fairly straightforward calculation, using proof by induction, leads to the following commutator identity for $H$ in the form (\ref{pham})
\begin{eqnarray}
\left[H,\,C^P(0)\right]= P Q^\dagger C^{P-1}(0)(GL-1-GP-2GN) 
\label{dressingcom}
\end{eqnarray} 
where 
\begin{eqnarray*}
Q^\dagger&=&\sum_{k=1}^L z_k b_k^\dagger. 
\end{eqnarray*}
If we partition the Hilbert space of states into subspaces defined by the eigenvalues $M'$ of the Cooper pair number operator $N$, there exists a subspace on which the commutator (\ref{dressingcom}) vanishes 
whenever 
$P=L-2M'-G^{-1}$ which is precisely equation (\ref{em}).
That is to say given any eigenstate $|\phi(Y')\rangle$ with $M'$ Cooper pairs, we can construct a new eigenstate $|\phi(Y)\rangle=C^P(0)|\phi(Y)\rangle$ of $M=M'+P$ Cooper pairs and the same energy as $|\phi(Y')\rangle$ provided (\ref{em}) holds. 
Whenever this is the case we say that the state $|\phi(Y)\rangle$ is a {\it dressing} of the state $|\phi(Y')\rangle$. 

To explore in more depth the consequences of this result, we first note that $P$ must necessarily be a non-negative integer.    
In terms of the filling fraction $x'=M'/L$, non-negativity of $P$ implies that 
$$ x' < \frac{1}{2}\left(1-\frac{1}{g}\right) $$
so that the state $|\phi(Y')\rangle$ belongs to the strong pairing phase. Letting $x=M/L=(M'+P)/L$ denote the filling of the dressed state we find 
$$x+x'=1-\frac{1}{g}$$
which is simply the duality relation (\ref{mfduality}) obtained from the mean-field theory. We can immediately conclude  that $|\phi(Y)\rangle$ belongs to the weak pairing phase. However one significant point about this derivation of the duality is that it only applies in particular cases when the coupling $g$ is {\it rational}. This aspect has some ramifications to which we will return later. 

For the moment, we will consider a simple example of dressing. Starting with the sector of the Hilbert space where the number of Cooper pairs is zero there is only one state, that being the vacuum $|0\rangle$. We can dress this state with $M$ additional Cooper pairs where, by equation (\ref{em}), we must choose 
\begin{eqnarray}
M=P=L-G^{-1}. 
\label{mprime}
\end{eqnarray}
There is no reason yet to expect that the dressed state will be the ground state in the sector with $M$ Cooper pairs. However we do recognise that (\ref{mprime}) is precisely the equation of the MR line (\ref{mrline}) found from mean-field theory, for which the mean-field ground-state energy is zero. The dressed state will have the same energy as the vacuum, which is also zero. Performing the dressing gives us  
\begin{eqnarray*}
C^{M}(0)|0\rangle &=& \left [ \sum_{k=1}^L \frac{b_k^\dagger}{z_k} \right]^{M}  |0\rangle 
=  \left [ \sum_{\bk\in{\bf K}_+} \frac{c_\bk^\dagger c^\dagger_{-\bk}}{k_x+ik_y}\right]^{M}  |0\rangle 
\end{eqnarray*}
which is precisely the MR state (\ref{mrstate}). Recall that the projected mean-field wavefunction gives the MR state as the ground state on the MR line with zero energy. We can show that this is indeed the ground-state. In fact, in all instances dressing a ground state in the strong pairing phase  leads to a ground state in the weak pairing phase, which we now show follows from the Perron-Frobenius theorem.

By adding an appropriate constant term to (\ref{pham}) we obtain an operator with strictly negative entries for each sector of $M'$ pairs. The Perron-Frobenius theorem tells us that the ground-state vector of this operator can be normalised such that all components of the vector are strictly positive. 
Applying the dressing operator   $C^P(0)$ with $P$ given by (\ref{em}) to the ground-state of $M'$ pairs leads to an eigenstate of (\ref{pham}) in the weak pairing phase, which also has strictly positive components. By the Perron-Frobenius theorem, this must be the ground state in the weak pairing phase for $M'+P$ pairs.     

It is of importance to compare this result with the previous discussions of the solution in the thermodynamic limit described in Subsection \ref{tdl}. There, it was concluded that on the MR line the ground-state roots formed a closed arc in the complex plane. In the finite system, we have now found that all ground-state roots are zero on the MR line. In both instances the ground-state energy was found to be zero. One way to resolve this apparent contradiction would be to suggest that the assumption used in Subsection \ref{tdl}, viz. that in the thermodynamic limit the ground-state roots form a dense arc in complex plane, was not valid.    
To investigate this matter further we next turn our attentions to the numerical solution of the Bethe anastz equations.

\subsubsection{Numerical solution of the Bethe ansatz equations for finite systems} \label{numerics}

First we will discuss the behaviour of the roots in the three different regions. Later we will discuss properties across all regions. The numerical solutions of the Bethe ansatz equations are obtained starting from the initial
condition $y_m \rightarrow (1 + G) \bk^2 \; (m = 1, \dots, M)$
as $G \rightarrow 0$, with the $\bk$ chosen to fill the Fermi sea. 
As $g$ increases, the roots $y_m$, closest to the
Fermi level become complex pairs.
When $g$ approaches the MR-line  the roots
bend towards the origin, and at the value $x = x_{MR}$
all the roots collapse onto the origin.
At larger values of $g$ one enters the weak pairing
phase where all the roots are non-zero, except at some rational
values of $g$ where a fraction of the roots collapse
again. Finally, in the strong pairing regime all the roots become real
and they belong to the interval $(a,b)$ on the negative real
axis. The RG-line is obtained when $b=0$, in which
case the quasiparticle energy (\ref{spectrum}) becomes gapless.

\begin{figure}[H]
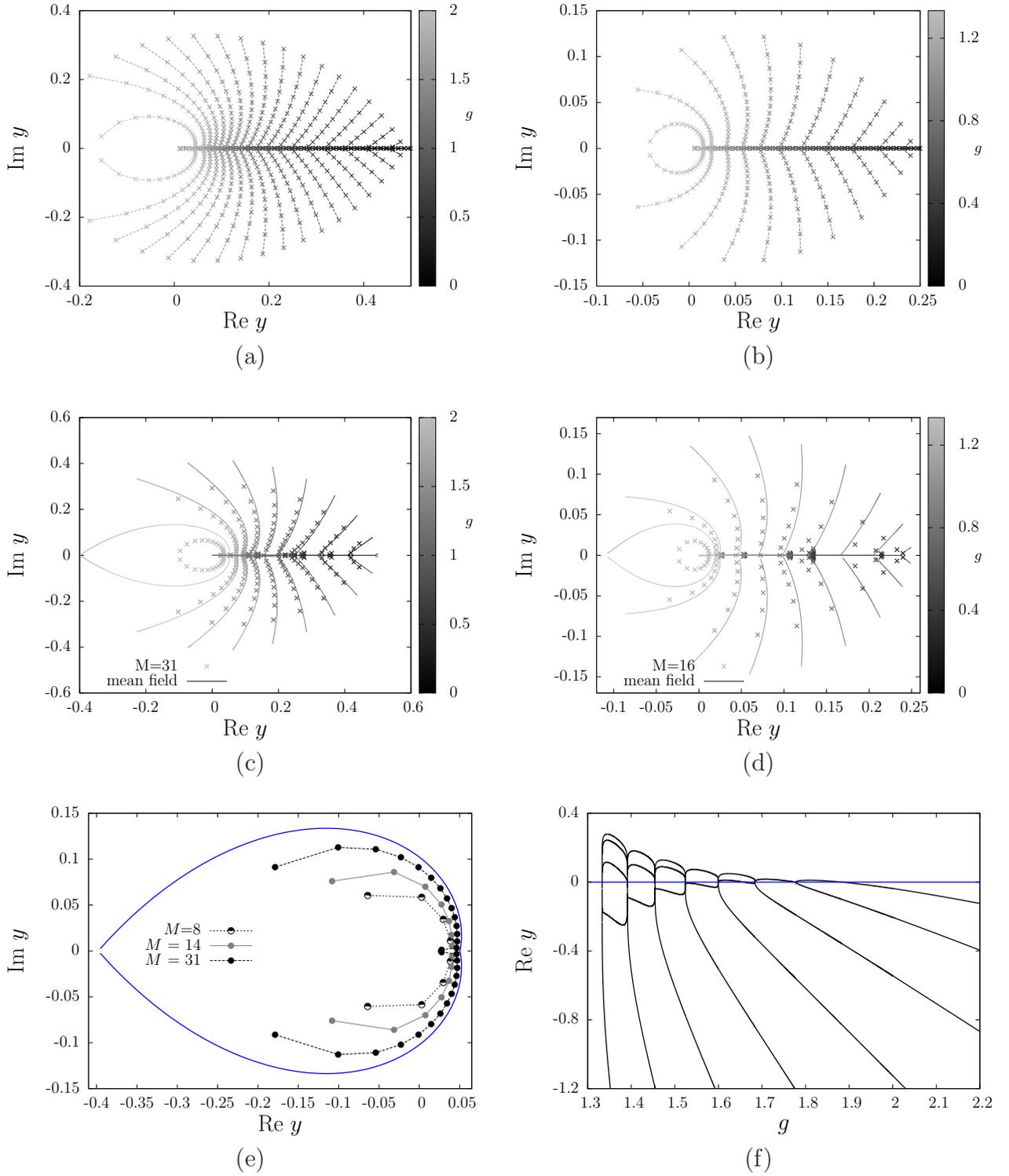

\begin{center}
$$
\begin{array}{cc} 
\scalebox{0.65}{\input{1Darcsx05.tex}} & 
\scalebox{0.65}{\input{1Darcsx025.tex}} \\
({\rm a}) & ({\rm b})
\end{array}
$$
$$
\begin{array}{cc} 
\scalebox{0.65}{\input{2Darcsx05.tex}} & 
\scalebox{0.65}{\input{2Darcsx025.tex}} \\
({\rm c}) & ({\rm d})
\end{array}
$$
$$
\begin{array}{cc} 
\scalebox{0.65}{\input{criticalArc.tex}} &
\scalebox{0.65}{\input{realswpM8.tex}} \\
({\rm e}) & ({\rm f})
\end{array}
$$
\end{center}
\caption{\footnotesize{Real and imaginary parts of the roots of the Bethe ansatz equations in several regimes for several values
of the coupling constant $g$. In the cases (c,d,e) the corresponding electrostatic arcs are also shown: 
(a)  1D model, $M=31, L=62$, 
(b)  1D model,  $M=16, L=64$, 
(c) 2D model,   $M=31, L=62$,
(d)   2D model,  $M=16, L=64$
(e)  2D model,  $g = 1.99$,  $M= 8,14,31,L=2M$. Notice that the MR point is $g=2$, $x=1/2$.
(f) Real part of the Bethe ansatz equation roots for 2D momentum distribution with  $M=8, L=32$. The range of $g$ covers both the weak pairing and strong pairing phases. The dressing points, in which some of the roots collapse to zero, can be seen to occur at regular intervals between $g_{MR}=4/3$ and $g_{RG}=2 $.
}}
\label{arcs}
\end{figure}

In the weak coupling regime the 
parameters $a,b$ appearing in (\ref{ab}) 
give the end points of the complex arcs shown
in Figs. \ref{arcs}(a) - \ref{arcs}(d) for the 1D and 2D models (recall Table 1). On the MR
line $a$ and $b$ become real and equal, giving $E=0$ from (\ref{abnrg}), and the complex arc
closes satisfying  eq. (\ref{a59}).

Fig. \ref{arcs}({e})  shows how the roots approach this
closed arc for increasing values of $M$ at $g = 1.99$
and $x= 1/2$. For $g=2$ and $x=M/L=1/2$ (MR line) 
all roots collapse to the origin in any finite system. The continuous arc is the solution of eq. (\ref{a59})
corresponding to $g=2, x=1/2$, where $L\rightarrow \infty$ is taken before $g\rightarrow 2$. 
However the exact solution
of the Bethe ansatz equations (\ref{pipbae}) is $y_m = 0, \forall m$, exactly giving the 
Moore-Read state \cite{mr91}  $C(0)^M |0 \rangle$ 
which also has  total energy $E=0$. In Fig. \ref{arcs}(f) we plot  the 
real part of the Bethe ansatz equations roots for a 2D momentum distribution in a system with $M=8, L=32$. 
The Moore-Read points, in which all the roots collapse to zero, are clearly  noticeable. 
This fact presents a paradoxical situation where 
the thermodynamic limit does not commute with the
limit where the coupling $g$ approaches the MR line from the weak coupling BCS phase.

In the strong pairing phase we expect from the calculations in the continuum limit that the ground-state roots should be real and negative. Moreover, the Perron-Frobenius theorem indicates that there can only be one such set of roots with this property. This fact makes the numerical analysis of the ground-state roots fairly efficient in the strong pairing phase,    
since the solution converges quickly to the unique real and negative solution set. In all instances we considered we always found a numerical  solution set of real and negative roots in this phase.  

\begin{figure}[h!]
\begin{center}
\includegraphics[height= 4.9cm,angle=0]{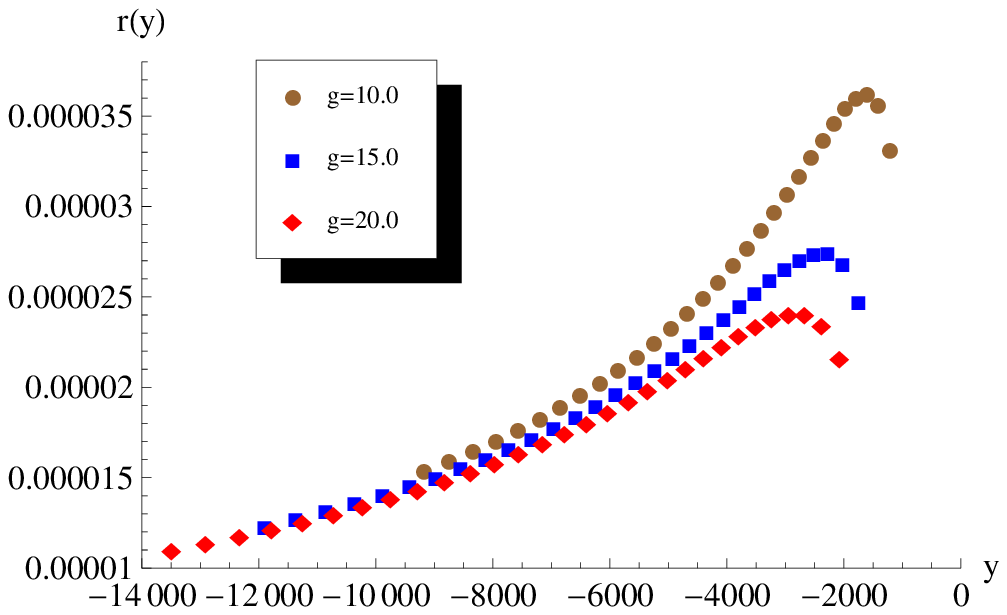}
\includegraphics[height= 4.9cm,angle=0]{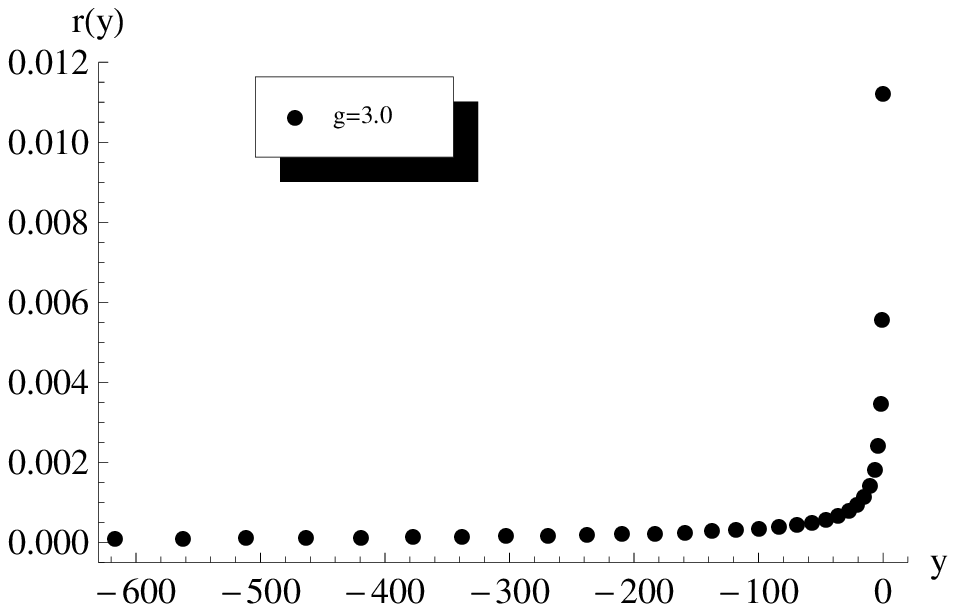}
\end{center}
\caption{\footnotesize{Discrete root densities as given by (\ref{drd}) for $L=450$ and $M=150$. The momenta have been chosen as $|\bk|=m,\,m=1,...,450$.  The left graph shows results for the coupling choices $g=10,15,20$, which correspond to the strong pairing phase. The support of the discrete root density lies on the negative real axis, and in each case the maximum value does not occur at the endpoint. For comparison, the right graph shows analogous results for $g=3$ corresponding to the Read--Green line. Here the maximum value of the discrete root density occurs at the endpoint of the support at $y=0$. Note also the different scale for the vertical axis. 
}}
\label{rootdensity}
\end{figure}

To illustrate the results for this phase we define the discrete root density to be 
\begin{eqnarray}
r(y_j)=\frac{1}{(M-1)(y_{j+1}-y_j)}.
\label{drd}
\end{eqnarray}
which we have plotted in Fig. \ref{rootdensity} for a system with $L=450$ and $M=150$. The results compare the cases $g=3$, the Read-Green line, and $g=20$. As discussed in Subsection \ref{spp} the continuum limit of the Bethe ansatz equations  predicts a divergence in the root density at the origin when the coupling is on the Read-Green line. The numerical results support this picture with the value of the root density at the origin for $g=3$ several times larger than for $g=20$.    

\begin{figure}[h!]
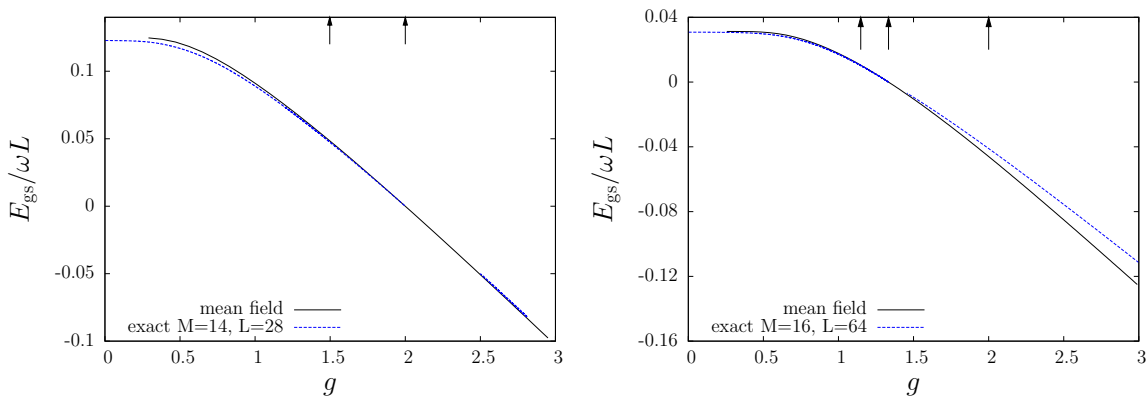

\begin{center}
\scalebox{0.6}{\input{totalEx05.tex}} 
\scalebox{0.6}{\input{totalEx025.tex}} 
\end{center}
\caption{\footnotesize{
Comparison between the exact and mean-field ground state energy in 2D for
$x=1/2, M=14, L=28$ (left) and  $x=1/4, M=16, L= 64$ (right). The arrows indicate, from left to right,
the values of the coupling $g$ corresponding to the Volovik, Moore-Read and Read-Green 
lines. For $x=1/2$ the latter point is absent. 
}}
\label{totalE}
\end{figure}

In Fig. \ref{totalE} we plot the exact and mean-field energy of the ground state  of a  2D system, showing results for the two 
filling fractions $x=1/2, 1/4$.  Observe that at the MR coupling the exact and mean-field energies
coincide, i.e. $E=0$. For $x=1/4$ there is a departure between the exact and mean-field results
in the strong pairing phase, which disappears for large systems. 

\begin{figure}[h!]
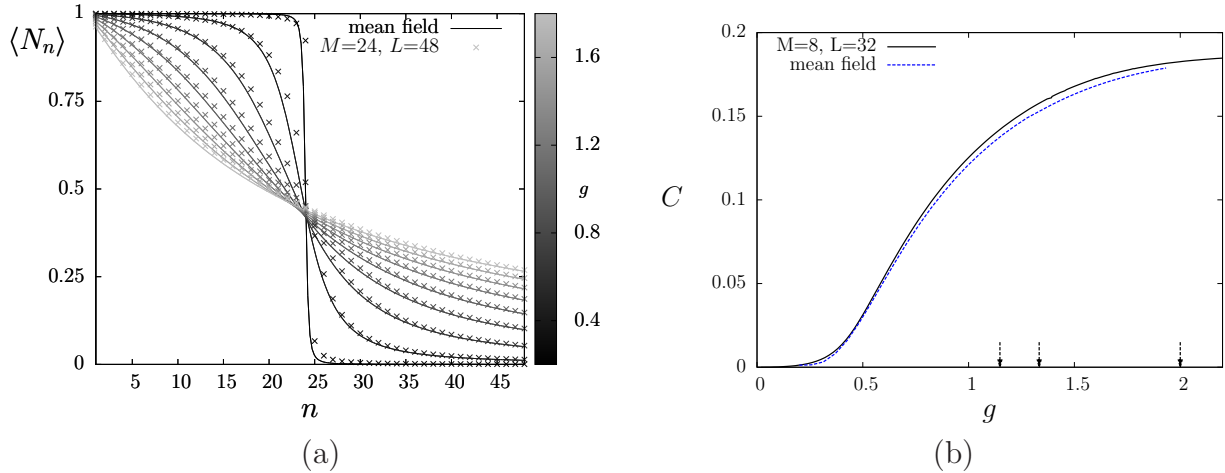

\begin{center}
$$
\begin{array}{cc}
\scalebox{0.65}{\input{Ns.tex}} &
\scalebox{0.62}{\input{Cparam.tex}} \\
({\rm a}) & ({\rm b})
\end{array}
$$
\end{center}
\caption{\footnotesize{a) Cooper pair expectation values of the $n$-th level, $\langle N_n\rangle$, for a 2D momentum distribution system with  $M=24$, $L=48$, for several values of $g$. The values of $g$ corresponding to the weak coupling BCS phase. The mean-field counterpart of this magnitude is also shown for each value of $g$ and $x$=1/2 (continuous lines). (b) The parameter $C$ as given by (\ref{Cparameter}), measuring fluctuations of the expected value of the ocupation number is plotted as a function of $g$. Results are shown using the roots of the Bethe ansatz equations for a $M$=8, $L$=32 system, and for mean-field case in the continuum limit with $x=1/4$. The arrows indicate the Volovik, Moore-Read and Read-Green points.}}
\label{cooper}
\end{figure}

In Fig. \ref{cooper}(a) we plot the expectation values of the Cooper pair density $\langle N_n \rangle$
(using  \ref{averagen}  in Appendix \ref{scalarproduct}). 
In Fig. \ref{cooper}(b) we plot the exact and mean-field  values of the 
the fluctuation parameter $C$ defined as
\begin{eqnarray}
C =  \frac{1}{L}  \sum_{j=1}^L \left( \langle N_j \rangle -  \langle N_j \rangle^2 \right).
\label{Cparameter}
\end{eqnarray}  
In both of these instances there is reasonably good agreement between the mean-field and the exact results.


%
%
%

\subsubsection{A zeroth-order quantum phase transition}

We have already observed that the MR line has the peculiar property that the behaviour of the roots of the Bethe ansatz equations in the limits $g \rightarrow g_{MR}$, $L\rightarrow\infty$ depends on the order in which the limits are taken. Another curious property  of the MR line is the  discontinuity 
of the ground-state energy $E(g,x)$ in the thermodynamic limit as the filling fraction
$x$ approaches the value $x_{MR}$ from the weak pairing region. 
To derive this result, for finite $L$
we take the one-pair state and dress it to give the dual 
GS in the weak pairing region. The filling $x_I$
of the dressed state is given by (\ref{mfduality}), setting  $x_{II} = 1/L$, i.e.  
%
$
x_I  = x_{MR}  - {1}/{L}, 
$
%
which approaches $x_{MR} = 1 - 1/g$
as $L \rightarrow \infty$.
Since the MR pairs carry no energy, the ground-state 
energy of the dressed state coincides with the one-pair energy. To compute this energy
we consider the Bethe ansatz equation  for one Cooper pair
and take the continuum limit (i.e. eq. (\ref{pipbae}) with $M=1$). For simplicity we take the momentum distribution to be that for free particles in 2D, i.e. $\rho(\varepsilon)=L/\omega$. This leads to   
%
$$
L - \frac{1}{G} - 1 =  
 \sum_{\bk \in {\bf K}_+} 
\frac{y}{y - {\bf k}^2}\; 
\Longrightarrow 
1 - \frac{1}{g} = y \; \int_0^\omega \frac{d \varepsilon}{\omega} 
\frac{ 1}{ y - \varepsilon}. 
$$
This equation
has a unique negative energy solution 
$y <0$ satisfying
%
$$1 - \frac{1}{g} = \frac{y}{\omega} \log \left(
\frac{ y}{y - \omega} \right) 
$$
which we denote as $y = {\mathcal{E}}(g)$. 
From here one derives the aforementioned discontinuity on the MR line $x_{MR}=1-g^{-1}$, 
%
$$\lim_{L \rightarrow \infty} E(g, x_I) = {\mathcal{E}}(g) 
\neq E(g,x_{MR}) = 0
$$
which may reasonably be termed a zeroth-order quantum phase transition. We note that in thermodynamic systems analogous zeroth-order transitions have been discussed in \cite{c02,ds02,m04}.

All of the above considerations point towards the MR line signifying a phase boundary which has significantly different properties to the RG line. In the latter case, both the excitation spectrum is gapless and the mean-field fidelity susceptibility is divergent. These are not properties associated with the MR line.

\subsubsection{Winding number of wavefunctions}

Here we formulate a description of the three phases in terms of the ground-state wavefunction topology. To this end we will use a winding number approach.



Let us  consider a continuous map $S^2 \rightarrow S^2$, given by
a function ${\mathfrak g}: (X,Y) \rightarrow  ({\mathfrak g}_X,{\mathfrak g}_Y)$ where $(X,Y)$ and $({\mathfrak g}_X,{\mathfrak g}_Y)$
are the stereographic coordinates of the spheres. 
The winding number is defined as
\begin{eqnarray}
w = \frac{1}{\pi} \int  
 \frac { d {\mathfrak g}_X \; d{\mathfrak g}_Y }{(1 + {\mathfrak g}_X^2 + {\mathfrak g}_Y^2)^2}
=  \frac{1}{\pi} \int_{\mathbb{R}^2}  dX \; dY 
 \frac{\partial J({\mathfrak g}_X, {\mathfrak g}_Y)}{\partial(X, Y)}
\frac{1}{(1 + {\mathfrak g}_X^2 + {\mathfrak g}_Y^2)^2}
\label{windingdef1}
\end{eqnarray}
The function  
$\displaystyle \frac{\partial J({\mathfrak g}_X, {\mathfrak g}_Y)}{\partial(X, Y)}$ is the Jacobian
of ${\mathfrak g}$. An alternative way to write (\ref{windingdef1}) is
\begin{eqnarray}
w 
=  \frac{1}{\pi} \int_{\mathbb{R}^2}  dX \; dY 
\left( \frac{\partial {\mathfrak g}}{\partial z}   \frac{\partial \bar{{\mathfrak g}} }{\partial \bar{z}}
-  \frac{\partial {\mathfrak g}}{\partial \bar{z}}   \frac{\partial \bar{{\mathfrak g}} }{\partial z}
\right) \frac{1}{(1 + {\mathfrak g} \bar{{\mathfrak g}})^2}
\label{windingdef2}
\end{eqnarray}
where ${\mathfrak g} = {\mathfrak g}_X + i {\mathfrak g}_Y$ and $z = X + i Y$.

To start, we consider the exact eigenstate of (\ref{pipham}) for one Cooper pair, as given by (\ref{pipstates}). In the limit $L\rightarrow \infty$ with one Cooper pair there are only two ground-state phases, the weak coupling BCS phase and the strong pairing phase, with the transition point given by $g=1$ at which the ground-state energy is zero. In first quantized formulation we have that the momentum-space wavefunction is 
$$
{\mathfrak g}({\bf k} ) = \frac{ k_x-ik_y}{ \bk^2 - E}
$$
where $E$ denotes the energy.
In terms of the complex variable $z=k_x+ik_y$, the Jacobian is given by
$$
J =  \frac{\partial {\mathfrak g}}{\partial z}   \frac{\partial \bar{{\mathfrak g}} }{\partial \bar{z}}
-  \frac{\partial {\mathfrak g}}{\partial \bar{z}}   \frac{\partial \bar{{\mathfrak g}} }{\partial z}
= \frac{ ( z \bar{z})^2 - E \bar{E}}{ ( z \bar{z} - E)^2 ( z \bar{z} - \bar{E})^2} .
$$
The winding number is then obtained as 
\begin{eqnarray*}
w 
&=&  \frac{1}{\pi} \int_{\mathbb{R}^2}  dk_x \; dk_y 
\frac{ ( z \bar{z})^2 - E \bar{E}}{ [ z \bar{z} +  ( z \bar{z} - E)  ( z \bar{z} - \bar{E})]^2} \\
&=& \int_0^\infty d u   \frac{ u^2 -  E \bar{E}}{[ u + (u - E) ( u - \bar{E})]^2} \\
&=& - \left[ \frac{ u}{u - ( E + \bar{E}) u + u^2 + E \bar{E}} \right]^\infty_0 
\end{eqnarray*}
which yields
\begin{eqnarray}
w = \left\{
\begin{array}{cc}
1, & E = 0, \\
0, & E \neq 0. 
\\
\end{array}
\right.
\label{wonepair}
\end{eqnarray}
In this manner we find a topological change in the wavefunction when the energy is zero, or equivalently $g=1$, consistent with the phase diagram Fig. \ref{volovik-phase}.  However in both the weak coupling BCS phase and the strong pairing phase the wavefunction topology is trivial. 

It is useful to contrast the above with the analogous result for the mean-field wavefunction. Up to a factor of $\hat{\Delta}^*$ which can be omitted, we have from (\ref{gees},\ref{ab})
$$
{\mathfrak g}(z,\bar{z}) =  \frac{ e(z, \bar{z})  - z \bar{z} + \mu}{ z},
\quad \;\; e(z, \bar{z}) =  \sqrt{ ( z \bar{z}  - a) ( z \bar{z} - b)}
$$
where $\mu^2  = a b$. The Jacobian is given by
$$
J = \frac{1}{ (2 z \bar{z}  \;  e(z, \bar{z}))^2} 
\left[ 
\left( ( a + b) z \bar{z} - 2 a b - 2 \mu \;  e(z,\bar{z}) \right)^2 - 
(z \bar{z})^2 \left(  2 z \bar{z} - a - b - 2 e(z, \bar{z}) \right)^2
\right].
$$
We have found numerically that
\begin{eqnarray}
w = \left\{
\begin{array}{cc}
1, & \mu > 0,   \\
a/(a - 1) , & \mu = 0,  \\ 
0,  & \mu < 0 \\  
\end{array}
\right.
\label{wBCS}
\end{eqnarray} 
where $\mu = 0$ corresponds to a choice $a < 0, b = 0$.
In this case the winding number is not an integer at $\mu=0$ . The reason for this is that setting $\mu=0$ gives  
$$ {\mathfrak g}(z, \bar{z}) = \sqrt{ \frac{\bar{z}}{z} } \left( 
\sqrt{z \bar{z} - a} - \sqrt{z \bar{z}} 
\right)  
$$
which is not continuous at $z=0$.  
The above shows very different topological predictions between the exact wavefunction and the mean-field wavefunction, specifically with the mean-field wavefunction having non-trivial topology in the weak coupling BCS phase. 

To gain a further understanding of the many-body system, next we consider the exact wavefunction for two Cooper pairs: 
\begin{eqnarray}
\psi(z_1,\bar{z}_1, z_2,\bar{z}_2) = \left\{ 
\begin{array}{cc} 
{\mathfrak g}(z_1,\bar{z}_1, E_1) {\mathfrak g}(z_2,\bar{z}_2, E_2) + {\mathfrak g}(z_1,\bar{z_1}, E_2) 
{\mathfrak g}(z_2,\bar{z}_2, E_1),
 & z_1 \neq z_2, \\
0, & z_1 =z_2 
\end{array}
\right.
\label{twopair}
\end{eqnarray}
where ${\mathfrak g}(z,\bar{z},E)$ is the one-pair wavefunction and $E_1,\,E_2$ are roots of the Bethe ansatz equations (\ref{pipnrg}), which can either be real or a complex conjugate pair. In order to apply the winding number approach described above we reduce  
(\ref{twopair}) to a complex-valued function by setting 
$$
{\mathfrak g}(z,E_1, E_2,  c) = {\mathfrak g}(z,E_2, E_1,  c) = \psi(z, z+c, E_1, E_2)
$$
for some $c\neq 0$. Numerically we find  
$$
w = \left\{
\begin{array}{ccc}
2, & E_1 = 0,  & E_2 = 0, \\
1, & E_1 = 0, & E_ 2  \neq 0, \\
0, & E_1 \neq 0, & E_ 2  \neq 0. \
\end{array}
\right.
$$
This result shows that the winding number coincides with the number of MR pairs. 
Equation (\ref{windingdef2}) is given an interesting interpretation
if  we parameterize 
${\mathfrak g}(z, \bar{z} )$ as in the BCS model 

\begin{eqnarray*}
{\mathfrak g}(z, \bar{z}) = \frac{v(z, \bar{z})}{u(z, \bar{z})}, \qquad u(z, \bar{z}) \bar{u}(z, \bar{z}) + v(z, \bar{z}) 
\bar{v}(z, \bar{z}) = 1 
\end{eqnarray*} 
where $u$ and $v$ are well defined functions in the complex plane. 
It can be shown  that 
\begin{eqnarray}
w 
&=&  \frac{1}{2 \pi i } \int_{\mathbb{R}^2}  dX \; dY 
\left[  \partial_X  ( \bar{u}\; \partial_Y  u + \bar{v}  \;  \partial_Y v) - 
\partial_ Y( \bar{u} \;  \partial_X  u + \bar{v}  \; \partial_X v) \right].
\end{eqnarray}
and from  the Gauss theorem 
\begin{eqnarray}
w 
&=&  \lim_{r \rightarrow \infty} \int_{0}^{ 2 \pi}   \frac{d \theta}{2 \pi i }  \; 
\left( \bar{u}\; \partial_\theta  u + \bar{v}  \;  \partial_\theta v \right) 
\label{windingdef3}
\end{eqnarray}
where $X+ i Y = r e^{i \theta}$. This expression indicates  that $w$ measures the angular
dependence of the wave funcion $ {\mathfrak g}(z, \bar{z})$. The result (\ref{wBCS}) for the 
BCS wavefunction   can be derived analytically from (\ref{windingdef3}),  by noticing that the proper definition
of  $u$ and $v$ depend on the sign of the chemical potential, i.e.
\begin{eqnarray*}
u & = &  \frac{ z \Delta}{ \sqrt{ 2 e(z,\bar{z}) (  e(z,\bar{z})  - z \bar{z} + \mu)}}, \qquad
v = \sqrt{  \frac{ e(z,\bar{z})  - z \bar{z} + \mu}{ 2 e(z,\bar{z}) }}, \qquad \mu > 0, \\
u & = & \sqrt{ \frac{ e(z,\bar{z}) + z \bar{z} - \mu}{ 2 e(z,\bar{z}) }} , \qquad
v = \frac{ \bar{z} \Delta}{ \sqrt{ 2 e(z,\bar{z}) (  e(z,\bar{z})  + z \bar{z} - \mu)}}, \qquad \mu <0
\end{eqnarray*}
which leads to
\begin{eqnarray*}
w & = & \lim_{ r \rightarrow \infty} |u(r,0)|^2 = 1, \quad \mu > 0 ,  \qquad 
w = -  \lim_{ r \rightarrow \infty} |v(r,0)|^2 = 0, \quad \mu < 0.
\end{eqnarray*}
Similarly, equation (\ref{wonepair}) can be derived from (\ref{windingdef3}) using the following expression of the
$u$ and $v$ functions  corresponding  to the one pair wavefunction
\begin{eqnarray*}
u & = &  \frac{ z}{ \sqrt{1+ z \bar{z}}} , \qquad\qquad\qquad\quad v =  \frac{ 1}{ \sqrt{1+ z \bar{z}}},  \qquad\qquad E  =  0,  \\
u & = &  \frac{ z \bar{z} - E }{ \sqrt{ |z \bar{z} - E|^2 + z \bar{z}}} , \qquad v =   \frac{ \bar{z}  }{ \sqrt{ |z \bar{z} - E|^2 + z \bar{z}}},  \qquad E  \neq   0 .
\end{eqnarray*}
In the general case of an exact wavefunction with $M$ pairs, having $P$ vanishing rapidities, the asymptotic expression of $u$ and $v$ immediately yields the expected value of $w$, i.e. 
\begin{eqnarray*}
u & \sim  & z^M \bar{z}^{M - P}, \quad v \sim \bar{z}^{M-P} 
\quad \Longrightarrow \quad w= P.
 \end{eqnarray*}
This result shows that non-trivial topology of the wavefunction can only be found in the weak pairing phase. 

As a final comment we note that the winding number $w$  appears in a completely
different context, namely the $O(3)$ non-linear sigma model  in 2D. The order parameter
of the latter model is a three component unit vector $\vec{n}$, which can be constructed
out of the two component spinor $\psi = (u,v)$, as $\vec{n} = \psi^\dagger \vec{\sigma} \psi$. 
Using  this parametrization the winding number $w$ can be written as
\begin{eqnarray*}
w =  \frac{1}{8 \pi} \int_{\mathbb{R}^2} \;  d^2 {\mathbf x}  \; \; 
\vec{n}  \cdot ( \partial_\mu  \vec{n}  \times  \partial_\nu  \vec{n} ) \; \epsilon^{\mu \nu}
\end{eqnarray*}
where $ \epsilon^{\mu \nu}$ is the 2D Levi-Civita tensor. 
This term appears in the action of the $O(3)$ non-linear sigma model
multiplied by a parameter $\theta$, which is defined modulo $2 \pi$.
In a famous work, Haldane mapped the 1D antiferromagnetic  Heisenberg model 
of spin $S$ into the $O(3)$ non-linear sigma model with $\theta = 2 \pi S$ \cite{h87}. 
The gapped nature of the integer spin chains then follows from the gapped
nature of the non-linear sigma model with $\theta = 0 \;  ({\rm mod}  \; 2 \pi)$.

\subsection{Vortex  wavefunctions}
\label{vortex}

So far we have discussed the $p+ip$ model for a constant value of the order parameter $\Delta$,
which describes the system in the bulk. It is however of  interest to study the effect of vortices. 
As shown by Read and Green \cite{rg00} there are vortices with zero energy when the chemical potential
$\mu$ is positive. We shall show below that the structure of the vortex  wavefunction for $\mu>0$
depends on the region of the phase diagram, with different behaviours
in the weak coupling BCS and weak pairing regimes.  However the transition between these two regimes
is smooth in the mean-field analysis. 

We shall consider the Bogolioubov-de Gennes (BdG) equations in the simplest
situation of a single vortex with rotational invariance characterized by an order
parameter
\begin{eqnarray*}
\Delta({\bf r}) 
&=& i e^{ i l \varphi} \; |\Delta(r)|
\end{eqnarray*}
where $l$ is the vorticity and $r, \varphi$ are the polar coordinates centered on the vortex.
(We shall closely  follow the results of references \cite{gr07,rg00,tdnzz06} with the appropiate notational
changes: the mass is equal to 1, the gap $\Delta$ and the chemical potential in this paper are twice those considered
in the cited references.) The BdG equations are a set of coupled equations for the functions $u_n({\bf r}), v_n({\bf r})$, corresponding
to energies $E_n$, which determine the Bogolioubov quasiparticle operators $\gamma_n$ by 
\begin{eqnarray*}
\gamma_n & = &  \int d^2 r  \left[ u^*_n({\bf r})  a({\bf r}) +  v^*_n({\bf r}) a^\dagger({\bf r}) \right] 
\end{eqnarray*}
where $ a({\bf r}),  a^\dagger({\bf r})$ are annihilation and creation operators of 
polarised electrons.  The zero energy solutions, i.e. $E_n = 0$, can be choosen to satisfy
the reality condition $u_n({\bf r}) = v_n({\bf r})$, in which case the  quasiparticle operator
$\gamma_n$ corresponds to a Majorana fermion. 
 For odd  vorticity, $l = 2 n -1$, the BdG equation
for $u({\bf r})$ can be most conveniently written in terms of a function $\chi(r)$ defined through
$$
u({\bf r}) 
=  \chi(r) \; {\rm exp} \left( - \frac{1}{2}  \int_0^r \; d r' \; |\Delta(r')| \right)
$$
and it reads
$$
-  \chi''  - \frac{\chi'}{ r} + \left( \frac{  |\Delta^2(r)|}{4} + \frac{n^2}{ r^2} \right) \chi = \mu \;  \chi.
$$
Let us suppose that the order parameter $|\Delta(r)|$ vanishes inside a core of radius $a$
and that  it reaches its bulk value, $\Delta_0$, outside i.e.
$$
|\Delta(r)| =
\left\{ \begin{array}{cc} 
0, & 0 < r  < a ,\\
\Delta_0,  &  a < r < \infty . 
\end{array}
\right.
$$
The previous equation becomes
\begin{eqnarray}
r^2  \chi''  + r \chi'  + \left[  \mu  r^2 - n^2 \right] \chi & =0, & 0 < r < a, \label{BdG1} \\
r^2  \chi''  + r \chi'  + \left[ \left( \mu -  \frac{  \Delta^2_0}{4} \right) r^2 - n^2 \right] \chi&  =0,  & 
a < r < \infty. 
\nonumber 
\end{eqnarray}
Notice that the coefficient multiplying the $r^2$ term in the second equation changes
sign when crossing the MR line $\mu = \Delta_0^2/4$. This fact leads to two different
type of equations  corresponding to the weak coupling and weak pairing regimes.
(We do not consider the strong coupling regime where the vortices have a trivial
topological structure  \cite{gr07,rg00,tdnzz06}.) 

\subsubsection{Weak coupling BCS phase} 

For  the weak coupling BCS phase $ \mu > \Delta_0^2/4$ we make the change of variables
$$
\mu = k_0^2, \quad \mu -  \frac{  \Delta^2_0}{4} = k^2, 
$$
Eq. (\ref{BdG1})  becomes 
\begin{eqnarray*}
x^2  \chi_{xx}   + x \chi_{x}  + \left[  x^2 - n^2 \right] \chi & =0, & x = k_0 r, \qquad 0 < r < a \\
y^2  \chi_{yy}   + y \chi_y  + \left[ y^2 - n^2 \right] \chi&  =0,  & y = k r, \qquad a < r < \infty 
\end{eqnarray*}
whose solutions are Bessel functions. 
Let us denote by  $I$  and $II$  the regions $ 0 < r < a$  and  $a < r < \infty $ respectively. Imposing that $\chi$ is
regular at $r=0$ and infinity one has
\begin{eqnarray*}
\chi_{I}  & =  & A J_n(k_0 r), \\
\chi_{II}  & =  & B J_n(k r) + C Y_n(k r). 
\end{eqnarray*}
The constants $A,B,C$ are found by matching the function $u(r)$ and its derivative $u'(r)$
at the boundary $r=a$.  The relation between $u$ and $\chi$ in the regions $I$ and $II$ is
\begin{eqnarray}
u_I ( r) & = & \chi_{I}(r) ,  \nonumber  \\
u_{II} (r) & = &  \chi_{II}(r) e^{ - \Delta_0 r/2}.   \label{uII}
\end{eqnarray}
Hence
\begin{eqnarray*}
u_I ( a) & = & u_{II}(a)  \Longrightarrow \chi_I(a) =  \chi_{II}(a) e^{ - a \Delta_0/2} ,  \\
u'_I ( a) & = & u'_{II}(a)  \Longrightarrow \chi'_I(a) =  (\chi'_{II}(a) - \frac{\Delta_0}{2} \chi_{II}(a))  e^{ - a \Delta_0/2}.
\end{eqnarray*}
Combining these two equations one can write 
\begin{eqnarray*}
 \chi_I(a)  & = &  \chi_{II}(a) e^{ - a \Delta_0/2} ,  \\
\chi'_I(a) + \frac{\Delta_0}{2} \chi_I(a) & = &   \chi'_{II}(a)   e^{ - a \Delta_0/2}
\end{eqnarray*}
with the explicit expressions 

\begin{eqnarray*}
 A J_n( k_0 a)   & = &  ( B J_n(k a) + C Y_n(k a) ) e^{ - a \Delta_0/2} ,  \\
 A \left( k_0 J'_n( k_0 a)  + \frac{\Delta_0}{2}   J_n(k_0 a) \right)  & = &  k ( B J'_n(k a) + C Y'_n(k a) ) e^{ - a \Delta_0/2}. 
\end{eqnarray*}
This then leads to 
\begin{eqnarray*}
\frac{B}{A}  & =  & \frac{1 }{ 2} a \pi e^{ a \Delta_0/2} \left[
k Y'_n(k a) J_n(k_0 a) - Y_n(k a) \left( k_0 J'_n( k_0 a) + \frac{\Delta_0}{2} J_n( k_0 a) \right) \right],  \\
\frac{C}{A}  & =  & \frac{1 }{ 2} a \pi e^{ a \Delta_0/2} \left[
- k J'_n(k a) J_n(k_0 a) + J_n(k a) \left( k_0 J'_n( k_0 a) + \frac{\Delta_0}{2} J_n( k_0 a) \right) \right]. 
\end{eqnarray*}
For an infinitesimal core,  $a \rightarrow 0$,  the zero mode is localized on an 
isolated vortex with wavefunction 
$$
u(r) = J_n( k r) \; e^{ - \Delta_0  r/2}, \qquad r >0 \qquad ( \mu > { \Delta^2_0}/{4} ) .
$$
This result reproduces the one found by Gurarie and Radzihovsky \cite{gr07}. 
The funcion $u(r)$ has an oscillatory and decaying asymptotic behaviour
dominated by $\Delta_0$, i.e. 
$$
u(r) \sim \frac{ \cos( k r + \phi_0)}{\sqrt{r}} e^{ - r   \Delta_0/2 } ,  \qquad r \rightarrow \infty. 
$$

\subsubsection{Weak pairing phase} 

\begin{figure}[t!]
\begin{center}
\includegraphics[height= 4.9cm,angle=0]{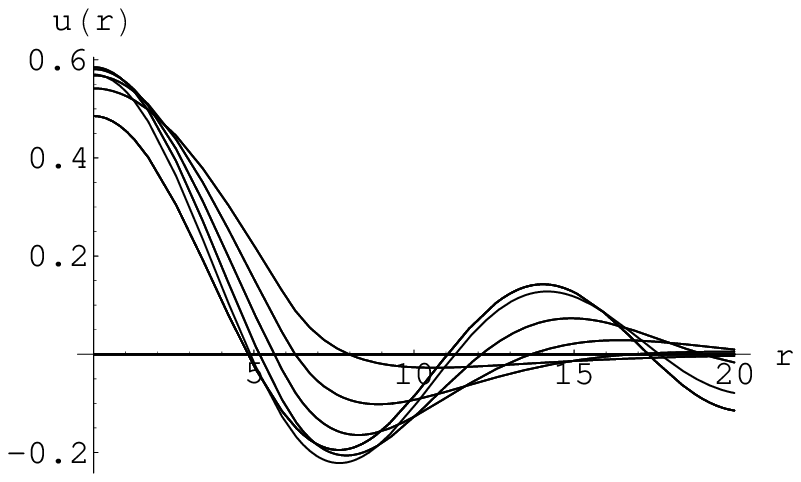}
\includegraphics[height= 4.9cm,angle=0]{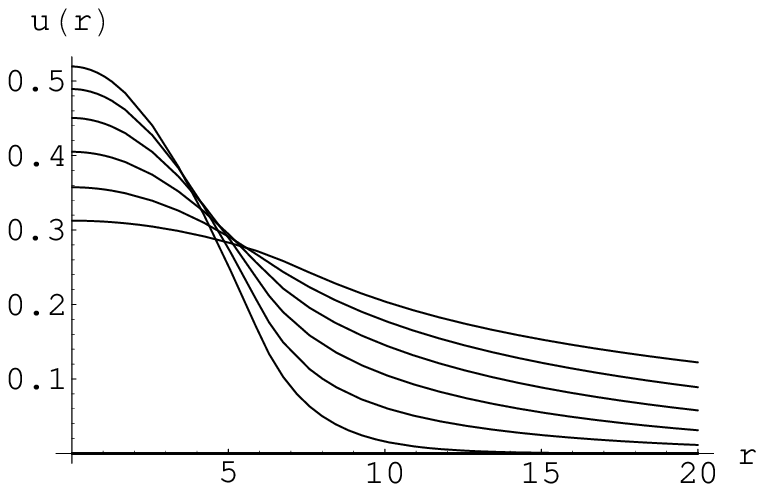}
\end{center}
\caption{\footnotesize{
The wavefunction $u(r)$ for the vortex solution with $l = -1$ for $x=1/4$ and several values
of the coupling constant $g$. The size of the core is $a= 6$ in units of $\sqrt{\omega}$.  
Left:  weak coupling BCS region for $g=0.3- 0.13$ in steps of $0.2$. 
Right: weak pairing region for $g=1.4- 1.9$ in steps of $0.1$.  
}}
\label{pwa6}
\end{figure}

To study the weak pairing case $0 <  \mu < \Delta_0^2/4$ we make the change of variables
$$
\mu = k_0^2, \quad -\mu +   \frac{  \Delta^2_0}{4} = k^2, 
$$
which corresponds to the replacement  $k \rightarrow i k$ in the previous case. 
The equation for $\chi$ becomes
\begin{eqnarray*}
x^2  \chi_{xx}   + x \chi_{x}  + \left[  x^2 - n^2 \right] \chi & =0, & x = k_0 r, \qquad 0 < r < a, \\
y^2  \chi_{yy}   + y \chi_y  - \left[ y^2 + n^2 \right] \chi&  =0,  & y = k r, \qquad a < r < \infty. 
\end{eqnarray*}
whose solutions are Bessel functions in region $I$ and modified Bessel functions in region $II$: 
\begin{eqnarray*}
\chi_{I}  & =  & A J_n(k_0 r), \\
\chi_{II}  & =  & B I_n(k r) + C K_n(k r).
\end{eqnarray*}
The computation of $A,B,C$ is similar to the previous one. 
The final solution is
\begin{eqnarray}
\frac{B}{A}  & =  & - a  e^{ a \Delta_0/2} \left[
k K'_n(k a) J_n(k_0 a) - K_n(k a) \left( k_0 J'_n( k_0 a) + \frac{\Delta_0}{2} J_n( k_0 a) \right) \right],  
\label{Bwp} 
\\
\frac{C}{A}  & =  & - a e^{ a \Delta_0/2} \left[
- k I'_n(k a) J_n(k_0 a) + I_n(k a) \left( k_0 J'_n( k_0 a) + \frac{\Delta_0}{2} J_n( k_0 a) \right) \right]. 
\nonumber 
\end{eqnarray}
In the limit $a \rightarrow 0$ one obtains
$$
u(r) = I_n( k r) \; e^{ - \Delta_0  r/2},  \qquad r >0 \qquad ( \mu < { \Delta^2_0}/{4} ) 
$$
in agreement with \cite{gr07}. 
It is important to notice that the exponential factor $e^{- \Delta_0 r/2}$ in equation (\ref{uII}) 
is essential to get a convergent  asymptotic behaviour of $u(r)$.  Assuming that the coefficient
$B$,  given in eq. (\ref{Bwp}),  is non zero one finds for $B\neq 0$ 
$$
u(r) \sim \frac{1}{\sqrt{r}} {\rm exp} \left[ - r  \left( \frac{ \Delta_0}{2} -
\sqrt{  \frac{ \Delta_0^2}{4} - \mu} \right) \right],  \qquad r \rightarrow \infty, \qquad 
$$
while if  $B=0$,  the asymptotic behaviour of $u(r)$ changes to  
$$
u(r) \sim \frac{1}{\sqrt{r}} {\rm exp} \left[ - r  \left( \frac{ \Delta_0}{2} +
\sqrt{  \frac{ \Delta_0^2}{4} - \mu} \right) \right],  \qquad r \rightarrow \infty. \qquad 
$$
which is a more localized vortex.  The condition for this to happen is
\begin{eqnarray}
B =0 \Longrightarrow 
k  \; \frac{ K'_n( k a)}{ K_n(k a)} - k_0  \; \frac{J'_n(k_0 a)}{ J_n(k_0 a)} = \frac{\Delta_0}{2}.
\label{amin}
\end{eqnarray} 
This equation has an infinite number of solutions for $a$ as a function of $\mu$ and $\Delta_0$.
In the case where $k a, k_0 a >>1$, the solutions of eq. (\ref{amin}) are
$$
k_0 a  \sim  \frac{\pi}{4} + \frac{ n \pi}{2}  + {\rm arctan}  \left( \frac{ k + \Delta_0/2}{k_0} \right)  \quad ({\rm mod} \;
\pi ).
$$
Hence there is a minimal core size in order to have these more localized vortices. 
In Fig. \ref{pwa6} we plot the function $u(r)$ for several values of the coupling constant $g$.
In the weak coupling regime the function $u(r)$ has an oscillatory 
behaviour, while in the weak pairing regime the oscillations are absent.  
We observe that the different qualitative behaviour
depends on the region of the phase diagram.

 \section{Conclusions}

In summary,  we have constructed an exactly solvable pairing model
which contains in various limits some of the most studied pairing models
of current studies such as: the Richardson model with $s$-wave symmetry,
the $p+i p$ BCS model, the Russian doll BCS model and the Gaudin models
with $s$-wave and $p$-wave symmetries. We have computed the wavefunctions
scalar products and one-point and two-point correlators. Futhermore,  we have
carried out an in depth study of the $p+ip$ model combining  the exact solution 
with mean-field methods. We have found the phase diagram of this  model characterizing the
various phases in terms of the fidelity susceptibility, topological invariants, and other quantities
such as the ground state energy, gap, chemical potential and occupation numbers. 
We have also studied the structure of the vortex wave  function which is shown
to depend on the region of the phase diagram. The richness of the $p + ip$ model
calls for further studies concerning its behaviour. This task
can be greaty facilitated by the existence of an exact solution for studies such as entanglement \cite{dlz05}, correlation functions \cite{ao02,fcc08} and quenching \cite{fcc09,fcc09a}.

~\\ 
 \noindent
{\it Acknowledgments}  
We thank I. Cirac,  J. Dukelsky, N. Read, S. Rombouts
and G.E. Volovik for helpful comments. 
 M.I. and G.S. are supported by the 
CICYT project FIS2006-04885. G.S. also acknowledges 
ESF Science Programme INSTANS 2005-2010. J.L. and 
S.-Y.Z. are funded by the Australian Research Council 
through Discovery Grant DP0663772.  C.D. and J.L. acknowledge the
Royal Society Travel Grants Scheme. 
 
 \appendix
 
 \section{Calculations for the anyonic pairing model}
 
\subsection{Cyclic renormalisation group} \label{cyclic}

In this Appendix we discuss how the Hamiltonian (\ref{ham}) is related to a known Hamiltonian, viz. the Galzek--Wilson model \cite{gw02,gw04}. This model was introduced as a simple example of a quantum mechanical system which admits a cyclic renormalisation group (RG) map. As will be shown below, the Glazek--Wilson is simply the one-body version of the anyonic pairing Hamiltonian (\ref{ham}). Cyclic behaviour of the RG  also occurs in the many-body system, and we will demonstrate how the RG equation is easily obtained from the Bethe ansatz equations.       

The model introduced in \cite{gw02,gw04}, which we will denote as $H_{GW}$, is defined by the matrix
\begin{eqnarray}
H_{GW}(n,m) = \sqrt{{\mathcal E}_n \; {\mathcal E}_m} ( \delta_{n,m} 
- g - i h \; {\rm sign}(n-m)) 
\label{gw}
\end{eqnarray}
where $\mathcal{M} < n,m < \mathcal{N}$. 
In \cite{gw02,gw04} the ${\mathcal E}_n$ are assumed to scale exponentially
$
{\mathcal E}_n = \mu^n, \, \mu > 1 
$
under which there is a discrete scale invariance $n \rightarrow n +1$
such that
\begin{eqnarray*}
H(n+1,m+1) = \mu  \; H(n,m). 
\label{scale}
\end{eqnarray*}
However this scale invariance is broken by the infra-red and ultra-violet cutoffs $\mathcal{M}$ and $\mathcal{N}$. 

In \cite{gw02,gw04} the RG is performed by a Gaussian elimination procedure for the highest component of the wavefunction in the Schr\"odinger equation, reducing $\mathcal{N}$ to $\mathcal{N}-1$. Under this process the result of the RG is that the coupling $h$ remains invariant while the coupling $g$ renormalises as 
\begin{eqnarray}
g_{\mathcal{N}-1}= \frac{g_\mathcal{N}+h^2}{1-g_\mathcal{N}}.    
\label{grg}
\end{eqnarray}
Reparameterising the couplings in terms of angles $\phi_\mathcal{N},\,\chi$ through 
\begin{eqnarray*}
\tan(\phi_\mathcal{N})&=&\frac{g_\mathcal{N}}{h}, \\
\tan(\chi)&=& h 
\end{eqnarray*}
the RG equation becomes 
\begin{eqnarray*}
\tan(\phi_{\mathcal{N}-1})=\frac{\tan(\phi_\mathcal{N})+\tan(\chi)}{1-\tan(\phi_\mathcal{N})\tan(\chi)}=\tan(\phi_\mathcal{N}+\chi)
\end{eqnarray*}
or equivalently, iterating the RG equation $p$ times,  
\begin{eqnarray*}
\phi_{\mathcal{N}-p}=\phi_\mathcal{N}+p\chi.
\end{eqnarray*}
A cycle occurs in the RG after $p$ steps if $\chi=\pi/p$ since
\begin{eqnarray*}
g_{\mathcal{N}-p}=h\tan(\phi_{\mathcal{N}-p})=h\tan(\phi_\mathcal{N}+p\chi)=h\tan(\phi_\mathcal{N}+\pi)=g_\mathcal{N}
\end{eqnarray*}
with the RG period given by 
\begin{eqnarray*}
p=\frac{\pi}{\arctan({h})}.
\end{eqnarray*}

Next we show that (\ref{gw}) is equivalent to the one-pair version of (\ref{ham}). The first thing to observe is that for the one-pair version of (\ref{ham}) the string operators (\ref{string}) appearing in the definition of the anyonic creation and annihilation operators can be removed by a unitary transformation. The Hamiltonian is then given in terms of the hard-core boson operators as
\begin{eqnarray*}
H_{\rm one-pair}&=&   \sum_{j=1}^L z_j^2 N_j -\frac{\sin(2\beta)}{\sin(\alpha-2\beta) }\sum_{k>r}^L z_k z_r 
\left(\exp(-i\alpha) 
 b_r^\dagger b_k + {\rm h.c.} \right)   
\label{hone}
\end{eqnarray*}
with the Hilbert space spanned by the one-body states
\begin{eqnarray*}
\left|j\right>=b^\dagger_j\left|0\right>,\qquad j=1,...,L.
\end{eqnarray*}
Here the energy is 
\begin{eqnarray*}
E_{\rm one-pair}&=& \frac{\sin(\alpha)}{\sin(\alpha-2\beta)} \,y^2
\label{nrgone}
\end{eqnarray*}
such that
\begin{eqnarray*}
\exp(-2i\alpha)\prod_{k=1}^L\frac{1-q^2 y^{-2}z^2_k}{1-q^{-2}y^{-2}z^2_k}
=1.
\label{1bodybae}
\end{eqnarray*} 
Setting 
\begin{eqnarray}
{\mathcal E}_k=z_k^2, \quad g= \tan(2\beta)\cot(\alpha), \quad h=\tan(2\beta) 
\label{gwsubs}
\end{eqnarray}
yields 
\begin{eqnarray*}
{H_{GW}}=(1-g)H_{\rm one-pair}, \qquad \mathcal{N}-\mathcal{M}=L.
\end{eqnarray*}
 
To derive the RG equations from the Bethe ansatz, we only assume that the variables
 $z_k$ are distinct and ordered $0<z_1<z_2<...<z_L$. We may then write 
\begin{eqnarray*}
\exp(-2i\alpha).\frac{1-q^2 y^{-2}z^2_L}{1-q^{-2}y^{-2}z^2_L}
.\prod_{k=1}^{L-1} \frac{1-q^2 y^{-2}z^2_k}{1-q^{-2}y^{-2}z^2_k}
=1 .
\label{bael}
\end{eqnarray*}
Assuming $|z_L|>>|y|$, which is a reasonable assumption for low energy states in a weakly coupled system,  we can eliminate the highest energy degree of freedom of the Hamiltonian by making the approximation
\begin{eqnarray*}
\frac{1-q^2 y^{-2}z^2_L}{1-q^{-2}y^{-2}z^2_L}&\approx& q^4, \\
\Rightarrow \qquad\quad \exp(-2i\alpha+4i\beta)\prod_{k=1}^{L-1} \frac{1-q^2 y^{-2}z^2_k}{1-q^{-2}y^{-2}z^2_k}
&\approx&1.
\end{eqnarray*}    
Thus for one RG step we have the renormalised coupling 
\begin{eqnarray}
\alpha_{\mathcal{N}-1}= \alpha_\mathcal{N}-2\beta,
\label{baerg}
\end{eqnarray} 
while $\beta$ remains invariant. Substituting back into (\ref{gwsubs}) shows that, in terms of the $H_{GW}$ coupling parameters, $h$ is invariant while 
\begin{eqnarray*}
g_{\mathcal{N}-1}&=& \tan(2\beta)\cot(\alpha_{\mathcal{N}-1}) \\
&=&\tan(2\beta)\cot(\alpha_{\mathcal{N}}-2\beta) \\
&=& \tan(2\beta)\frac{\cot(\alpha_\mathcal{N})+\tan(2\beta)}{1-\tan(2\beta)\cot(\alpha_\mathcal{N})}\\
&=& \frac{g_\mathcal{N}+h^2}{1-g_\mathcal{N}}
\end{eqnarray*}
in agreement with (\ref{grg}). 

For the many-body Hamiltonian (\ref{ham}), a completely analogous approach can be taken to establish the existence of the cyclic RG equation. Starting with the Bethe ansatz equations (\ref{bae})
we can make the same approximation as in the one-pair case by eliminating the highest energy level, i.e. if $|z_L|>>|y_m|$ for all $m=1,...,M$ then 
\begin{eqnarray*}
\frac{1-q^2 y_m^{-1}z^2_L}{1-q^{-2}y_m^{-1}z^2_L}&\approx& q^4 \qquad\qquad \forall \,m=1,...,M 
\end{eqnarray*} 
and (\ref{bae}) gives
\begin{eqnarray*}
\exp(-2i\alpha+4i\beta)\prod_{k=1}^{L-1}\frac{1-q^2 y_m^{-2}z^2_k}{1-q^{-2}y_m^{-2}z^2_k}
=\prod_{j\neq m}^{M} \frac{ 1- q^4 y_m^{-2}y_j^2 }{1- q^{-4} y_m^{-2}y_j^2 }
\qquad\qquad m=1,...,M .
\end{eqnarray*} 
This leads to exactly the same RG equation (\ref{baerg}) for $\alpha$, with $\beta$ again invariant.

\subsection{Wavefunction scalar product and correlation functions} 
\label{scalarproduct}

One of the byproducts of formulating the model via the QISM is that it readily provides access to the evaluation of correlation functions. Again, we will only outline these calculations. We will follow the general procedures of \cite{kmt99}, where explicit details of derivations can be found. 
   
It is  convenient to first rescale the $L$-operator as 
\begin{eqnarray} 
\tilde{L}(x)
&=& \left(\begin{array}{ccccc} 
1 &0&|&0&0 \\
0&\frac{q^{-1}x^{}-qx^{-1} }{qx-q^{-1}x^{-1}}&|&\frac{q^2-q^{-2}}{qx-q^{-1}x^{-1}}&0 \\ 
-&-&~&-&- \\ 
0&\frac{q^2-q^{-2}}{qx-q^{-1}x^{-1}}&|&\frac{q^{-1}x^{}-qx^{-1} }{qx-q^{-1}x^{-1}}&0 \\
0&0&|&0&1 \end{array} \right)  
\label{tildelax}
\end{eqnarray}
such that it has the properties 
\begin{eqnarray*} 
\left.\tilde{L}(x)\right|_{q=1}&=&    I\otimes I \\
\left.\tilde{L}(x)\right|_{x=q}&=& P 
\end{eqnarray*}
with $P$ the permutation operator. We also rescale the matrix ${U}$ to be 
$$
\tilde{U}= \left( \begin{array} {cc} 
\exp(-2i\alpha') & 0 \\
0 & 1 
\end{array}  \right). 
$$
These rescalings have the net effect of working with rescaled elements of the monodromy matrix
\begin{eqnarray*}
\tilde{T}^i_j(x)&=&\exp(-i\alpha')\left(\prod_{k=1}^L\frac{1}{qxz_k^{-1}-q^{-1}x^{-1}z_k}\right)    T^i_j(x) 
\end{eqnarray*}
The transfer matrix eigenvalues of $\tilde{t}(x)=\tilde{T}^1_1(x)+\tilde{T}^2_2(x)$ are now 
\begin{eqnarray*}
\tilde{\Lambda}(x) &=& \tilde{a}(x) 
\prod_{j=1}^M \frac{ q^2 x y_j^{-1} - q^{-2}y_j x^{-1} }{ xy_j^{-1} -  y_jx^{-1} } +\tilde{d}(x))\prod_{j=1}^M \frac{ q^{-2} x y_j^{-1}- q^2y_j x^{-1}   }{  xy_j^{-1} -y_jx^{-1} }
\end{eqnarray*}
with
\begin{eqnarray*}
\tilde{a}(x)&=&\exp(-2i\alpha')\prod_{k=1}^L\frac{q^{-1}xz_k^{-1}-qx^{-1}z_k}{qxz_k^{-1}-q^{-1}x^{-1}z_k} \\
\tilde{d}(x)&=1
\end{eqnarray*}
One reason for working with the the rescaled monodromy matrix is that it provides a compact solution of the ``inverse problem'' of expressing the local operators $b_j,\,b_j^\dagger,\,N_j$ in terms of the elements of the monodromy matrix. The result, which is obtained by a minor generalisation of the methods used in \cite{dl04,kmt99,mt00}, is 
\begin{eqnarray*}
b_j=\left(\prod_{k=1}^{j-1} \tilde{t}(qz_k)\right) \tilde{T}^1_2(qz_j)\left(\prod_{k=1}^j \tilde{t}^{-1}(qz_k)\right),\\
b^\dagger_j=\left(\prod_{k=1}^{j-1} \tilde{t}(qz_k)\right) \tilde{T}^2_1(qz_j)\left(\prod_{k=1}^j \tilde{t}^{-1}(qz_k)\right),\\
I-N_j=\left(\prod_{k=1}^{j-1} \tilde{t}(qz_k)\right) \tilde{T}^2_2(qz_j)\left(\prod_{k=1}^j \tilde{t}^{-1}(qz_k)\right).
\end{eqnarray*}
Note that the operators $\{\tilde{t}(qz_j):j=1,...,L\}$ are conserved. Explicitly these are 
\begin{eqnarray}
\tilde{t}(qz_k) = \tilde{L}_{k(k-1)}(qz_kz^{-1}_{k-1}) ...\tilde{L}_{k1}(qz_k z^{-1}_{1}) \tilde{U}_k \tilde{L}_{kL}(qz_kz^{-1}_L )
...\tilde{L}_{k(k+1)}(qz_kz^{-1}_{k+1}) 
\label{conserved}
\end{eqnarray}
where
\begin{eqnarray*}
\tilde{L}_{kl}(qz_kz_l^{-1})&=& \frac{qz_kz_l^{-1}+q^{-1}z_k^{-1}z_l}{ (q-q^{-1}) (q^2z_kz_l^{-1}-q^{-2}z_k^{-1}z_l) }\left( q^{2N_k+2N_l-2} + q^{2-2N_k-2N_l} \right) \\
&&\qquad\qquad
+\frac{q^2-q^{-2}}{q^2z_kz_l^{-1}-q^{-2}z_k^{-1}z_l}\left( b_k^\dagger b_l + b_k b_l^\dagger\right) \\ 
&&\qquad\qquad -\frac{(q+q^{-1})(z_kz_l^{-1}+z_k^{-1}z_l)}{(q-q^{-1}) (q^2z_kz_l^{-1}-q^{-2}z_k^{-1}z_l) } I .
\end{eqnarray*} 
The eigenvalues of $\tilde{t}(qz_k)$ are 
\begin{eqnarray}
\tilde{\Lambda}(qz_k) &=& \prod_{j=1}^M \frac{ q^{-1} z_k y_j^{-1}- qy_j z_k^{-1}   }{  qz_ky_j^{-1} -q^{-1}y_jz_k^{-1} }.
\label{consevalues}
\end{eqnarray}

Next we consider the scalar products between the states $\langle\tilde{\Phi}(W)|,\,
|\tilde{\Phi}(Y)\rangle$, where
\begin{eqnarray}
\langle\tilde{\Phi}(W)|&=&\left<0\right|\prod_{j=1}^L \tilde{T}^1_2(w_j), \label{left}\\
|\tilde{\Phi}(Y)\rangle&=&\prod_{j=1}^L \tilde{T}^2_1(y_j)\left|0\right>. \label{right}
\end{eqnarray}
If $W=\{w_i\}$ is a set of solutions of the Bethe ansatz equations (\ref{bae}), and $Y=\{y_j\}$ are arbitrary parameters, then the Slavnov formula for the the scalar product is \cite{s89} 
\begin{eqnarray}
\langle \tilde{\Phi}(W) | \tilde{\Phi}(Y)\rangle =  \mathcal{C}(W,Y)
    \det (F(W,Y))
\label{slavnov}
\end{eqnarray}
where elements of the matrix $F(W,Y)$ are given by
\begin{eqnarray}
F(W,Y)_{ij}&=&\frac{(q^2-q^{-2})\tilde{d}(w_i)}{ (y_jw_i^{-1}-y_j^{-1}w_i)} 
 \left(\tilde{a}(y_j)\prod_{m\neq i}^M(y_jw_m^{-1}q^2-y_j^{-1}w_mq^{-2})\right. \nonumber\\
 &&~~~~~~~~~~~~~~~~~~~~~~~~~~~~\left. +\tilde{d}(y_j)\prod_{m\neq i}^M({y_jw_m^{-1}q^{-2}-y_j^{-1}w_mq^2})\right),
\label{fmatrix}
\end{eqnarray}
and 
\begin{eqnarray}
\mathcal{C}(W,Y)=\frac{1}
{\prod_{k>l}^M(y_ky_l^{-1}-y_k^{-1}y_l)(w_lw_k^{-1}-w_l^{-1}w_k)}.
\label{c}
\end{eqnarray}
Note that 
$$\mathcal{C}(Y,Y)=\frac{1}{\prod_{p=1}^M\prod_{q\neq p}^M(y_py_q^{-1}-y_p^{-1}y_q)}. $$ 
We remark that the Slavnov formula (\ref{slavnov}) is a {\it scalar} product as opposed to an {\it inner} product. However for the anyonic pairing model we expect, due to a lack of symmetries, that generically there are no degeneracies in the energy spectrum for fixed particle number. In such an instance the 
eigenstates must be real, up to an overall phase factor, in which case the scalar product between any two eigenstates is equivalent to the inner product up to a phase.  

An application of the Slavnov formula (\ref{slavnov}) is that it allows us to compute the normalised wavefunction overlap between two states with {\it different} values of the coupling parameter $\alpha$. Specifically this permits us to compute the {\it fidelity}, as defined in \cite{zp06,zb08}, which has been proposed as a technique to identify quantum phase transitions (for a review see \cite{g08}). To make this point transparent, we start with the explicit form of the monodromy matrix elements in terms of the elements of the $L$-operators which is  
\begin{eqnarray*}
\tilde{T}^i_j(x)=\sum_{k_1,...,k_L=1}^L \tilde{g}^{k_{L}}_{j}\left[\tilde{L}^{k_{L-1}}_{k_{L}}(xz_L^{-1})\right]_L ...\left[\tilde{L}_{k_2}^{k_1}(xz_2^{-1})\right]_2\left[\tilde{L}_{k_1}^i(xz_1^{-1})\right]_1.
\end{eqnarray*}
Since $\tilde{g}$ is a diagonal matrix we can then write
\begin{eqnarray}
\tilde{T}^i_j(x)=\tilde{g}^{j}_{j}\sum_{k_1,...,k_{L-1}=1}^L \left[\tilde{L}^{k_{L-1}}_{j}(xz_L^{-1})\right]_L ...\left[\tilde{L}_{k_2}^{k_1}(xz_2^{-1})\right]_2\left[\tilde{L}_{k_1}^i(xz_1^{-1})\right]_1
\label{texpand}
\end{eqnarray}
and each of the operators
\begin{eqnarray}
\left[\tilde{L}^{k_{j-1}}_{k_j}(xz_j^{-1})\right]_j
\label{lax-elements}
\end{eqnarray}
is not explicitly dependent on $\alpha$. Consequently the explicit $\alpha$-dependence of $\tilde{T}^i_j(x)$ is only in terms of an overall phase. Thus  we can take the set $W$ to be a solution of the Bethe ansatz equations for a given coupling, say $\alpha_1$, and the set $Y$, which in  general may be arbitrary, to be a solution to the Bethe anstaz equations with a second choice of the coupling, say $\alpha_2$. This is not to say that the eigenstates of the Hamiltonian (\ref{ham}) are independent of $\alpha$, but rather the dependence is implicit through the solution sets $W$ and $Y$ of the Bethe ansatz equations. In this way the normalised wavefunction overlaps between ground states with different values of the coupling $\alpha$ can be computed, which is in essence the computation of the fidelity.\footnote{In the context of the $s$-wave model this property has been exploited in \cite{fcc09,fcc09a} to study quenching dynamics.} Note however we cannot compute the overlaps between ground states with different values of $q$ nor the $z_k$, since these parameters appear explicitly in the operator elements (\ref{lax-elements}).

Using the Slavnov formula and the solution of the inverse problem the form factors (matrix elements) for the local operators $b_j,\,b_j^\dagger$ can also be written, although we will not give details here (however see Appendix \ref{2cf} for an example). Our main objects of interest for now are the one-point correlation functions, or expectation values, of the Cooper pair 
number operators $N_j$ . We begin with the following  
form factor between two Bethe states \cite{kmt99}
\begin{eqnarray*}
&&\langle \tilde{\Phi}(W)|\tilde{T}^2_2(x) |\tilde{\Phi}(Y)\rangle\nonumber\\
 &=&\mathcal{C}(W,Y)
  \tilde{d}(x)\theta(x)
  \prod_{j=1}^M{(xy_j^{-1}q^{-2}-x^{-1}y_jq^{2})\over
                (xy_j^{-1}-x^{-1}y_j)}
   \,\, \det (F(W,Y)+Q(W,y;x)),
\end{eqnarray*}
where 
$$\theta(x)=\prod_{k=1}^M\frac{xw_k^{-1}-x^{-1}w_k}{xy^{-1}_k-x^{-1}y_k}  , $$
and $Q(W,Y;x)$ is a rank-one matrix with elements
\begin{eqnarray*}
Q(W,Y;x)_{ij}&=&{(q^2-q^{-2})\tilde{d}(w_i)\tilde{d}(y_j)\over
 (xw_i^{-1}-x^{-1}w_i)(xw_i^{-1}q^{-1}-x^{-1}w_iq)}
               \\
 &&\times \prod_{k=1}^M(y_jy_k^{-1}q^{-2}-y_j^{-1}y_kq^{2})
 \left(1-\frac{\tilde{a}(x)}{ \tilde{d}(x)}\prod_{k\neq i}^M
 \frac{xw_k^{-1}q^{2}-x^{-1}w_kq^{-2}}{
  xw_k^{-1}q^{-2}-x^{-1}w_kq^{2}}\right).
\end{eqnarray*}
This expression enables us to compute the
expectation value
\begin{eqnarray}
\langle N_m\rangle&=& \frac{\langle\tilde{\Phi}(Y)|N_m|\tilde{\Phi}(Y)\rangle}{\langle\tilde{\Phi}(Y)|\tilde{\Phi}(Y)\rangle} \nonumber \\
&=& 1-t^{-1}(qz_m)\frac{\langle\tilde{\Phi}(Y)|\tilde{T}^2_2(qz_m)|\tilde{\Phi}(Y)\rangle}
{\langle\tilde{\Phi}(Y)|\tilde{\Phi}(Y)\rangle} \nonumber\\ 
&=& 1-\frac{{\det}(F(Y,Y)+Q(Y,Y;qz_m))}{\det F(Y,Y)}.
\label{averagen}
\end{eqnarray}
Note that for the present case because $\tilde{a}(qz_m)=0,\,\tilde{d}(x)=1$ we have the following simplification for the matrix $Q(Y,Y;qz_m)$: 
\begin{eqnarray*}
Q(Y,Y;qz_m)_{ij}&=&\frac{(q^2-q^{-2})}{
 (qz_my_i^{-1}-q^{-1}z_m^{-1}y_i)(z_my_i^{-1}-z_m^{-1}y_i)}
    \prod_{k=1}^M(y_jy_k^{-1}q^{-2}-y_j^{-1}y_kq^{2}).
\end{eqnarray*}

\subsection{Duality  and the analogue of the Moore-Read state}

The $p+ip$ model was derived from the anyonic pairing model in a limit
where the parameters $\alpha, \beta$ go to zero while keeping their
ratio fixed,  which gives  the coupling constant 
$G = 2/(\alpha \beta^{-1} -2)$ (recall  (\ref{gcoupling})). One may ask wether the
results obtained in this section have an analogue in the anyonic pairing model. 
To answer this question we first look for a solution of the Bethe ansatz equations 
(\ref{bae}) with the property that  all the roots vanish, i.e. $y_j \rightarrow 0 \; (j=1, \dots, M)$.
From eq. (\ref{bae}) this implies
\begin{eqnarray*}
\exp(4 i  \beta( L - M - \frac{\alpha}{2 \beta}  +1)) = \lim_{ y_j \rightarrow 0} 
\prod_{j\neq m}^{M} \frac{ q^{-2} y_m^2 - q^2 y_j^2 }{q^2 y_m^2 - q^{-2} y_j^2 }
\qquad\qquad m=1,...,M. 
\label{baeanyon}
\end{eqnarray*} 
The RHS of this equation is identically equal to 1 in the limit $\beta \rightarrow 0$, which
is why there is such a solution in the $p+ip$ model. One can readily check that this result
remains true for any $\beta$ provided 
\begin{eqnarray*}
y_m^2 \propto  e^{ 2 \pi i m/M}, \qquad\qquad m=1,...,M,
\end{eqnarray*} 
which  implies
\begin{eqnarray*}
M = L - �\frac{\alpha}{2 \beta} +1 =  L - G^{-1}.
\end{eqnarray*} 
This equation coincides with the condition (\ref{mprime}) for the existence of a MR state in the $p+i p$ model. 
In the latter equation the parameter $G$ should not be confused with the coupling constant
of the anyonic pairing  model  given by $G_{AP} = \sin(2 \beta)/\sin(\alpha- 2 \beta)$, except in the limit
$\alpha, \beta \rightarrow 0$.  The corresponding eigenstate of the anyonic pairing model  has zero energy
and it is an anyonic deformation of the MR state. Explicitly the state is  
$$\left|AMR\right> = \left(\sum_{j=1}^L\frac{d_j^\dagger}{z_j}  \right)^M\left|0\right>. $$

As in the $p+ip$ model we can look for a set of roots $Y$, where a subset $Z$ of $P$ roots
vanish, while the other set $Y'$ with $M'$ roots do not. The Bethe ansatz equations  for the roots in $Z$ yield
\begin{eqnarray*}
\exp(4 i  \beta( L - P - 2 M'  - \frac{\alpha}{2 \beta}  +1)) = \lim_{ y_j \rightarrow 0} 
\prod_{y_j \in Z, y_j \neq y_m}^{M} \frac{ q^{-2} y_m^2 - q^2 y_j^2 }{q^2 y_m^2 - q^{-2} y_j^2 }
\qquad\qquad y_m \in Z
\end{eqnarray*} 
while for the roots in $Y'$ the Bethe ansatz equations   become
\begin{eqnarray*}
\exp(-2i\alpha)\prod_{k=1}^L\frac{1-q^2 y_m^{-2}z^2_k}{1-q^{-2}y_m^{-2}z^2_k}
=\prod_{y_j \in Y' , y_j\neq m}^{M} \frac{ 1- q^4 y_m^{-2}y_j^2 }{1- q^{-4} y_m^{-2}y_j^2 }
\qquad\qquad y_m \in Y'. 
\end{eqnarray*} 
Hence the roots in $Y'$ satisfy the same Bethe ansatz equations  as the original model, provided
\begin{eqnarray*}
 P = L  - 2 M'  - \frac{\alpha}{2 \beta}  +1 = L  - 2 M'  - G^{-1}
\end{eqnarray*} 
which is the same condition as found in (\ref{em}). 
Hence we expect that the duality symmetry of the $p+ip$ model
to also be a property of the anyonic pairing model. 
  
\section{Calculations for the $p+ip$ model}

\subsection{Solution of the gap and chemical potential
equations} \label{mfnum}

In this appendix we collate some results for the dimensionless parameters $\bar{a}$ and $\bar{b}$ which are solutions of equations (\ref{gapcont},\ref{chempotcont}). In all instances the results for the strong pairing phase can be deduced from those of the weak pairing phase through use of the duality relation (\ref{mfduality}). 
Below, the convention for the elliptic integrals are as in \cite{gr00}:
\begin{eqnarray*}
E(\phi, k) &=& \int_0^\phi dx \, \sqrt{ 1 - k^2 \sin^2 x}, 
 \\
F(\phi, k) &=& \int_0^\phi dx \, \frac{1}{\sqrt{ 1 - k^2 \sin^2 x}}.
\end{eqnarray*}
With the density functions as given by (\ref{dens1},\ref{dens2}), performing  the integrals in (\ref{gapcont},\ref{chempotcont}) one finds the following 
results.

\subsubsection*{1D case}

Weak coupling BCS phase: ($\bar{a}, \bar{b} = \bar{\epsilon}  \mp i \bar{\delta}$)

\begin{eqnarray*}
\frac{2}{g}& = & \sqrt{R} \left( F(\phi,k) - 2 E(\phi,k) \right)  
+ 2 \frac{ \sqrt{ 1 - 2 \bar{\epsilon} + R^2}}{1 + R},  \\
4 \left( x - \frac{1}{2} + \frac{1}{2 g}\right) 
&=& \sqrt{R} F(\phi, k) 
\end{eqnarray*}
where
\begin{eqnarray*}
R = \sqrt{ \bar{\epsilon}^2 + \bar{\delta}^2}, \;\; 
\phi = 2 \arctan( 1/\sqrt{R}), \; \;
k^2 = \frac{1}{2} \left( 1 + \frac{ \bar{\epsilon}}{R}\right). 
\end{eqnarray*}

\noindent
Moore--Read line: ($\bar{a} = \bar{b} < 0, \bar{a} = - |\bar{a}|$)
\begin{eqnarray*}
\frac{1}{g_{MR}} = 1 - \sqrt{ |\bar{a}|} \arctan( 1/ \sqrt{ |\bar{a}|}).
\end{eqnarray*}

\noindent
Weak pairing phase: ($\bar{a} = - |\bar{a}|, \bar{b} = - |\bar{b}| , |\bar{a}| > |\bar{b}|$)
\begin{eqnarray*}
\frac{1}{g} &=& \sqrt{ \frac{ 1 + |\bar{a}|}{ 1 + |\bar{b}|} } - 
\sqrt{|\bar{a}|} \;  E(\psi,m),  \\
2 \left( x - \frac{1}{2} + \frac{1}{2 g}\right) 
&=& \sqrt{|\bar{b}|} \;  F(\psi, m) 
\label{a27}
\end{eqnarray*}
where
\begin{eqnarray*}
\psi = \arcsin( 1/\sqrt{1 + |\bar{b}|}), \; \;
m^2 = 1 - \frac{ |\bar{b}|}{|\bar{a}|}. 
\end{eqnarray*}

\noindent
Read--Green line: ($\bar{a} = - |\bar{a}| < 0, \bar{b} = 0$)
\begin{eqnarray*}
\frac{1}{g_{RG}} =  \sqrt{1+  |\bar{a}|} - \sqrt{|\bar{a}|}.
\end{eqnarray*}

\subsubsection*{2D case}

\noindent
Weak coupling BCS: ($\bar{a}, \bar{b} = \bar{\epsilon}  \mp i \bar{\delta}$)
\begin{eqnarray*}
\frac{1}{g} &=& \sqrt{ 1+ R^2 - 2 \bar{\epsilon}} - R 
+ \bar{\epsilon} \; \log \left( \frac{
1 - \bar{\epsilon} +   \sqrt{ 1+ R^2 - 2 \bar{\epsilon}}}{
- \bar{\epsilon} + R}
\right), \\ 
2 \left( x - \frac{1}{2} + \frac{1}{2 g}\right) 
&=& R \;  \log \left( \frac{
1 - \bar{\epsilon} +   \sqrt{ 1+ R^2 - 2 \bar{\epsilon}}}{
- \bar{\epsilon} + R}
\right),
\end{eqnarray*}
where
\begin{eqnarray*}
R = \sqrt{ \bar{\epsilon}^2 + \bar{\delta}^2}.
\end{eqnarray*}

\noindent
Moore--Read line: ($\bar{a} = \bar{b} < 0, \bar{a} = - |\bar{a}|$)
\begin{eqnarray*}
\frac{1}{g_{MR}} = 1 -  |\bar{a}| \log(1 +  1/ |\bar{a}|).
\end{eqnarray*}

\noindent
Weak pairing phase: ($\bar{a} = - |\bar{a}|, \bar{b} = - |\bar{b}| , |\bar{a}| > |\bar{b}|$)

\begin{eqnarray*}
\frac{1}{g} &=& \sqrt{ (1 + |\bar{a}|)( 1 + |\bar{b}|) } - 
\sqrt{|\bar{a} \bar{b}|} + ( |\bar{a}| + |\bar{b}|) 
\log \left(
\frac{ \sqrt{ |\bar{a}|} + \sqrt{|\bar{b}|} }{ \sqrt{1+ |\bar{a}|} + \sqrt{1 + |\bar{b}|} } 
\right), \\ 
 \left( x - \frac{1}{2} + \frac{1}{2 g} \right) 
&=& \sqrt{| \bar{a} \bar{b}|} \; 
\log \left(
\frac{ 
\sqrt{1+ |\bar{a}|} + \sqrt{1 + |\bar{b}|} } 
{ \sqrt{ |\bar{a}|} + \sqrt{|\bar{b}|} }
\right).
\end{eqnarray*}

\noindent
Read--Green line: ($\bar{a} = - |\bar{a}| < 0, \bar{b} = 0$)
\begin{eqnarray*}
\frac{1}{g_{RG}} =  \sqrt{1+  |\bar{a}|} - |\bar{a}| 
\log \left( \frac{1}{\sqrt{ |\bar{a}|}} + \sqrt{ 1 + \frac{1}{|\bar{a}|}}\right).
\end{eqnarray*}


\subsection{Ground-state wavefunction: exact results versus mean-field results}
\label{gs-meanf}

%
%
%
%
%
We first recall from eq. (\ref{pipstates}) that for the $p+ip$ model 

\begin{eqnarray*}
C(y) = \sum_{\bk \in {\mathbf K}_+}
\frac{k_x - i k_y}{\bk^2 - y } 
c^\dagger_\bk  c^\dagger_ {-\bk },
\end{eqnarray*}
\noindent
while in 1D this simplifies to  
\begin{eqnarray*}
C(y) = \sum_{k>0}
\frac{k}{k^2 - y } 
c^\dagger_k  c^\dagger_ {-k }.
\end{eqnarray*}
In the latter formulae we extended the domain of the momentum variables
to include negative values of $k_x$ which is done to properly account for the antisymmetry of the wavefunction.
Now we will compare the exact wavefunction with the
mean-field one. Recalling (\ref{pgs}) we have 

\begin{eqnarray*}
| \psi  \rangle = \left( 
\sum_{\bk} {\mathfrak g}(\bk)  \;  c^\dagger_\bk  c^\dagger_ {-\bk } 
\right)^M |0\rangle
\end{eqnarray*}

\noindent
This expression can be viewed as an  average of the exact wavefunction 
(\ref{pipstates}) which involves a product of pair operators
characterized by the roots $y_m$. 
Read and Green have shown \cite{rg00} that in the limit where $\bk \rightarrow 0$
the BCS wavefunction behaves as

\begin{eqnarray*}
{\mathfrak g}(\bk) \sim 
\left \{
\begin{array}{ccc}
k_x - i k_y,  &   \mu < 0, &  ({\rm strong-coupling}), \\
1/( k_x + i k_y), & \mu > 0, &  ({\rm weak-coupling}). \\
\end{array} 
\right. 
\end{eqnarray*}

\noindent 
This behaviour can be compared with the 
wavefunction associated to the roots appearing 
in (\ref{pipstates}), 

\begin{eqnarray*}
{\mathfrak g}(\bk,y) = \frac{k_x - i k_y}{\bk^2 - y } \sim
\left \{
\begin{array}{cc}
 k_x - i k_y, & \,{\rm if} \; |\bk| << \sqrt{|y|}, \\
1/( k_x + i k_y), & {\rm if} \; |\bk| >> \sqrt{|y|} \\
\end{array} 
\right. 
\end{eqnarray*}

\noindent
which in real space leads to

\begin{eqnarray*}
\hat{{\mathfrak g}}(\br,y) = \int_{- \infty}^\infty \int_{- \infty}^\infty  d^2 \bk  \; 
e^{ i \bk \cdot \br} \;  {\mathfrak g}(\bk,y)
\sim \frac{ {\bar z}}{|z|} K_1( |z|\sqrt{-y} )
\end{eqnarray*}

\noindent
where $z\in\mathbb C$ and $K_1(z)$ is a Bessel function. 
Using the asymptotic results $K_1(z) \sim 1/z \; ( |z| <<1)$
and $K_1(z) \sim   e^{-z}\sqrt{{\pi}/{2 z}} \; (|z| >>1)$
one finds

\begin{eqnarray*}
\hat{{\mathfrak g}}(\br,y) \sim 
\left \{
\begin{array}{cc}
 { \bar{z}}{|z|^{-3/2}} e^{ - |z| \sqrt{-y} }, & |z| >> 1/\sqrt{|y|}, \\
1/z, &  |z| << 1/\sqrt{|y|}. \\
\end{array} 
\right. 
\end{eqnarray*}
Hence at short distances all the pairs behave as in the Moore-Read wavefunction and at large distances they
show the typical behaviour associated to 
localized BCS pairs with a correlation length

\begin{eqnarray}
\xi_y = \frac{1}{{\rm Re} ( \sqrt{- y} )}.
\label{e10}
\end{eqnarray}
If most of the roots 
$y$ lie near the origin then the overall behaviour
of the exact wavefunction will be well described by the
Moore-Read wavefunction.

In 1D the real-space wavefunction associated to a root $y$ is given by

\begin{eqnarray*}
{\mathfrak g}(x,y) = \int_{- \infty}^\infty   dk \; e^{i k x} 
\frac{ k}{k^2 - y} = i \pi {\rm sign}(x) \; e^{ - |x| \sqrt{-y}}
\end{eqnarray*}

\noindent
which decays exponentially with a correlation length given by
(\ref{e10}). 
 
\subsection{Wavefunction scalar product and one-point correlation functions}  \label{1cf}

The wavefunction scalar product and one-point correlation functions of the $p+ip$ model can be obtained directly as a limiting case of the results in Appendix \ref{scalarproduct}.
The elements of the $L$-operator (\ref{tildelax}) act locally, to leading order in $\gamma$,  as 
\begin{eqnarray*}
\tilde{L}^1_1(x)&\sim&I-2p\gamma\frac{x+x^{-1}}{x-x^{-1}}(I-N),    \\
\tilde{L}^1_2(x)&\sim& \frac{4p\gamma}{x-x^{-1}} b ,   \\
\tilde{L}^2_1(x)&\sim& \frac{4p\gamma}{x-x^{-1}} b^\dagger  , \\
\tilde{L}^2_2(x)&\sim& I-2p\gamma\frac{x+x^{-1}}{x-x^{-1}}N  . 
\end{eqnarray*}
Now using (\ref{texpand}) we  have to leading order
\begin{eqnarray*}
\tilde{T}^2_1(x)&\sim& 4p\gamma \sum_{j=1}^L \frac{1}{xz_k^{-1}-x^{-1}z_k} b_j^\dagger \\
&=& 4xp \gamma \left(\sum_{j=1}^L \frac{z_j b_j^\dagger}{x^2-z^2_j}\right).
\end{eqnarray*}
We define the operator (cf. (\ref{pipstates}))
\begin{eqnarray}
C(x)=\sum_{j=1}^L \frac{z_j b_j^\dagger}{x^2-z^2_j}
\label{creation}
\end{eqnarray}
such that the states of the system are of the form
\begin{eqnarray}
\left|\phi(Y)\right>=\prod_{j=1}^M C(y_j)\left|0\right>
\label{pstates}
\end{eqnarray}
where $Y=\{y_j\}$ are a solution set of the Bethe ansatz equations (\ref{pbae}). 
The formula for the scalar products of the states (\ref{pstates}) follows from taking the appropriate limit of (\ref{slavnov}). For the limit of the matrix (\ref{fmatrix}) the result is 
\begin{eqnarray*}
F(W,Y)_{ij}
&\sim&\frac{4p\gamma^2\prod_{l\neq i}^M(y_jw_l^{-1}-y_j^{-1}w_l)}{ (y_jw_i^{-1}-y_j^{-1}w_i)}
 \left(-2t-2p\sum_{r=1}^L\frac{y_j^2+z^2_r}{y_j^2-z_r^2}    +4p\sum_{m\ne i}^M\frac{y_j^2+w_m^2}{y_j^2-w_m^2}
  \right).
\end{eqnarray*}
Recalling the Bethe ansatz equations in the form (\ref{triggaudin}),
we can substitute into the above to eliminate $t$: 
\begin{eqnarray*}
F(W,Y)_{ij}
&\sim&4p\gamma^2y_j w_i\prod_{l\neq i}^M(y_jw_l^{-1}-y_j^{-1}w_l) \\
 &&~~~~~~~~~~~\times \left(4p\sum_{k=1}^L \frac{z_k^2}{(w_i^{2}-z_k^2)(y_j^2-z_k^2)}
  +8p\sum_{m\neq i}^M\frac{w^2_m}{(w_m^2-w_i^2)(y_j^2-w_m^2)}  
  \right).  
\end{eqnarray*}
To simplify the calculation we introduce the diagonal matrices 
\begin{eqnarray*}
\Gamma(W,Y)_{ij}&=&\delta_{ij}\left(\frac{w_j}{y_j}\right)\frac{y_jw_j^{-1}-y_j^{-1}w_j}{\prod_{k=1}^M(y_jw_k^{-1}-y_j^{-1}w_k)},  \\
\Pi(W,Y)_{ij}&=& \delta_{ij} \frac{1}{16p^2\gamma^2w^2_j}   
\end{eqnarray*}
such that 
\begin{eqnarray*}
\det(\Pi(W,Y))&=& \left(\frac{1}{16p^2\gamma^2}\right)^M\left(\prod_{j=1}^M\ w_j^{-2} \right),  \\
\det(\Gamma(W,Y))&=&  \prod_{j=1}^M\left(\frac{w_j}{y_j}\right)\Upsilon(W,Y) 
\end{eqnarray*}
 where
 \begin{eqnarray*}
\Upsilon(W,Y)&=&\frac{\prod_{j=1}^M (y_jw_j^{-1}-y_j^{-1}w_j)}{\prod_{j=1}^M\prod_{k=1}^M(y_jw_k^{-1}-y_j^{-1}w_k)}.
\end{eqnarray*}
Note that 
\begin{eqnarray*}
\Upsilon(Y,Y)&=& \mathcal{C}(Y,Y)
\end{eqnarray*}
with $\mathcal{C}(W,Y)$ given by (\ref{c}). 
Now defining 
\begin{eqnarray*}
G(W,Y)&=&\lim_{\gamma\rightarrow 0}\Pi(W,Y) F(W,Y) \Gamma(W,Y) 
\end{eqnarray*}
leads to
\begin{eqnarray*}
G(W,Y)_{ij}
&=&\frac{(y^2_j-w^2_j)}{(y^2_j-w^2_i)}
 \left(\sum_{k=1}^L \frac{z_k^2}{(w_i^{2}-z_k^2)(y_j^2-z_k^2)}  
 +2\sum_{m\neq i}^M\frac{w^2_m}{(w_m^2-w_i^2)(y_j^2-w_m^2)}  
  \right).   
  \end{eqnarray*}  
We will also need  
\begin{eqnarray*}
\frac{\mathcal{C}^2(W,Y)}{\Upsilon^2(W,Y)}
&=& \frac{\prod^M_{j\neq k}(y_jw_k^{-1}-y_j^{-1}w_k)}
{\prod^M_{m>l}(y_my_l^{-1}-y_m^{-1}y_l)(w_lw_m^{-1}-w_l^{-1}w_m)} \\
&&~~~~~~~~~~~~~\times
\frac{\prod^M_{a\neq b}(y_aw_b^{-1}-y_a^{-1}w_b)}
{\prod^M_{p>q}(y_py_q^{-1}-y_p^{-1}y_q)(w_qw_p^{-1}-w_q^{-1}w_p)}\\
&=&\prod_{k=1}^M\prod^M_{j\neq k}\frac{(y_j^2-w_k^2)^2}{(y_j^2-y_k^2)(w_j^2-w_k^2)}.
\end{eqnarray*}  
Keeping in mind that $C(x)$ differs from the leading term in the expansion of $\tilde{T}^2_1(x)$ by a scale factor, we now have 
\begin{eqnarray*}
\langle\phi(W)|\phi(Y)\rangle=\frac{\prod_{k=1}^M \prod_{j\neq k}^M(y^2_j-w^2_k)}{\prod_{j<k}^M(y^2_j-y^2_k)(w^2_k-w^2_j)} \det(G(W,Y)).
\end{eqnarray*}
Finally we can make explicit the square of the normalised wavefunction scalar product  $\mathcal{F}(W,Y)$ (equivalent to the square of the fidelity) between two states as a function of the Bethe roots, but independent of the coupling $G$. The result is 
\begin{eqnarray*}
\mathcal{F}(W,Y)^2&=&
\frac{\langle\phi(W)|\phi(Y)\rangle^2}{\langle\phi(W)|\phi(W)\rangle\langle\phi(Y)|\phi(Y)\rangle} \\
&=&\frac{\mathcal{C}^2(W,Y)}{\Upsilon^2(W,Y)} \frac{\det(G(W,Y))^2}{\det(G(W,W))\det(G(Y,Y))}
\\
&=&\prod_{k=1}^M\prod_{j\neq k}^M\frac{(y_j^2-w_k^2)^2}{(y_j^2-y_k^2)(w_j^2-w_k^2)}
\frac{\det(G(W,Y))^2}{\det(G(W,W))\det(G(Y,Y))}
\end{eqnarray*}
 and we note
\begin{eqnarray}
{G}(Y,Y)_{ii}&=& \sum_{k=1}^L \frac{z_k^2}{(y^2_i-z_k^2)^2}  
 -2\sum_{m\neq i}^M\frac{y^2_m}{(y^2_i-y^2_m)^2},  
     \nonumber \\
{G}(Y,Y)_{ij}&=& \frac{2y^2_j}{(y^2_j-y^2_i)^2}, \qquad\quad i\neq j.  
\label{norm}
\end{eqnarray}   

In a similar manner we obtain the formula for the one-point correlation functions from the limit of (\ref{averagen}). The result is
\begin{eqnarray}
\left<N_m\right> =1-\frac{\det({G}(Y,Y)-z_m{\mathcal{W}}_m(Y))}{\det({G}(Y,Y))}
\label{p1cf}
\end{eqnarray}
with $G(Y,Y)$ given by (\ref{norm}) and 
\begin{eqnarray}
\left({\mathcal{W}}_m(Y)\right)_{ij}=\frac{z_m}{(y^2_i-z_m^2)^2}.
\label{W}
\end{eqnarray}

\subsection{Two-point correlation functions} \label{2cf}

Finally we turn our attention to the more technically demanding problem of calculating two-point correlation functions. We will derive formulae for both the off-diagonal case $\langle b^\dagger_m b_n \rangle$ and the diagonal case $\langle N_m N_n\rangle$. For the calculation of these results we extend the methods used in \cite{fcc08,o03,zlmg02} which were developed for the $s$-wave model.    

First we consider the off-diagonal case. For any fixed $n$ we may express the state $|\phi(Y)\rangle$ as 
\begin{eqnarray}
&&|\phi(Y)\rangle=\prod_{\beta=1}^M C(y_\beta)|0\rangle
=\prod_{\beta=1}^M\left(\tilde{C}_n(y_\beta)+A^\beta_n b^\dag_n\right)|0\rangle,
\end{eqnarray}
where
$$ \tilde{C}_n(y_\beta)=\sum_{l\neq n}^M {z_l b^\dag_l\over y_\beta^2-z_l^2}, \quad\quad
 A^\beta_n=\frac{z_n}{y_\beta^2-z_n^2}. $$
Keeping in mind that $\left(b^\dag_n\right)^2=0,\, \, b_n|0\rangle=0$,
we find  
 \begin{eqnarray}
b_n|\phi(Y)\rangle
&=& \sum_{\beta=1}^M A^\beta(n)\prod_{\alpha\neq\beta}^M \tilde{C}_n(y_\alpha)|0\rangle\nonumber\\
&=&\sum_{\beta=1}^M A^\beta_n\prod_{\alpha\neq\beta }^M\left(C(y_\alpha)-A^\alpha_n b^\dag_n\right)|0\rangle\nonumber\\
&=&\sum_{\beta=1}^M A^\beta_n\prod_{\alpha\neq \beta}^M C(y_\alpha)|0\rangle
   -\sum_{\beta=1}^M\sum_{\alpha\neq\beta}^M A^\alpha_n A^\beta_n b^\dag_n\prod_{\gamma\neq \alpha,\beta}^M C(y_\gamma)|0\rangle.
   \label{eq:b+}
\end{eqnarray}
With the help of (\ref{eq:b+}) we have that for $m\neq n$ the unnornormalised two-point off-diagonal correlation functions are  
\begin{eqnarray}
\langle\phi(Y) |b_m^\dagger b_n|\phi(Y)\rangle 
&=&\sum_{\beta=1}^M A^\beta_n \langle \phi(Y)|b^\dag_m \prod_{\alpha\neq \beta}^M C(y_\alpha)|0\rangle
   \nonumber\\ &&\mbox{}\quad
 -\sum_{\beta=1}^M \sum_{\alpha\neq\beta}^M A^\alpha_n A^\beta_n
  \langle \phi(Y)|b^\dag_m b^\dag_n\prod_{\gamma\neq \alpha,\beta}^M C(y_\gamma)|0\rangle 
\nonumber \\
&=&\sum_{\beta=1}^M A^\beta_n \langle \phi(Y)|b^\dag_m \prod_{\alpha\neq \beta}^M C(y_\alpha)|0\rangle
  \nonumber\\ &&\mbox{}\quad
-2 \sum_{\beta=1}^M\sum_{\alpha<\beta}^M A^\alpha_n A^\beta_n
 \langle \phi(Y)|b^\dag_m b^\dag_n\prod_{\gamma\neq \alpha,\beta}^M C(y_\gamma)|0\rangle   \label{de:CFmn}
\end{eqnarray}
where in the last step we use the fact that the terms in the double sum are symmetric with respect to interchange of the labels $\alpha$ and $\beta$.
Now the two-point correlation function has been reduced to a sum of one-point form factors of $b^\dagger_m$ and a double sum over  
the two-point form factors of $b^\dag_m b^\dag_n$. 
Recalling the definition (\ref{creation}), for the $p+ip$ model the inverse problem is easily solved as
\begin{eqnarray*}
 b^\dag_m=\lim_{y\rightarrow z_m}\frac{ y^2-z_m^2}{ z_m}C(y). 
 \end{eqnarray*}
Using this expression and the wavefunction scalar product formula we obtain the matrix elements of both $b^\dagger_m$ and $b^\dagger_m b_n$. 
For the one-point case the result is  
\begin{eqnarray*}
\langle \phi(Y)|b^\dag_m\prod_{\alpha_\neq\beta}^M C(y_\alpha)|0\rangle 
&=&\lim_{v\rightarrow z_m}\frac{v^2-z_m^2}{z_m}
   \langle \phi(Y)|C(v)\prod_{\alpha\neq\beta}^MC(y_\alpha)|0\rangle \nonumber\\
&=&   \mathcal{K}^\beta_m
 \mbox{det}({\mathcal T}^{\beta}_m)\end{eqnarray*}
where $\mathcal{T}^{\beta}_m$ is the $M\times M$ matrix with elements 
\begin{eqnarray*}
\left({\mathcal{T}}^\beta_m\right)_{ij}&=&{G}_{ij},
    \quad\quad\qquad\qquad j\neq \beta,\\
\left({\mathcal{T}}^\beta_m\right)_{i\beta}&=&\left(\mathcal{W}_m\right)_{i\beta}, 
\end{eqnarray*}
$\mathcal{K}^\beta_m=(y_\beta^2-z_m^2)$, 
$G\equiv G(Y,Y)$ is given by (\ref{norm}), and $\mathcal{W}_m\equiv \mathcal{W}_m(Y)$ is given by (\ref{W}).  
(To keep the notation compact, we hereafter omit the $Y$ dependence on all such matrices.)
For the two-point form factors  we have
\begin{eqnarray}
&&\langle \phi(Y)|b^\dag_m b^\dag_n\prod_{\gamma\neq \alpha,\beta}^M C(y_\gamma)|0\rangle
 \nonumber\\ &&
 =\lim_{u\rightarrow z_m}\lim_{v\rightarrow z_n}\frac{(u^2-z_m^2)(v^2-z_n^2)}{z_mz_n} 
 \langle \phi(Y)|C(u)C(v)\prod^M_{\gamma\ne \alpha,\beta}C(y_\gamma)|0\rangle
  \nonumber\\ &&
=  {K}^{\alpha\beta}_{mn}
 \mbox{det}({T}^{\alpha\beta}_{mn}) 
\end{eqnarray}
where $T^{\alpha\beta}_{mn}$ is the $M\times M$ matrix with the elements
\begin{eqnarray*}
\left(T^{\alpha\beta}_{mn}\right)_{ij}&=&{G}_{ij}, \qquad\qquad\qquad j\neq \alpha,\beta,\\
\left({T}^{\alpha\beta}_{mn}\right)_{i\alpha}&=& \left(\mathcal{W}_n\right)_{i\alpha} 
  ,\nonumber\\
\left({T}^{\alpha\beta}_{mn}\right)_{i\beta}&=& \left( \mathcal{W}_m\right)_{i\beta}
  ,\nonumber\\
\end{eqnarray*}
and 
\begin{eqnarray*}
K^{\alpha\beta}_{mn}=\frac{(y_\alpha^2-z_m^2)(y_\alpha^2-z_n^2)(y_\beta^2-z_m^2)(y_\beta^2-z_n^2)}{(y_\alpha^2-y_\beta^2)(z_m^2-z_n^2)}.
\end{eqnarray*}
This leads to 
\begin{eqnarray*}
\langle\phi(Y)|b_m^\dagger b_n|\phi(Y)\rangle&=&\sum_{\beta=1}^M \mathcal{J}^\beta_{mn} \mbox{det}(\mathcal{T}^\beta_m)) -\sum_{\beta=1}^M\sum_{\alpha<\beta}^M  \mathcal{J}^\beta_{mn} J^{\alpha\beta}_{mn} \mbox{det}(T^{\alpha\beta}_{mn})
\end{eqnarray*}
where
\begin{eqnarray*}
\mathcal{J}^\beta_{mn}&=& A^\beta_n\mathcal{K}^\beta_m \\
&=& \frac{z_n(y_\beta^2-z_m^2)}{y^2_\beta-z_n^2} \\
J^{\alpha\beta}_{mn}&=& \frac{2A^\alpha_n K^{\alpha\beta}_{mn}}{\mathcal{K}^\beta_m} \\
&=& \frac{2z_n(y_\alpha^2-z_m^2)(y_\beta^2-z_n^2)}{(y_\alpha^2-y_\beta^2)(z_m^2-z_n^2)}.
\end{eqnarray*} 
We can conveniently express the determinants of $\mathcal{T}^\beta_m$ and $T^{\alpha\beta}_{mn}$ in a vector notation. Defining the $M$-dimensional vectors $\vec{G}_j,\,j=1,...,M$ and $\vec{\mathcal{W}}_m,\,m=1,...,M$ to have entries 
\begin{eqnarray*}
(\vec{G}_j)_i= G_{ij}  \\
(\vec{\mathcal{W}}_m)_i= \left(\mathcal{W}_m\right)_{ij} 
\end{eqnarray*}
we can write 
\begin{eqnarray*}
\mbox{det}(\mathcal{T}^{\beta}_m)&=&\left|\vec{G}_{1},...,\vec{G}_{\beta-1},\vec{\mathcal{W}}_m,\vec{G}_{\beta+1},...,  \vec{G}_M\right|, \\
\mbox{det}(T^{\alpha\beta}_{mn})&=&\left|\vec{G}_{1},...,\vec{G}_{\alpha-1},\vec{\mathcal{W}}_n,
\vec{G}_{\alpha+1},...\vec{G}_{\beta-1},\vec{\mathcal{W}}_m,\vec{G}_{\beta+1},...,  \vec{G}_M\right|.
\end{eqnarray*}
Using standard properties of determinants we have 
\begin{eqnarray*}
\mbox{det}(\mathcal{T}^{\beta}_{m})&=&\left|\vec{G}_{1}-\frac{J^{1\beta}_{mn}}{J^{2\beta}_{mn}}\vec{G}_2,...,\vec{G}_{\beta-2}-\frac{J^{(\beta-2)\beta}_{mn}}{J^{(\beta-1)\beta}_{mn}}\vec{G}_{\beta-1},\vec{G}_{\beta-1,}\vec{\mathcal{W}}_m,\vec{G}_{\beta+1},... , \vec{G}_M\right|, \nonumber \\
\sum_{\alpha<\beta}^MJ^{\alpha\beta}_{mn}\mbox{det}(T^{\alpha\beta}_{mn}) 
&=&
J^{(\beta-1)\beta}_{mn}\left|\vec{G}_{1}-\frac{J^{1\beta}_{mn}}{J^{2\beta}_{mn}}\vec{G}_2,...,\vec{G}_{\beta-2}-\frac{J^{(\beta-2)\beta}_{mn}}{J^{^(\beta-1)\beta}_{mn}}\vec{G}_{\beta-1},\vec{\mathcal{W}}_n,\vec{\mathcal{W}}_m,\vec{G}_{\beta+1},...,  \vec{G}_M\right|.
\end{eqnarray*}
Combining these  we have a single determinant expression 
\begin{eqnarray*}
&&\mbox{det}(\mathcal{T}^{\beta}_m) - \sum_{\alpha<\beta}^MJ^{\alpha\beta}_{mn}\mbox{det}(T^{\alpha\beta}_{mn}) \\
&&\qquad=
\left|\vec{G}_{1}-\frac{J^{1\beta}_{mn}}{J^{2\beta}_{mn}}\vec{G}_2,...,\vec{G}_{\beta-2}-\frac{J^{(\beta-2)\beta}_{mn}}{J^{^(\beta-1)\beta}_{mn}}\vec{G}_{\beta-1},\vec{G}_{\beta-1}-J^{(\beta-1)\beta}_{mn}\vec{\mathcal{W}}_n,\vec{\mathcal{W}}_m,\vec{G}_{\beta+1},...,  \vec{G}_M\right|.
\end{eqnarray*}
This allows us to express the off-diagonal two-point correlation functions as 
\begin{eqnarray*}
\langle b^\dagger_m b_n \rangle &=& \frac{\langle \phi(Y)|b^\dagger_m b_n|\phi(Y) \rangle}
{\langle \phi(Y)|\phi(Y)\rangle} \nonumber \\
&=&\frac{1}{\mbox{det}( G)}\sum_{\beta=1}^M  \mathcal{J}^\beta_{mn} \mbox{det}(\mathcal{D}^\beta_{mn})
\end{eqnarray*}
where 
\begin{eqnarray*}
\left(\mathcal{D}^\beta_{mn}\right)_{ij}&=& G_{ij}-\frac{(y^2_{j+1}-y_\beta^2)(y_j^2-z_m^2)}{(y_j^2-y_\beta^2)(y^2_{j+1}-z_m^2)}G_{i(j+1)}, \qquad \qquad j< \beta-1,  \\
\left(\mathcal{D}^\beta_{mn}\right)_{i(\beta-1)}  &=&G_{i(\beta-1)} -\frac{2z_n^2(y_{\beta-1}^2-z_m^2)(y_\beta^2-z_n^2)}{(y_{\beta-1}^2-y^2_\beta)(z_m^2-z_n^2)(y_i^2-z_n^2)^2},  \\
\left(\mathcal{D}^\beta_{mn}\right)_{i\beta} &=&  \frac{z_m}{(y_i^2-z_m^2)^2},\\
\left(\mathcal{D}^\beta_{mn}\right)_{ij} &=& G_{ij} , \qquad \qquad j>\beta.
\end{eqnarray*}

Next we consider the computation of the diagonal two-point correlation functions. 
To facilitate this,  we first note the 
commutation relation 
\begin{eqnarray}
\left[N_m,\,C(v)\right]=\frac{z_m}{v^2-z_m^2}b_m^\dagger.  
\label{dcomm}
\end{eqnarray} 
With the help of the commutation relation (\ref{dcomm}), and the fact that each $N_m$ vanishes on the vacuum, we can write for $m\neq n$
\begin{eqnarray*}
 \langle \phi(Y)| N_m N_n |\phi(Y)\rangle&=& \langle \phi(Y)| N_m N_n \prod_{\beta=1}^M C(y_\beta)|0\rangle  \\
 &=& \sum_{\beta=1}^M \frac{z_m}{y_\beta^2-z_m^2}\langle \phi(Y) |N_n b_m^\dagger \prod_{\alpha\neq\beta}^M C(y_\alpha)|0\rangle  \\
&=& \sum_{\beta=1}^M \frac{z_m}{y_\beta^2-z_m^2}\sum_{\alpha\neq\beta}^M\frac{z_n}{y_\alpha^2-z_n^2}\langle \phi(Y) |b^\dagger_m b_n^\dagger \prod_{\gamma\neq\alpha,\beta}^M C(y_\alpha)|0\rangle \\
&=& \sum_{\beta=1}^M \sum_{\alpha\neq\beta}^M\frac{z_m}{y_\beta^2-z_m^2}  \frac{z_n}{y_\alpha^2-z_n^2} K^{\alpha\beta}_{mn}
\mbox{det}(T^{\alpha\beta}_{mn}) \\ 
&=& \frac{z_m}{2} \sum_{\beta=1}^M \sum_{\alpha\neq\beta}^M J^{\alpha\beta}_{mn}
\mbox{det}(T^{\alpha\beta}_{mn}) .   
\end{eqnarray*}
Using similar techniques as before, we can reduce the above to a sum of $M$ determinants. Doing this leads to the final result 
\begin{eqnarray*}
\langle N_m N_n \rangle&=&  
\frac{\langle\phi(Y)|N_m N_n |\phi(Y)\rangle}{\langle\phi(Y)|\phi(Y)\rangle } \\
&=&\frac{z_m}{2\mbox{det}(G)}\left(\sum_{\beta=1}^{M-1} J^{M\beta}_{mn} \mbox{det}({D}^\beta_{mn})+ J_{mn}^{(M-1)M}\mbox{det}(E_{mn})\right)
\label{pdcf}
\end{eqnarray*}
where 
\begin{eqnarray*}
\left({D}^\beta_{mn}\right)_{ij}&=& G_{ij}-\frac{(y^2_{j+1}-y_\beta^2)(y_j^2-z_m^2)}{(y_j^2-y_\beta^2)(y^2_{j+1}-z_m^2)}G_{i(j+1)}, \qquad \qquad j\neq \beta-1,\beta,M,  \\
\left({D}^\beta_{mn}\right)_{i(\beta-1)}&=& G_{i(\beta-1)}-
\frac{(y^2_{\beta+1}-y_\beta^2)(y_{\beta-1}^2-z_m^2)}{(y_{\beta-1}^2-y_\beta^2)(y^2_{\beta+1}-z_m^2)}G_{i(\beta+1)},   \\
\left({D}^\beta_{mn}\right)_{i\beta} &=&  \frac{z_m}{(y_i^2-z_m^2)^2}, \\
\left({D}^\beta_{mn}\right)_{iM} &=& \frac{z_n}{(y_i^2-z_n^2)^2}, \\
\left({E}_{mn}\right)_{ij}&=& G_{ij}
-\frac{(y^2_{j+1}-y_M^2)(y_j^2-z_m^2)}{(y_j^2-y_M^2)(y_{j+1}^2-z_m^2)}G_{i(j+1)}, \qquad \qquad j\neq M-1,M,  \\
\left({E}_{mn}\right)_{i(M-1)} &=&  \frac{z_n}{(y_i^2-z_n^2)^2}, \\
\left({E}_{mn}\right)_{iM} &=& \frac{z_m}{(y_i^2-z_m^2)^2}.
\end{eqnarray*}

\end{document}